\documentclass[useAMS,usenatbib]{mnras}

\usepackage{graphicx}
\usepackage[T1]{fontenc}
\usepackage{aecompl}
\usepackage[]{amsmath}
\usepackage{color}
\usepackage{xspace}

\newcommand{\msun}{\mathrm{M}_\odot}
\graphicspath{{figures/}}

\newcommand{\mytab}{\begin{table}[htb]}
\newcommand{\myfig}{\begin{figure}[htbp]}

\newcommand{\sextractor}{\textsc{Sextractor}\xspace}

\def\lsim{ \lower .75ex \hbox{$\sim$} \llap{\raise .27ex \hbox{$<$}} }

\graphicspath{{figures/}}

\title[]{The stellar halo of isolated central galaxies in the Hyper Suprime-Cam imaging survey
}

\author[Wang et al.]{Wenting Wang$^{1}$\thanks{wenting.wang@ipmu.jp, bilinxing.wenting@gmail.com}, 
Jiaxin Han$^{2,1}$, Alessandro Sonnenfeld$^{3,1}$, Naoki Yasuda$^{1}$, \newauthor 
Xiangchong Li$^{1}$, Yipeng Jing$^{2}$, Surhud More$^{4,1}$, Paul A. Price$^{5}$, Robert Lupton$^{5}$, \newauthor
Eli Rykoff$^{6}$, David V. Stark$^{1}$, Ting-Wen Lan$^{1}$, Masahiro Takada$^{1}$, Song Huang$^{7}$,\newauthor
Wentao Luo$^{1}$, Neta A. Bahcall$^{5}$, Yutaka Komiyama$^{8,9}$ \\
{}$^{1}$ Kavli IPMU (WPI), UTIAS, The University of Tokyo, Kashiwa, Chiba 277-8583, Japan \\
{}$^{2}$ Department of Astronomy, Shanghai Jiao Tong University, Shanghai 200240, China \\
{}$^{3}$ Leiden Observatory, Leiden University, Niels Bohrweg 2, 2333 CA Leiden, the Netherlands \\
{}$^{4}$ The Inter-University Centre for Astronomy and Astrophysics, Post bag 4, Ganeshkhind, Pune 411007, India \\
{}$^{5}$ Department of Astrophysical Sciences, Princeton University, 4 Ivy Lane, Princeton, NJ08544, USA\\
{}$^{6}$ Stanford University, Palo Alto, CA94305, USA\\
{}$^{7}$ Department of Astronomy and Astrophysics, University of California Santa Cruz, 1156 High St., Santa Cruz, CA 95064, USA\\
{}$^{8}$ National Astronomical Observatory of Japan, 2-21-1 Osawa, Mitaka, Tokyo 181-8588, Japan\\
{}$^{9}$ Graduate University for Advanced Studies (SOKENDAI), 2-21-1 Osawa, Mitaka, Tokyo 181-8588, Japan
  }

\begin{document}

\maketitle

\begin{abstract}
We study the faint stellar halo of isolated central galaxies, by stacking galaxy images in the HSC survey and 
accounting for the residual sky background sampled with random points. The surface brightness profiles in HSC 
$r$-band are measured for a wide range of galaxy stellar masses ($9.2<\log_{10}M_\ast/M_\odot<11.4$) and 
out to 120~kpc. Failing to account for the stellar halo below the noise level of individual images will lead to 
underestimates of the total luminosity by $\leq 15\%$. Splitting galaxies according to the concentration parameter 
of their light distributions, we find that the surface brightness profiles of low concentration galaxies drop 
faster between 20 and 100~kpc than those of high concentration galaxies. Albeit the large galaxy-to-galaxy scatter, 
we find a strong self-similarity of the stellar halo profiles. They show unified forms once the projected distance 
is scaled by the halo virial radius. The colour of galaxies is redder in the centre and bluer outside, with high 
concentration galaxies having redder and more flattened colour profiles. There are indications of a colour minimum, 
beyond which the colour of the outer stellar halo turns red again. This colour minimum, however, is very sensitive 
to the completeness in masking satellite galaxies. We also examine the effect of the extended PSF in the measurement 
of the stellar halo, which is particularly important for low mass or low concentration galaxies. The 
PSF-corrected surface brightness profile can be measured down to $\sim$31~$\mathrm{mag}/\mathrm{arcsec}^2$ at 
3-$\sigma$ significance. PSF also slightly flattens the measured colour profiles.
\end{abstract}   


\section{Introduction}
\label{sec:intro}
In the structure formation paradigm of $\Lambda$CDM, galaxies form by the 
cooling and condensation of gas at centres of an evolving population of dark matter 
haloes \citep{1978MNRAS.183..341W}. Dark matter haloes grow in mass and size through 
both smooth accretion of diffuse matter and from mergers with other haloes spanning 
a very wide range in mass \citep[e.g.][]{2011MNRAS.413.1373W}. Smaller haloes having 
their own central galaxies fall into larger haloes and become ``subhaloes'' and 
``satellites'' of the galaxy at the centre of the dominant host halo. Orbiting 
around the central galaxy and undergoing tidal stripping, these satellites and 
subhaloes lose their mass. Stripped stars form stellar streams, which then gradually 
mix in phasespace, losing their own binding energy and sinking to the centre due to 
dynamical frictions. These stars form the diffuse light or the stellar halo around 
the central galaxy \citep[e.g.][]{2005ApJ...635..931B,2010MNRAS.406..744C}. In the 
end, satellite galaxies and stripped material from these satellites merge with the 
central galaxy and contribute to its growth. 

Theoretical studies on the formation of extended stellar haloes involve a few 
different approaches including analytical models \citep[e.g.][]{2007ApJ...666...20P}, 
numerical simulations \citep[e.g.][]{2010ApJ...725.2312O,2012MNRAS.425..641L,
2014MNRAS.444..237P,2016MNRAS.458.2371R,2018arXiv180810454K} and semi-analytical approaches 
of particle painting/tagging method \citep[e.g.][]{2013MNRAS.434.3348C,2015MNRAS.451.2703C}. 
In these studies, it is demonstrated that galaxy formation involves two phases, an early 
rapid formation of ``in-situ'' stars through gas cooling and a later phase of mass 
growth through accretion of smaller satellite galaxies. Accreted stellar material 
typically lies in the outskirts of galaxies and are more metal poor than ``in-situ'' 
stars. The fraction of accreted stellar mass with respect to the total mass of galaxies 
is higher for more massive galaxies and for elliptical galaxies. 

Observationally, with the advent of large telescopes and deep imaging, galaxy stellar 
haloes and the connection to galaxy mergers in both our Milky Way and nearby individual 
galaxies have already been detected and studied \citep[e.g.][]{1980ApJ...237..303S,
1983ApJ...274..534M,1992AJ....104.1039S,2005ApJ...631L..41M,2009AJ....138.1417T,
2010AJ....140..962M,2014ApJ...782L..24V,2018ApJ...861...81G,2018JKAS...51...73A}. 
Tidal structures have been observed in individual galaxies, which supports the above 
structure formation paradigm and theoretical studies. In particular, for our Milky way 
and very nearby disk and lenticular galaxies, the stellar population can be resolved 
and the stellar halo can be studied through star counts \citep[e.g.][]{2008ApJ...680..295B,
2013ApJ...766..106M,2014ApJ...780..128I,2015ApJ...800...13P,2015MNRAS.454.3613S}.

The intensity or surface brightness, $I$, of extended objects drops with distance, in 
a relationship with redshift, $z$, as $I\propto(1+z)^{-4}$. Hence for more distant galaxies, 
their faint stellar haloes can be only a few percent or even less than the sky background. 
This makes it relatively difficult to study distant stellar haloes individually. However,
deep imaging data still makes it possible to look at the outer stellar halo, tidal 
structures and mass accretion through cosmic time for massive individual galaxies up 
to redshift $z\sim1$ \citep[e.g.][]{2017MNRAS.466.4888B}.

Alternatively, images of a large sample of galaxies with similar properties can be stacked 
together to achieve the average extended light distribution of galaxies and their stellar haloes 
\citep[e.g.][]{2004MNRAS.347..556Z,2005MNRAS.358..949Z,2011ApJ...731...89T,2014MNRAS.443.1433D}. 
Although stacking smooths out delicate structures and the fine shape of galaxies, it is a powerful 
approach that enables studying the averaged smooth light distribution of the faint stellar haloes 
for more distant galaxies. The stacking approach helps to increase the signal-to-noise level 
compared with individual images and the stacked surface brightness profiles can be measured for 
galaxies spanning a wide range in luminosity or stellar mass, covering those smaller than the 
Milky Way ($\log_{10}M_\ast/M_\odot\sim10$) to massive cD galaxies of groups and clusters. So 
far, existing observations are generally consistent with theoretical studies.

There are claims that the surface brightness of the faint stellar halo can be measured down 
to $\sim$33~$\mathrm{mag}/\mathrm{arcsec}^2$ \citep[e.g.][]{2016ApJ...823..123T}, based on the long 
cumulative exposure on large telescopes. However, even by stacking images to push beyond the noise 
level of individual images, robust measurements of the surface brightness fainter than 31~$\mathrm{mag}/\mathrm{arcsec}^2$ 
is challenging, given a series of possible systematics in the data, including those arising from 
the residual sky background and incomplete masking of companion sources. 

We are now in the era of big data. More and more deep imaging data are being or will be 
collected through collaborative surveys using advanced cameras on large telescopes, such as the 
Hyper Suprime-Cam Subaru Strategic Program Survey \citep[HSC-SSP or HSC;][]{2018PASJ...70S...4A},
the Dark Energy Survey (DES), the Dark Energy Camera Legacy Survey (DECaLS), and the planned Large 
Synoptic Survey Telescope \citep[LSST;][]{2008arXiv0805.2366I} survey. These new imaging surveys 
can help to measure very faint diffuse structures, but the measurement should be based on the 
prerequisites of properly controlling different sources of systematics and uncertainties. In this 
paper, we are going to look at the stellar haloes of galaxies using data from the ongoing and deep 
HSC imaging survey. The coadd images of the HSC survey based on multiple exposures have effectively 
achieved a much deeper depth than many other contemporary surveys, including the Sloan Digital Sky 
Survey (SDSS), DECaLS and DES. 

The deep imaging data of HSC has already been proved powerful in studying properties of faint individual 
objects. It has been used to detect new dwarf galaxies in the local universe~\citep[e.g.][]{2018ApJ...866..112G} 
and satellite galaxies in our Milky Way \citep[e.g.][]{2018PASJ...70S..18H}, to study low surface brightness 
galaxies \citep[e.g.][]{2018ApJ...857..104G} and to investigate the stellar density profile of our Milky Way 
out to large distances \citep[e.g.][]{2018PASJ...70...69F}. The surface brightness profiles of individual 
brightest cluster galaxies (BCGs) in galaxy clusters ($\log_{10}M_\ast/\msun>11.4$) have been measured up to 
$\sim$100~kpc at $0.3<z<0.5$ \citep{2018MNRAS.475.3348H} using HSC. Tidal features such as stellar shells and 
streams can be detected and measured for galaxies from $z\sim0.05$ up to $z\sim 0.45$ \citep{2018ApJ...866..103K} 
using this data.

Instead of looking at individual galaxies, we aim to stack the HSC data around a sample of 
isolated central galaxies which are brighter than all the other local companions at $z\sim0.1$. 
Tested against a mock galaxy catalogue based on cosmological simulations, these galaxies 
are mostly central galaxies of dark matter haloes. They span a wide range in stellar 
mass, which enables us to investigate the faint stellar halo of smaller galaxies and 
out to larger radial scales. This study can be further extended with data from the future 
LSST survey which promises an even deeper depth and a much larger sky coverage.

For observational results, we adopt as our fiducial cosmological model the first-year 
Planck cosmology \citep{2014A&A...571A..16P}, with present values of the Hubble constant 
$H_0=67.3\mathrm{km s^{-1}/Mpc}$, the matter density $\Omega_m=0.315$ and the cosmological 
constant $\Omega_\Lambda=0.685$.

\section{data}
\label{sec:data}
\subsection{Isolated central galaxies}
\label{sec:isogal}

\begin{figure} 
\includegraphics[width=0.49\textwidth]{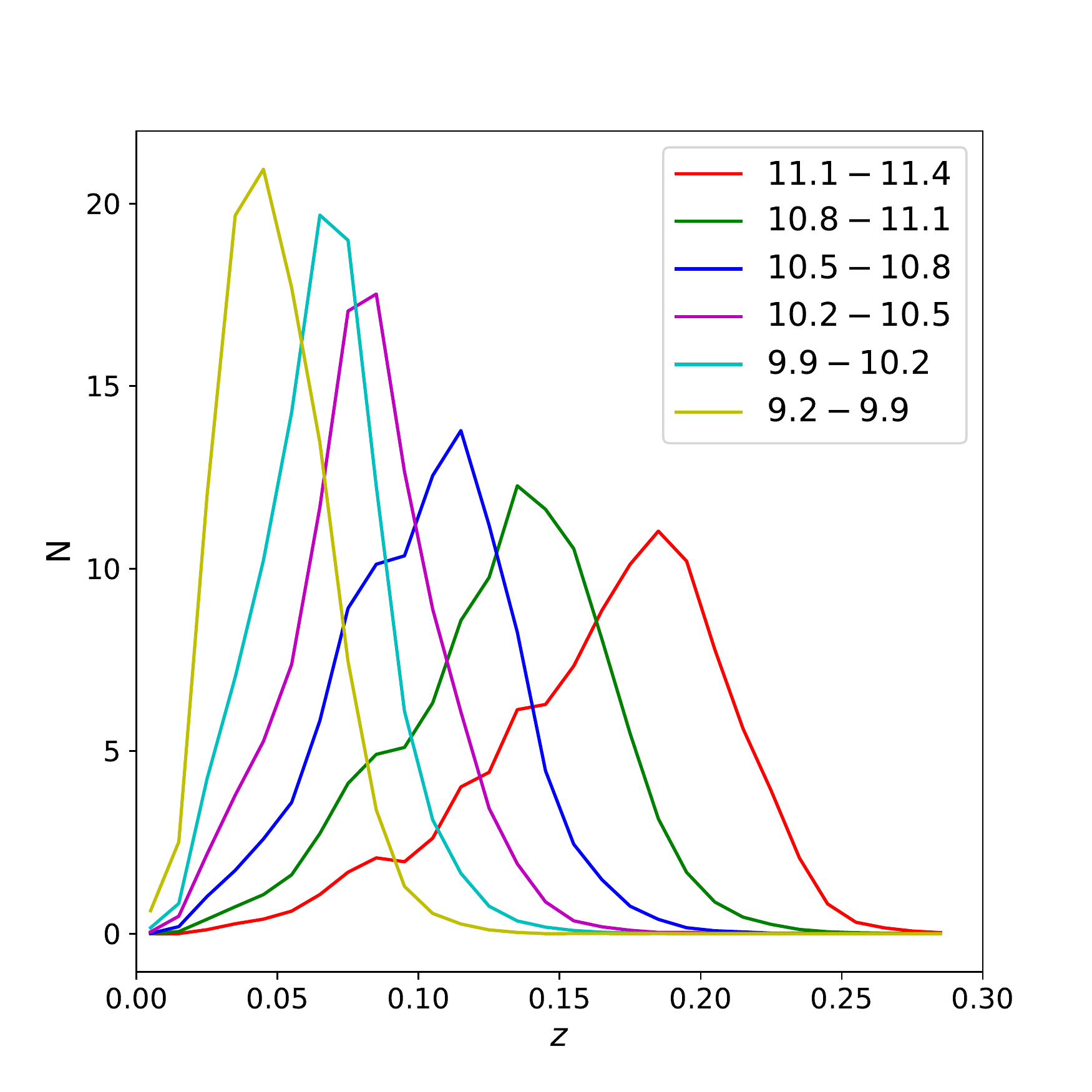}
\caption{Normalised redshift distributions of isolated central galaxies in six log stellar 
mass bins ($\log_{10}M_\ast/M_\odot$). More massive galaxies extend to higher redshifts.
}
\label{fig:redshifthist}
\end{figure}

To identify a sample of galaxies with a high fraction of central galaxies in dark matter 
haloes (purity), we select galaxies that are the brightest within given projected and line-of-sight 
distances. The parent sample used for this selection is the NYU Value Added Galaxy Catalogue 
\citep[NYU-VAGC;][]{2005AJ....129.2562B}, which is based on the spectroscopic Main galaxy sample 
from the seventh data release of the Sloan Digital Sky Survey \citep[SDSS/DR7;][]{2009ApJS..182..543A}. 
The sample includes galaxies in the redshift range of $z=0.001$ to $z=0.4$, which is flux 
limited down to an apparent magnitude of 17.77 in SDSS $r$-band, with most of the objects
below redshift $z=0.25$. Stellar masses for this sample were estimated from the K-corrected  
galaxy colours by fitting a stellar population synthesis model~\citep{2007AJ....133..734B} 
assuming a \cite{2003PASP..115..763C} initial mass function.

Following \cite{2014MNRAS.443.1433D}, we at first exclude galaxies whose minor to major axis 
ratios are smaller than 0.3, which are likely edge-on disc galaxies. \cite{2008MNRAS.388.1521D} 
pointed out that the scattered light through the far wings of point spread function (PSF) from 
edge-on disc galaxies can potentially contaminate the stellar haloes.

We require that galaxies are brightest within the projected virial radius, $R_{200}$, of their 
host dark matter haloes\footnote{$R_{200}$ is defined to be the radius within which the average 
matter density is 200 times the mean critical density of the universe.} and within three times 
the virial velocity along the line of sight. Moreover, these galaxies should not be within the 
projected virial radius (also three times virial velocity along the line of sight) of another 
brighter galaxy. The virial radius and velocity are derived through the abundance matching 
formula between stellar mass and halo mass\footnote{We have also tested the stellar mass and 
halo mass relation derived through Halo Occupation Modelling of \cite{2010MNRAS.402.1796W}, 
and it gives very similar results in terms of the sample selection.} of \cite{2010MNRAS.404.1111G}. 
The selection criteria have been adopted in \cite{2013MNRAS.428..573S}, and based on mock 
catalogues it was demonstrated that the choice of three times virial velocity along the 
line of sight is a safe criterion that identifies all true companion galaxies.

The SDSS spectroscopic sample suffers from the fiber-fiber collision effect that two fibers 
cannot be placed closer than 55$\arcsec$. As a result, galaxies in dense regions such as 
galaxy clusters and groups could miss spectroscopic measurements. To avoid the case when a 
galaxy has a brighter companion but this companion does not have available redshift and is 
hence not included in the SDSS spectroscopic sample, we use the SDSS photometric catalogue 
to make further selections. The photometric catalogue is the value-added Photoz2 catalogue 
\citep{2009MNRAS.396.2379C} based on SDSS/DR7, which provides photometric redshift probability 
distributions of SDSS galaxies. We further discard galaxies that have a photometric companion 
whose redshift information is not available but is within the projected separation of the 
given selection criterion, and the photoz probability distribution of the photometric 
companion gives a larger than 10\% of probability that it shares the same redshift as the 
central galaxy, based on the spectroscopic redshift of the central galaxy.

Fig.~\ref{fig:redshifthist} shows the redshift distribution of selected galaxies in a few 
different stellar mass bins, indicated by the legend. The distribution spans from $z=0$ to 
slightly above $z=0.25$. Due to the cosmic redshift and time-dilution effect, the observed 
bands are redder for galaxies at higher redshifts. In principle, to ensure fair comparisons 
for galaxies at different redshifts, proper K-correction is needed to transfer observed-frame 
magnitudes and colours to rest-frame quantities. However, K-correction often relies on 
modelling of galaxy photometry. Model templates for the faint stellar halo in outskirts 
of galaxies is theoretically not well studied, whereas applying templates of central 
galaxies to the outer stellar halo might be dangerous, which may potentially introduce 
additional uncertainties. So instead of involving K-corrections, we choose to use galaxies 
in a narrow redshift range of $0.05<z<0.16$ for our analysis. The amount of K-correction 
is negligible compared with the difference among the surface brightness profiles of 
galaxies in different stellar mass bins. To measure colour profiles which are more sensitive 
to the amount of K-correction, we further split the sample into two subsamples with $0.05<z<0.1$ 
and $0.1<z<0.16$, and calculate the colour profiles accordingly.

\cite{2018MNRAS.475.3348H} converted the observed surface brightness to stellar mass assuming 
that the massive galaxies can be well described by an average stellar mass to light ratio. 
\cite{2018MNRAS.475.3348H} achieved SED fitting and K-correction using the five-band HSC cModel 
magnitudes. In our analysis, we choose to focus on the surface brightness instead of looking 
at the stellar mass mainly because of the following reasons. First of all, we aim to push to 
less massive galaxies, which are composed of more complicated stellar populations. The average 
stellar mass to light ratio cannot be trivially applied to the whole galaxy and the faint 
stellar halo in outskirts. Secondly, the colour profiles of galaxies are not constants, which 
vary with radius, and thus using fixed and radius-independent magnitudes for SED fitting would 
introduce additional uncertainties. We choose to avoid this in our analysis. In principle, we 
can model the multi-band magnitudes as a function of radius, but we postpone this to our future 
studies and in this paper we simply focus on the surface brightness. Lastly, as we have mentioned 
above, model templates for central galaxies might not be directly applicable to the extended 
stellar halo.

The number of selected galaxies in different mass ranges and within the HSC footprint (the internal 
S18a data release) is summarised in the second column of Table~\ref{tbl:isos}. In the next subsection, 
we investigate the sample purity, completeness and average halo virial radius, using a mock galaxy 
sample. We will show the selected sample has a purity of about 85\% true halo central galaxies, and 
hence we call this sample of galaxies as isolated central galaxies.

\begin{table*}
\caption{Total number of isolated central galaxies within the S18a footprint, average halo virial radius ($R_{200}$, 
based on isolated central galaxies in a mock galaxy catalogue rather than direct abundance matching) and image size 
(number of pixels) for six stellar mass bins considered in our study. We also provide the information for a broader 
bin of $9.2<\log_{10}M_\ast/M_\odot<10.2$, which is a combination of the two least massive bins above, in order to 
give better signals for results in Sec.~4.5 and Sec.~4.6.}
\begin{center}
\begin{tabular}{lrrrrrr}\hline\hline
$\log M_*/M_\odot$ & \multicolumn{1}{c}{$N_\mathrm{galaxy}$} & \multicolumn{1}{c}{$R_\mathrm{200}$ [kpc]} & \multicolumn{1}{c}{image size [pixel$\times$pixel]}   \\ \hline
11.1-11.4  & 1438 & 459.08 & 1523$\times$1523  \\
10.8-11.1 & 5068 & 288.16 & 941$\times$941   \\
10.5-10.8 & 5572 & 214.80 & 698$\times$698   \\
10.2-10.5 & 3331 & 173.18 & 558$\times$558  \\
9.9-10.2 & 1536 & 142.85 & 394$\times$394   \\
9.2-9.9 & 801 & 114.64 & 464$\times$464   \\
9.2-10.2 & 2337 & 120.76 & 373$\times$373   \\
\hline
\label{tbl:isos}
\end{tabular}
\end{center}
\end{table*}

\subsection{Purity and completeness implied from a mock galaxy catalogue}

\begin{figure} 
\includegraphics[width=0.49\textwidth]{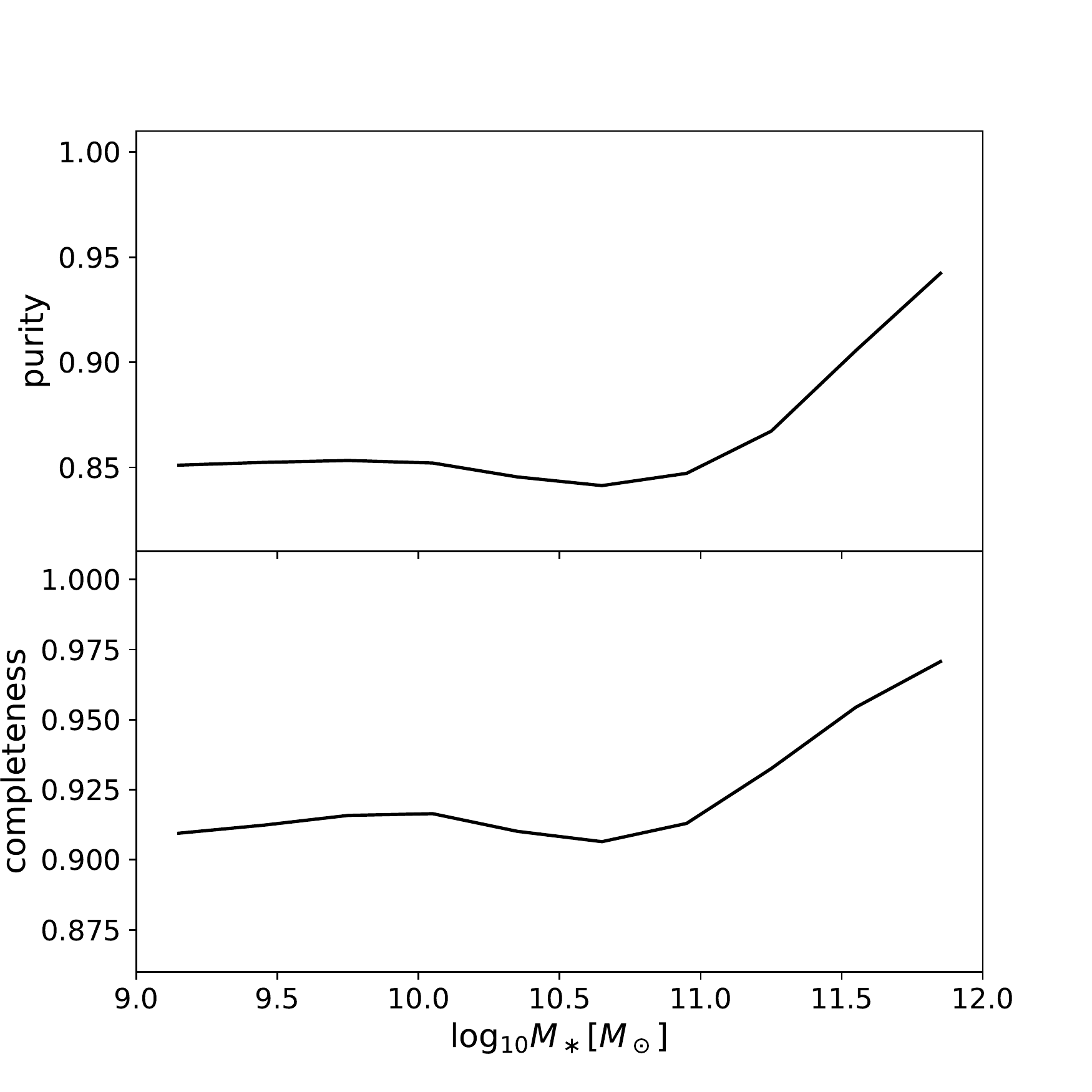}
\caption{The purity of true halo central galaxies of isolated central galaxies (upper panel), and the 
completeness of isolated central galaxies with respect to all central galaxies (lower panel), reported 
as a function of stellar mass. The purity and completeness are based on a mock galaxy catalogue. 
}
\label{fig:puritycomplete}
\end{figure}

Applying the same selection criteria to simulated galaxies in a mock catalogue, we investigate 
the sample purity and completeness. The mock galaxy catalogue is based on dark matter halo 
merger histories from the cosmological Millennium simulation, galaxy formation and evolution 
are modelled following the physics of \cite{2011MNRAS.413..101G}. It matches well the observed 
properties of real galaxies in the local universe, including luminosity, stellar mass 
distributions and clustering. The Millennium simulation is based on the first-year data of 
WMAP \citep{2003ApJS..148..175S}.

To select a sample of isolated central galaxies in analogy to SDSS, we project the $z=0$ output 
of the simulation box along $z$-axis. Each galaxy in the simulation is assigned a redshift 
based on its $z$ coordinate and its velocity along the $z$ direction. Selections are then 
made based on the projected separation and redshift difference in the same way as for  
observational data\footnote{Instead of using the true virial radius in the simulation, we 
estimate the virial radius from the stellar mass through abundance matching, to be consistent 
with the observational data.}. However, it fails to include observational effects such as the 
flux limit of the real survey, the K-corrections to obtain rest-frame  magnitudes, and the 
incompleteness of close pairs caused by fibre collisions and the complex geometry of SDSS. 
Using a full light-cone mock catalogue, \cite{2012MNRAS.424.2574W} and \cite{2014MNRAS.442.1363W} 
have compared satellite properties based on such direct projections and found that the direct 
projection gives unbiased results.

The purity and completeness are shown in Fig.~\ref{fig:puritycomplete}. The purity is above 95\% 
at the massive end, and drops to almost a constant fraction of $\sim$85\% at $9.2<\log_{10}M_\ast/M_\odot<11$. 
The completeness fraction is about 96\% at the massive end, and drops to slightly above 90\% at 
 $9.2<\log_{10}M_\ast/M_\odot<11.2$. 

In a few previous studies which probe the gas content of galaxies through Sunyaev-Zeldovich effect 
\citep{2013A&A...557A..52P,2016A&A...586A.140P,2015PhRvL.115s1301H}, X-ray \citep{2015MNRAS.449.3806A}, 
calibration of the scaling relations between SZ signal/X-ray luminosity and halo mass through 
weak gravitational lensing \citep{2016MNRAS.457.3200M,2016MNRAS.456.2301W}, we select galaxies 
which are brightest within a projected separation of 1~Mpc and within 1000~km/s along the line of sight. 
1~Mpc is larger than the mean halo virial radius at $\log_{10}M_\ast/M_\odot<11.5$, and 1000~km/s 
is comparable to three times the mean virial velocity for galaxies with $\log_{10}M_\ast/M_\odot\sim11.1$. 
Thus for galaxies smaller than $\log_{10}M_\ast/M_\odot=11.1$, the 1~Mpc (1000~km/s) selection is 
more rigorous than the selection criteria introduced earlier in this section.

For the purpose of this study, we need to balance between a large enough sample size in order to 
obtain good signals and a high enough fraction of true halo central galaxies to avoid possible 
contamination from nearby massive galaxies. So we focus on our current sample. The less stringent 
selection gives a larger sample size, which helps us to push to smaller stellar mass ranges and 
larger radial scales of the surface brightness profiles. In Appendix~\ref{app:isocmp}, we make 
comparisons to the sample of galaxies selected by the 1~Mpc criteria in those previous studies, 
to show the robustness of our results to the sample selection and to the purity of central galaxies. 

The third column from the left of Table~\ref{tbl:isos} provides the average virial radius, $R_{200}$, 
for isolated central galaxies in different stellar mass ranges of the mock catalogue. The average $R_{200}$ 
of isolated central galaxies can be biased from that of all central galaxies in the corresponding 
stellar mass bin. Thus, although we have used the virial radius estimated from abundance matching to 
select our sample of isolated central galaxies, we will use the $R_{200}$ values in Table~\ref{tbl:isos} 
to determine the corresponding size of image cutouts. Details are provided in Sec.~\ref{sec:method}.

\subsection{HSC photometry and data reduction}
\label{sec:step}
HSC-SSP \citep{2018PASJ...70S...4A} is based on the new prime-focus camera, the Hyper Suprime-Cam 
\citep{2012SPIE.8446E..0ZM,2018PASJ...70S...1M,2018PASJ...70S...2K,2018PASJ...70S...3F} 
on the 8.2-m Subaru telescope. It is a three-layer survey, aiming for a wide field of 1400 deg$^2$ 
with a depth of $r\sim26$, a deep field of 26 deg$^2$ with a depth of $r\sim27$ and an ultra-deep 
field of 3.5 deg$^2$ with one magnitude fainter. In this work we use the wide field data.
HSC photometry covers five bands, namely HSC-$grizy$. The transmission and wavelength range for 
each of the HSC $gri$-bands are almost the same as those of SDSS \citep{2018PASJ...70...66K}. 

HSC-SSP data is processed using the HSC pipeline. The pipeline is an enhanced version of the 
LSST \citep{2010SPIE.7740E..15A,2017ASPC..512..279J} pipeline code, specialised for HSC. 
Details about the HSC pipeline are available in the pipeline paper \citep{2018PASJ...70S...5B}, 
and here we only introduce the main data reduction steps and corresponding data products of HSC. 

HSC has 104 main science CCDs, which are arranged on the focal plane and provide a 1.5 deg 
field of view in diameter. The size of each CCD is $2048\times4176$ pixels, with an average 
pixel size of 0.168$\arcsec$. Gaps exist between CCDs, and there are two different gap size 
\citep{2018PASJ...70S...2K}, approximately 12$\arcsec$ and 53$\arcsec$ between neighbouring 
CCDs. In the context of HSC, a single exposure is called one ``visit'' with a unique ``visit'' 
number. The same sky field is observed or ``visited'' multiple times, and hence for the same 
object, it can appear for multiple times on different CCDs (or different locations of the 
focal plane). The HSC pipeline involves four main steps: (1) processing of single 
exposure/visit image (2) joint astrometric and photometric calibration (3) image coaddition 
and (4) coadd measurement. 

In the first step, bias, flat field and dark flow are corrected for. Bad pixels, pixels hit 
by cosmic rays and saturated pixels are masked and interpolated. The sky background
is estimated and subtracted before source detections (see Sec.~\ref{sec:S18baksub} for 
more details). Detected sources are matched to external reference catalogues in 
order to calibrate the zero point and a gnomonic world coordinate system (TAN-SIP) for each 
CCD. After galaxies and blended objects are filtered out, a secure sample of stars are used 
to construct the PSF model. Background-subtracted images together with the subtracted sky 
backgrounds are both stored to the disc, which enables the users to recover the original 
images by adding back the subtracted sky backgrounds if needed. The outputs of this step are 
called Calexp images (means ``calibrated exposure''). They are given on individual exposure 
basis.

In HSC, four lamps in the dome are used for the flat field. Flat fielding with the dome 
flats is a necessary and crucial step which helps to flatten the image and aids to fit and 
remove the sky background. However, the effective temperature 
are not the same for different lamps, and the camera vignetting \citep{2018PASJ...70S...1M}, 
which is a reduction of the brightness or saturation for the periphery of images compared 
to the image centre, also couples individual lamps to particular areas on the focal plane, 
which prevent the flat field from being ideally flat. In addition, the pixel size of CCDs 
varies \citep{2018PASJ...70S...1M}. Pixels near the edge of the CCD plate can be $\sim$10\% 
smaller in area than pixels sits at the centre of the plate. Since pixel values in both 
data images and the flat fields are flux instead of surface brightness or intensity (not 
divided by the pixel size), after correcting the flat field, pixel area variations over 
the CCD plate are interpreted as quantum efficiency and divided out. However, as we will 
describe in the third coaddition step, the relative area variation between input and output 
pixels is still considered for resampling before coaddition. To correct for the nonuniform 
flat field and put back pixel size variations, a mosaic correction step is run by including 
the Jocabian matrix that reflects pixel area variations and correcting for the 
flux variations across images using a seventh order polynomial, in order to make the measured 
flux of sources consistent with that of the reference stars. This corrects the nonuniform 
dome flats, and improves both the astrometry and photometry.

In the second joint calibration step, the astrometric and photometric calibrations are refined 
by requiring that the same source appearing on different locations of the focal plane during 
different visits should give consistent positions and fluxes. Readers can find details in 
\cite{2018PASJ...70S...5B}. The joint calibration step improves the accuracy of both
astrometry and photometry. The difference for the same sources on different CCDs peaks at 
35 mas without the joint calibration step, while it decreases to 10 mas after joint calibration. 

In the third step, the HSC pipeline resamples images to the pre-defined output skymap  
(warping) by properly considering the relative area variations and the overlapping fraction 
for pixels between inputs and outputs. It involves resampling of both the single exposure 
images and the PSF model \citep{2011PASP..123..596J} to the common output skymap using a 
3rd-order Lanczos kernel \citep[e.g.][]{Turkowski:1990:FCR:90767.90805}. All warped/resampled 
images are combined together (coaddition). The inverse of the average values of the variance 
in each image are used as weights for coaddition. Compared with direct averages or single 
long-time exposures, the weighted average gives better signal-to-noise and also helps to 
avoid pixel saturation. Images produced through warping and coaddition are called coadd 
images. The number of single exposures that contribute to the final coadd varies 
over the position on the sky, ranging from 1 to about 10 images. On average, about 5 
images contribute to each coadd.

In the last step, objects are detected, deblended and measured from the coadd images. 
For our analysis in this paper, we focus on image-level analysis without referring to 
the HSC source catalogue, so this last step is not directly related to our science. 
We refer the readers to the pipeline paper for more details \citep{2018PASJ...70S...5B}.


\subsection{Improved sky background and instrumental pattern subtraction}
\label{sec:S18baksub}

\begin{figure*} 
\includegraphics[width=0.9\textwidth]{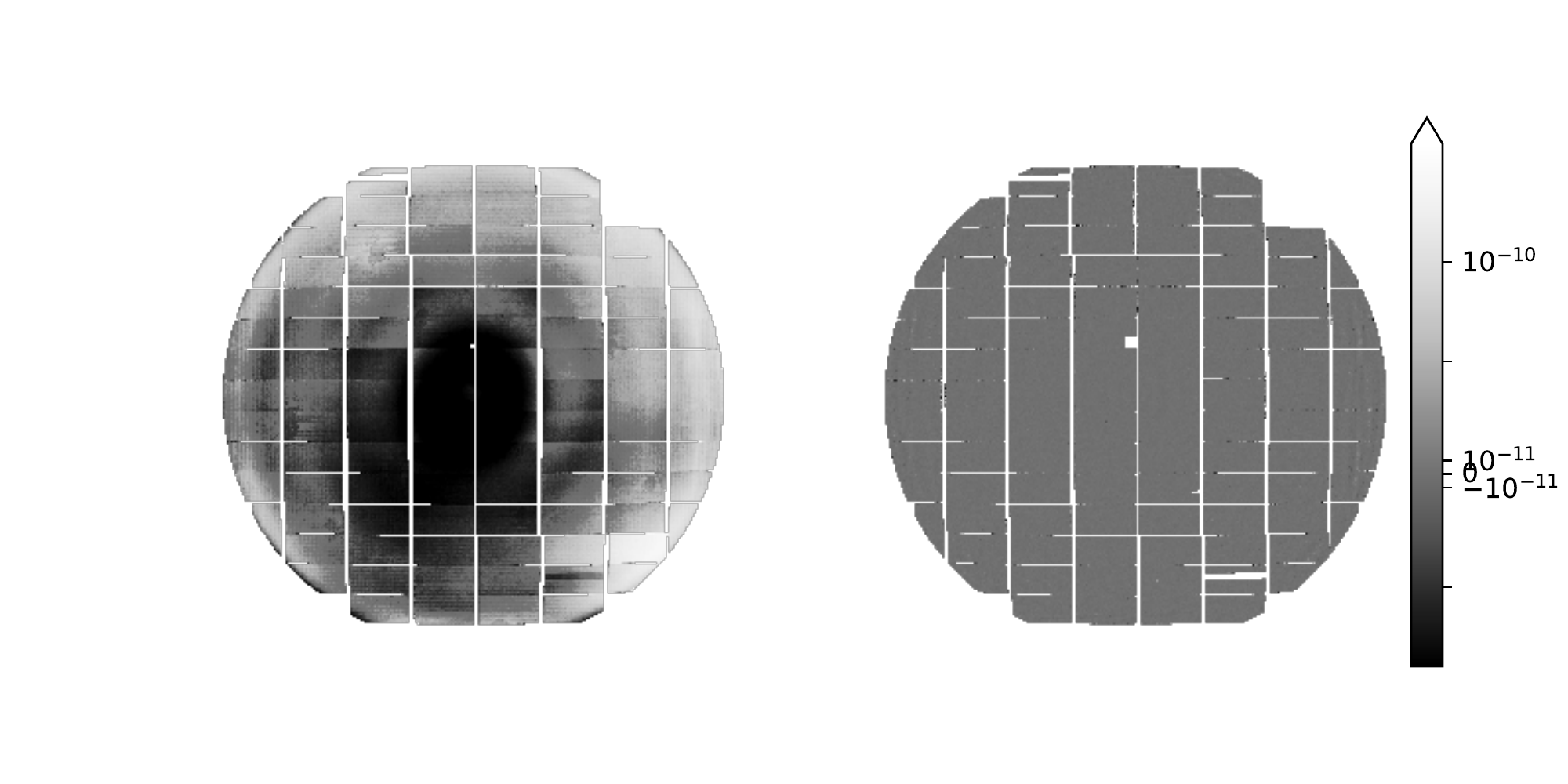}
\caption{{\bf Left:} Stack of Calexp images in HSC-$r$ filter of the S15b internal data 
release (609 visits in total), before any sky background and instrumental features have been subtracted 
by the pipeline. We estimate a mean constant background for each plate and subtract it before stacking, and 
hence there are negative pixel values. Pixel values are intensity, $I$, divided by zero point intensity, $I_0$. 
Note $-2.5\log_{10}I/I_0$ gives the surface brightness in unit of magnitudes. To show negative values, regions 
of $I/I_0>10^{-11}$ and $I/I_0<-10^{-11}$ are displayed in log scales for the absolute values, while the region 
of $-10^{-11}<I/I_0<10^{-11}$ is in linear scale. Sources are masked before stacking. The large scale and ring-like 
structures are due to the radial-dependence of the filter response. {\bf Right:} Stack of the same set of 
S15b visits, but are reprocessed in the S18a data release, with the sky background and instrumental features 
removed. The colour scale is exactly the same as that for the left plot. The ring-like structure disappears, 
and the stacked image becomes uniform. 
}
\label{fig:plate}
\end{figure*}

The faint stellar halo can be less than a few percent of the mean sky background, and thus 
it is very important to properly model and subtract the sky background, which is a challenging 
task. One complication comes from the fact that the true sky background is often mixed with other 
factors such as the scattered light from bright objects and  instrumental features. For example, 
the filter response curve shows strong radial dependence \citep{2018PASJ...70...66K} in both 
HSC-$r$ and HSC-$i$ filters\footnote{New filters of HSC-$r2$ and HSC-$i2$ have been procured 
\citep{2018PASJ...70...66K}, which do not have these features.}, which brings in ring-like 
structures crossing all CCDs \footnote{The ring-like structures are not due to nonuniform 
dome flats, which have been corrected for through the mosaic correction step in Sec.~\ref{sec:step}.} 
(see the left panel of Fig.~\ref{fig:plate}). These features are mixed with the true sky background.

The HSC internal data releases S15, S16 and S17 use a 6th-order Chebyshev polynomial to fit individual 
CCD images and model the sky background plus instrumental features together\footnote{The public first 
data release is similar to the internal S15b data release.}. It over-subtracts the light around 
bright sources and leaves a dark ring structure. The over-subtraction is mainly 
caused by the scale of the background model (or order of the polynomial fitting) and unmasked outskirts 
of bright objects. It is difficult to know how extended objects are before coaddition. 

The new version of HSC pipeline (v6.5.3) and the latest internal data release of S18a implement a 
significantly improved background subtraction approach\footnote{The up-coming second public data 
release of HSC will be based on the internal S18a release.} (HSC Collaboration et al., 2019, in 
preparation). An empirical background model crossing all CCDs are used to model the sky background, meaning 
that discontinuities at CCD edges are avoided. A scaled ``frame'', which is the mean response of the 
instrument to the sky for a particular filter, is used to correct for static instrumental features that 
have a smaller scale than the empirical background model. As shown in Fig.~\ref{fig:plate}, the ring-like 
structures apparent in the left panel are successfully removed in the right panel following the use of 
the S18a data, leaving uniform images (see also HSC collaboration et al., 2019, in preparation). In addition, 
the latest pipeline adopts a larger scale of $\sim$1000 pixels to model the sky background, which minimises 
over-fitting due to small scale fluctuations (see details in Appendix~\ref{app:comprelease}).

In our analysis throughout the main text of this paper, we focus on results based on the coadd images 
of the S18a release, with the sky background and instrumental features subtracted by the pipeline. We 
introduce our methodology of processing these coadd images in the next section. In Appendix~\ref{app:comprelease}, 
we show a comparison between results based on S15b and S18a coadd images, to demonstrate the significant 
improvement of S18a in avoiding over-subtracting extended emissions of bright sources.

\section{Methodology}
\label{sec:method}

In the following, we describe the steps of processing S18a coadd images. 

\subsection{Image cutouts and zero point correction of flux}
Given the celestial coordinates of our galaxy sample and the average virial radius, $R_{200}$, in different 
mass bins, we extract image cutouts, which are approximately square sky regions centred on each galaxy with 
a side length of $2.6R_{200}$ (or a radius of $1.3R_{200}$). The factor of 1.3 is chosen to have the 
image radius slightly larger than $R_{200}$. $R_{200}$ is transformed to angular scales at the redshift of 
the galaxy. Pixel values of each image is divided by the zero point flux, which is produced by the pipeline 
with reference stars (see Sec.~\ref{sec:step}). 

\subsection{Image resampling}

The number of pixels within a given 
physical scale can vary significantly for galaxies at different redshifts. To stack images in physical coordinates,  we need 
to resample the image cutouts to a common grid of pixels with the same physical size. This is achieved by the warping module 
of the HSC pipeline, which performs very well in terms of flux conservation. It ``warps'' input images to a pre-defined 
output WCS, image size and pixel size. Basically, for a given output pixel and its central coordinate, the module at first 
resamples the input pixels at the location of the output pixel. The procedure is accomplished through Lanczos sampling, i.e., 
the pixel value at position $x$ is given by the convolution between discrete pixel values and the third order Lanczos function. 
The Lanczos function serves as a filter to reconstruct pixel values at any given position, according to the Nyquist sampling 
theorem. The resampled value is then corrected for the change in the pixel area from the input to the output image. 

After resampling, the number of pixels is exactly the same for all images centred on galaxies in the same stellar mass 
bin, and these resampled images will be stacked afterwards. Each pixel in the resampled images corresponds to a physical 
scale of 0.8~kpc, independent of the redshift of the central galaxy\footnote{The CCD pixel size for HSC is $\sim$0.168$\arcsec$, 
which corresponds to 0.8~kpc at redshift $z\sim0.23$, and hence for our sample of galaxies at $0.05<z<0.16$, we are sampling 
pixels with an interval larger than the input pixel size.}. The image size (number of pixels) are provided in the fourth 
column of Table~\ref{tbl:isos}. We have carefully tested that changing the image and pixel size within a reasonable range 
does not affect the final stacked surface brightness or colour profiles of galaxies and stellar haloes.

\subsection{Cosmic dimming correction}

The pixel values are in unit of flux, and we divide them by the corresponding pixel area 
(solid angle), which gives the surface brightness or intensity in each pixel. The surface brightness is a 
conserved quantity that does not vary with the distance in Euclid space, but in the expanding universe it 
scales with redshift, $z$, in the manner of $(1+z)^{-4}$, which is called ``cosmic dimming''\footnote{Strictly 
speaking, the scaling with redshift of cosmic dimming is perfectly valid for bolometric luminosity or a fixed 
band. For an identical object at different redshifts, we are observing different bands. Proper K-correction 
is necessary for fair comparison of objects at different redshifts and for proper ``cosmic dimming'' corrections. 
But as we have discussed in Sec.~\ref{sec:isogal}, K-corrections of the outer stellar halo might introduce 
additional uncertainties, and we choose to focus on a narrow redshift range to avoid K-corrections.}. To 
correct for the effect, each image is multiplied by $(1+z)^4/1.1^4$, i.e., correcting the cosmic dimming to 
$z=0.1$.

\subsection{Source masking}
\label{sec:masking}
Bad pixels such as those which are saturated, close to the CCD edge, outside the footprint with available data, hit 
by cosmic rays and so on, are masked by the pipeline. Moreover, to investigate the smooth stellar halo of the central 
galaxy, we need to mask all the companion sources including their extended emissions. To create deep masks, 
we at first resample and stack all coadd images in HSC $g$, $r$ and $i$-bands. We then run \sextractor on these images, 
using detection thresholds of 1.5, 2 and 3 times the background noise of the image, respectively. \sextractor 
outputs ``segments'' of detected sources, which are the detected footprints of sources above the given thresholds of 
background noise level. The segments can be used to mask out extended regions of pixels associated with companion 
sources including stars, satellite galaxies and projected foreground/background objects. We successively apply the 
1.5, 2 and 3-$\sigma$ detection outputs by \sextractor to the image cutout of each galaxy, to mask the segments associated 
with all companion sources, but not the central galaxy. We choose such a combination of detection thresholds to avoid the 
failure of deblending companion sources, which we discuss and test in detail in Appendix~\ref{app:maskrandom} and briefly 
explain the motivation below. 

A high detection threshold gives more robust detections of real sources, but a smaller segment for detection, whereas 
a low detection threshold might introduce fake detections which might be background noise, but the footprint associated 
with each detection is larger and helps to safely remove their extended emissions. To find an optimal way of source detection 
and creating safe masks, we provide in Appendix~\ref{app:maskrandom} detailed tests based on a series of different source 
detection thresholds. We discuss how the choice of detection thresholds can potentially affect our results, including the 
issues of masking background and foreground objects, masking physically associated satellite galaxies and deblending 
of companion sources.

\subsection{Clipping and stacking}

For each pixel in the final stacked image, pixels at the corresponding location of all input images are stacked 
by taking their mean value\footnote{We do not weight the input images by the inverse of the variance map, 
because the the correlated errors across pixels have been ignored by the pipeline to produce resampled 
variance maps.}. As mentioned above, pixels associated with companion sources are masked out. Besides, 
for regions that do not contain available data, such as CCD gaps and edges, the pixels are masked as well. 
Hence these masked pixels do not contribute to the stack. As a result, the true number of input pixels 
contributing to the final stack varies over the output image by about 30\% to 40\% from the centre to the 
periphery of image cutouts. For each pixel in the output, we also clip the sample of all useful input pixels 
by discarding 10\% pixels at the two ends of the distribution tail. We have checked that varying this fraction 
between 1\% and 10\% does not bias the stacked light profiles, and it makes the stacked images more smooth.

In the end, we note that some previous studies \citep[e.g.][]{2014MNRAS.443.1433D,2018MNRAS.475.3348H} derived 
surface brightness profiles using isophotal ellipses centred on galaxies. In this study, we will not rotate or 
align galaxies, and hence the surface brightness profiles derived in this study are simply circularly averaged 
profiles. We provide in Appendix~\ref{app:ell} comparisons between circularly averaged profiles 
and profiles based on stacking major axis aligned galaxies with elliptical isophotal binning. 

\begin{figure*} 
\includegraphics[width=0.9\textwidth]{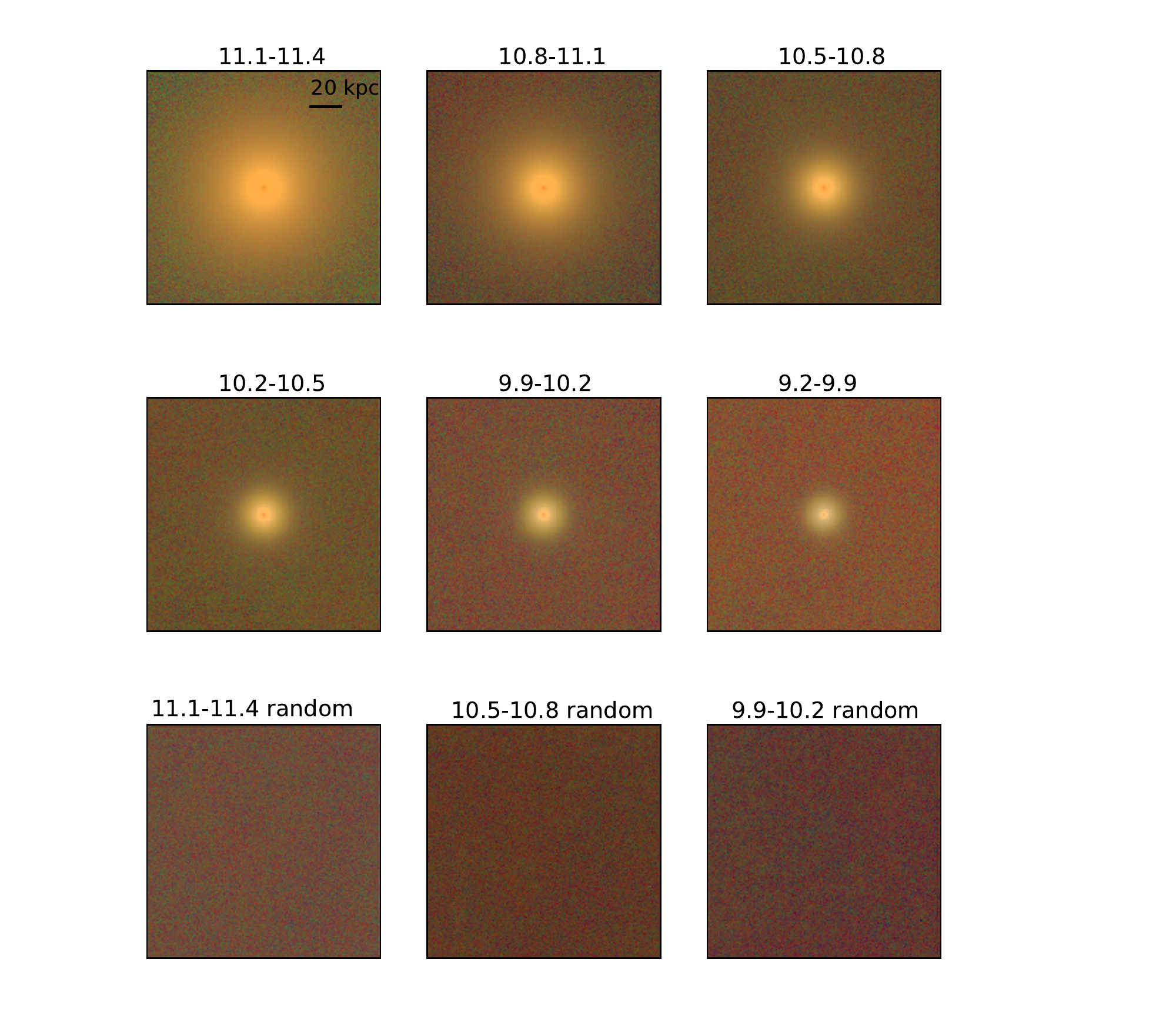}%
\caption{From top left to middle right are stacked and residual sky background corrected images of isolated central 
galaxies in RGB colour, obtained through images in HSC $g$, $r$ and $i$-bands. The stellar mass range of galaxies 
used for stacking in each panel is indicated by the text on top. The three images in the bottom are stacked images 
on random points that share the same assigned redshift and image size distributions as real galaxies in three out 
of the six stellar mass bins (see the text on top). The colour scales of all six panels are exactly the same. Image 
edge lengths are 160~kpc, i.e., ranging from -80~kpc to 80~kpc. No PSF deconvolution or correction has been made in 
these images.
}
\label{fig:imagedeep}
\end{figure*}

\begin{figure} 
\includegraphics[width=0.49\textwidth]{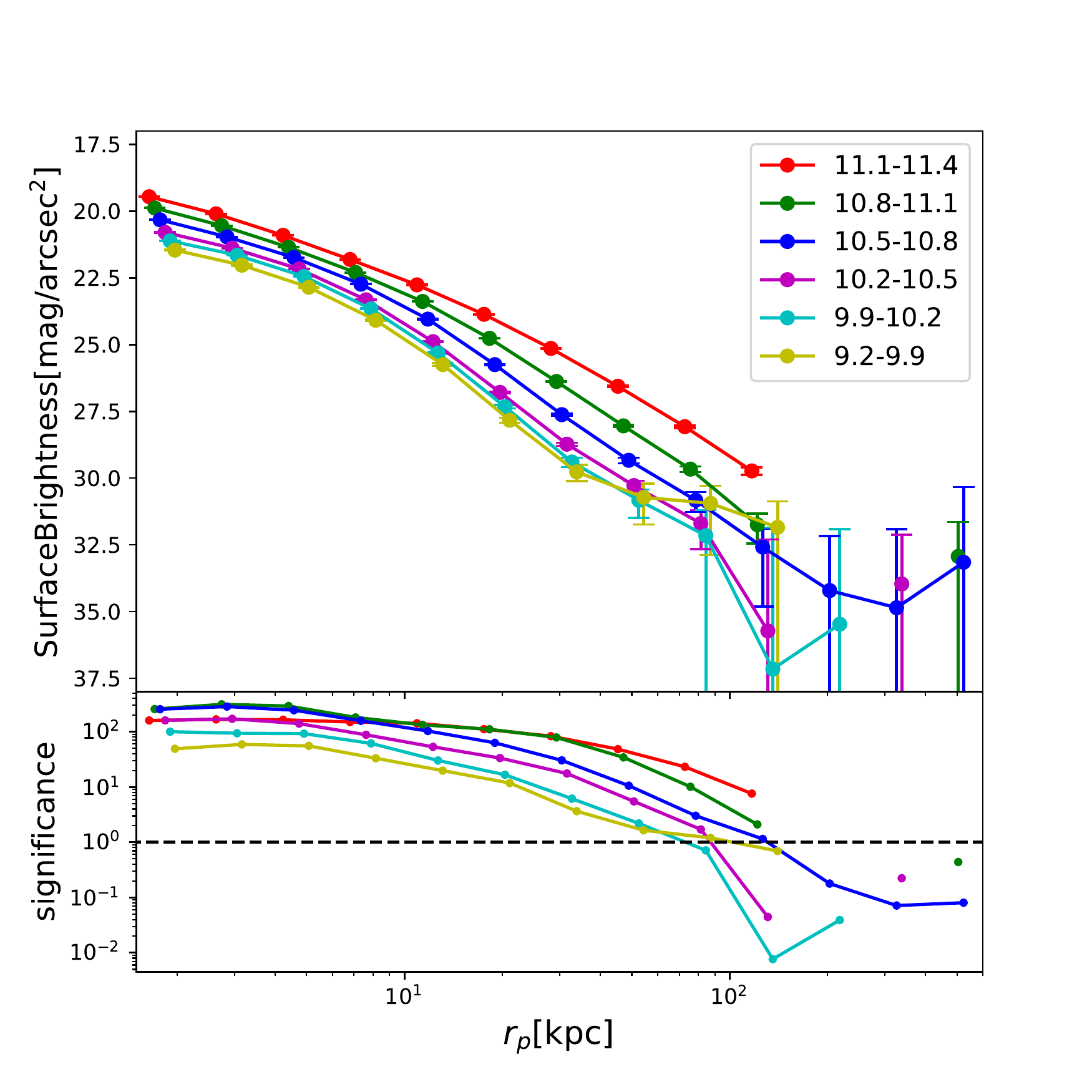}%
\caption{Average surface brightness profiles in HSC $r$-band and stacked on isolated central galaxies in six stellar 
mass bins, as indicated by the legend. Errorbars show the 1-$\sigma$ scatter of the stacked profiles based on 50 
boot-strap resampled realisations. Small horizontal shifts have been added to the second to least massive bins, 
in order to better display the errors. No PSF deconvolution or corrections have been made. The lower panel shows 
the signal to noise ratio of the measurements.
}
\label{fig:profmagr}
\end{figure}

\begin{figure} 
\includegraphics[width=0.49\textwidth]{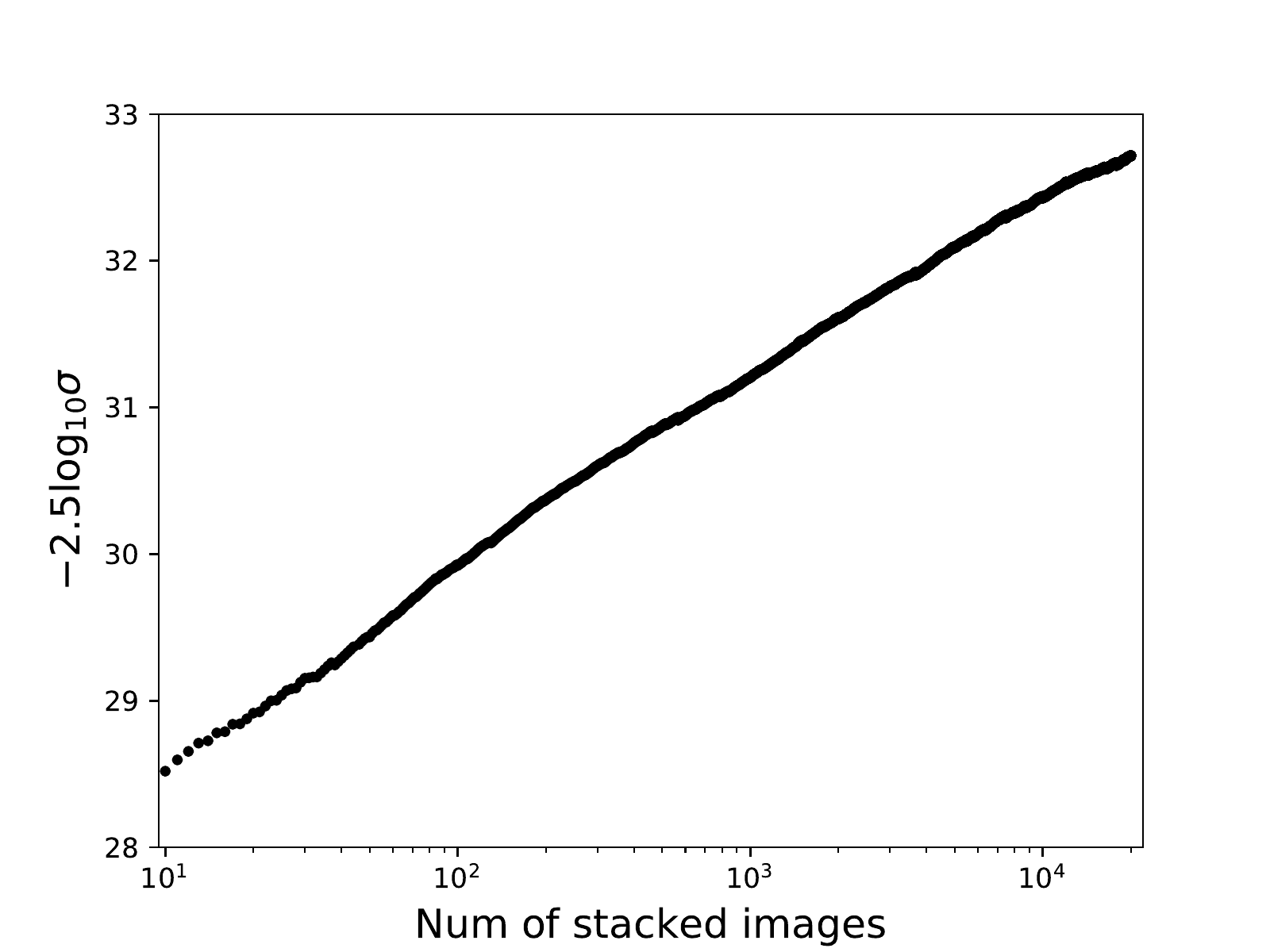}%
\caption{The standard deviation of the mean values in patches of $16\times 16$ pixels after source masking, 
as a function of the number of stacked images centred on a sample of random points within the HSC footprint. 
}
\label{fig:numstack}
\end{figure}

\subsection{Random sample correction for residual sky background}
\label{sec:ranresidual}
The sky background and instrumental features have already been modelled and subtracted off by the pipeline for 
coadd images, but there might be residual sky background remained to a small level, which can be either positive 
or negative. The amount of the residual sky background is very small, which is only a few percent level of the 
total sky background, but it is comparable to the brightness of the stellar halo in outskirts of galaxies and 
hence has to be corrected.  

To achieve the correction, we repeat the same steps for a sample of random points, whose sky coordinates are 
randomly distributed within the survey footprint. The random sample size is chosen to be comparable to that of 
our isolated central galaxies. In addition, we force the random sample to have exactly the same stellar mass and 
redshift distributions as that of isolated central galaxies, to ensure the same angular size distributions for images. 
Note for each individual image, its edge length (diameter) is the angular scale that corresponds to $2.6R_{200}$
(see Table~\ref{tbl:isos}) at the redshift of the central galaxy. The stacked image or surface brightness profiles 
centred on these random points are used as an estimate of the residual sky background, and are further subtracted from 
the stacks centred on real isolated central galaxies. 

In Appendix~\ref{app:maskrandom} and \ref{app:comprelease}, we show that the stacked profiles 
on random points are very close to be flat. The estimated residual sky background on random points is 
sensitive to the detailed choice of source detection thresholds used for masking companion sources\footnote{The 
random points are completely randomly generated within the HSC footprint, without avoiding footprints of detected 
sources. Thus incomplete source masking and scattered light from nearby very bright stars can contaminate the 
estimated residual sky background, but as we will show in Appendix~\ref{app:maskrandom}, this is not a problem 
of this paper. The latest pipeline has defined sky objects, which are placed to avoid footprints 
of bright sources and are not completely random. The estimated residual sky background level on these sky objects 
is much smaller.}. On large scales, the random stacks and real galaxy stacks agree with each other in amplitude, 
while there seem to be some small levels of over-subtraction in the outskirts of more massive galaxies 
($\log_{10}M_\ast/M_\odot>10.8$) in S18a. Although the estimated absolute residual sky background level depends 
on how sources are masked, we show in Appendix~\ref{app:maskrandom} that as long as we apply exactly the same 
source detection threshold and use it to create masks for companions in image cutouts centred on both real galaxies 
and random points, the large scale profiles centred on galaxies and random points change in the same direction, 
and the difference between the two is much less sensitive to the choice of source detection thresholds. 

\section{Results}

\subsection{surface brightness profiles split by stellar mass}
\label{sec:overallprof}

The stacked images of isolated central galaxies in RGB colour are presented in Fig.~\ref{fig:imagedeep}, for six stellar 
mass bins of Table~\ref{tbl:isos}. In addition, We also provide three images stacked on random points. We choose 
not to show all random stacks to simplify and shorten the figure. The RGB images are based on the stacked images in HSC 
$g$, $r$ and $i$-bands, and mapped to the RGB colour following the colour mapping strategy of \cite{2004PASP..116..133L}.

From the most massive population of galaxies in the top left to the least massive galaxies in the middle right, 
the galaxy size decreases, and the surface brightness on a given radius to image centre decreases as well. 
The colour transition is also obvious that more massive galaxies are redder. It is very encouraging 
that the stacked images centred on random points are smooth and uniform, which proves that our image 
processing is successful. 

Binned in projected radial distance to the galaxy centre, $r_p$, the one-dimensional surface brightness profiles of 
isolated central galaxies and their stellar haloes are presented in the top panel of Fig.~\ref{fig:profmagr} for HSC 
$r$-band. The errorbars are based on 50 boot-strap samples. Each boot-strap sample is generated by randomly selecting 
galaxies from the original sample with repeats, and hence the boot-strap samples have exactly the same sample size 
as the original one. The standard deviation of the 50 samples gives the boot-strap error. 

We have positive measurements of the surface brightness out to about 120~kpc. The faintest magnitudes at which 
we can still have positive values range from $\sim$29.7~$\mathrm{mag}/\mathrm{arcsec}^2$ for the most massive bin to 
$\sim$37.2~$\mathrm{mag}/\mathrm{arcsec}^2$ for smaller stellar mass bins.

The third stellar mass bin (blue dots connected by lines) shows the most extended profile, with positive measurements 
out to $\sim$500~kpc. This is benefited from the large number of stacked images ($>5000$) in that bin. The second 
massive bin (green) also has $\sim$5000 number of images, but the measured profiles are not as extended, which is 
likely due to the slight over-subtraction in extended emissions for massive objects (see Appendix~\ref{app:maskrandom}
for details). 

Positive values are not equivalent to robust measurements or detections. We further investigate the significance 
of our measurements in the lower panel of Fig.~\ref{fig:profmagr}, which shows the ratios between the surface 
brightness profiles and their boot-strap errors as a function of the projected distance to the galactic centre. At 
$\sim$70~kpc, measurements of the four most massive bins are larger than the uncertainty level, whereas the 
fifth bin drops below unity.

The least massive bin of galaxies in our analysis shows some flattening beyond 70~kpc, and the significance is 
above unity at 70~kpc. The flattening does not disappear with different choices of source detection thresholds 
to mask companions. In addition, we show in Appendix~\ref{app:isocmp} that a more strictly selected sample of isolated 
central galaxies, which have a higher purity, does not help to reduce the flattening. Thus the flattening is unlikely 
to be caused by satellite contamination in our sample of isolated central galaxies. One possible explanation is the 
contribution from the environment or neighbouring massive halos. Because these smallest halos are more likely to have 
massive companions, the stacked contribution from the outer haloes of these companions can boost the light profile 
near the boundary of these small galaxies. Another possibility is the contamination from massive galaxies due to 
stellar mass error. Because the stellar halo of a more massive galaxy is brighter and more extended, massive galaxies 
mis-classified into this low mass bin could lead to a bloated outer profile. However, because this flattening is only 
identified by three data points with large errors, and the focus of this work is to present the observational results, 
we refrain from further quantitative interpretation of it in the current paper.

At 120~kpc, only the three most massive bins have significances above their uncertainty levels (7.41, 2.11 and 1.14). 
The corresponding surface brightness are $\sim$29.7, 31.5 and 32.5~$\mathrm{mag}/\mathrm{arcsec}^2$. The significance 
level tells that by stacking $\sim$5000 galaxy images in the second massive bin, we can marginally measure the surface 
brightness profile down to $\sim$31.5~$\mathrm{mag}/\mathrm{arcsec}^2$, with a 2-$\sigma$ of significance. For the 
three less massive bins, the significance drops below unity at 120~kpc, despite the fact the data values stay positive. 
Note the galaxy sample size of the third massive bin is also larger than 5000, but the significance drops because the 
surface brightness on the same physical scale is fainter for smaller galaxies.

In Fig.~\ref{fig:numstack}, we provide one more independent test of the background noise. Fig.~\ref{fig:numstack} shows 
the background noise level as a function of the total number of stacked images. Basically, we create image cutouts, resample 
pixel size, mask detected sources and stack images centred on $2\times 10^{4}$ random points, following the approach in 
Sec.~\ref{sec:method}. Similar to \cite{2014MNRAS.443.1433D}, we calculate the standard deviation of the mean values in 
patches of $16\times 16$ pixels, which is is an estimate of the background noise level. We report the standard deviation 
as a function of the number of stacked images. It clearly decreases with the increase of the number of stacked images. 
When the number of stacked images reaches 5000, the background noise level is about 32~$\mathrm{mag}/\mathrm{arcsec}^2$, 
which is slightly lower than our marginal measurement of 31.5~$\mathrm{mag}/\mathrm{arcsec}^2$. 

Stacking SDSS images, \cite{2014MNRAS.443.1433D} measured the surface brightness profiles in $r$-band out to slightly 
beyond 100~kpc for galaxies in the mass range of $10<\log_{10}M_\ast/M_\odot<11.4$, and the faintest magnitude measured 
is claimed to be about 32~$\mathrm{mag}/\mathrm{arcsec}^2$. Here we have made careful investigations of the uncertainties 
in the residual sky background and the robustness of our results to different choices of source detection thresholds  
(see Appendix~\ref{app:maskrandom}). Despite the fact that the total sky footprint of the S18a internal release of HSC is 
at least twenty times smaller\footnote{Our isolated central galaxies are selected from the VAGC catalogue of SDSS/DR7, and 
about one-third of the S18a footprint does not overlap with SDSS/DR7. Hence the effective area can be even smaller.} than 
SDSS/DR7, we can marginally measure the surface brightness profiles up to similar scales and depth as \cite{2014MNRAS.443.1433D}, 
thanks to the significantly deeper HSC survey with high image qualities, which is very encouraging.

However, so far we cannot rule out the possibility that our measurements of the outer stellar halo could be affected 
by scattered light from the central galaxies through extended wings of the PSF. We move on to investigate the PSF in the 
next subsection.

\subsection{PSF effect on surface brightness profiles}
\label{sec:psfdeconv}

\begin{figure} 
\includegraphics[width=0.49\textwidth]{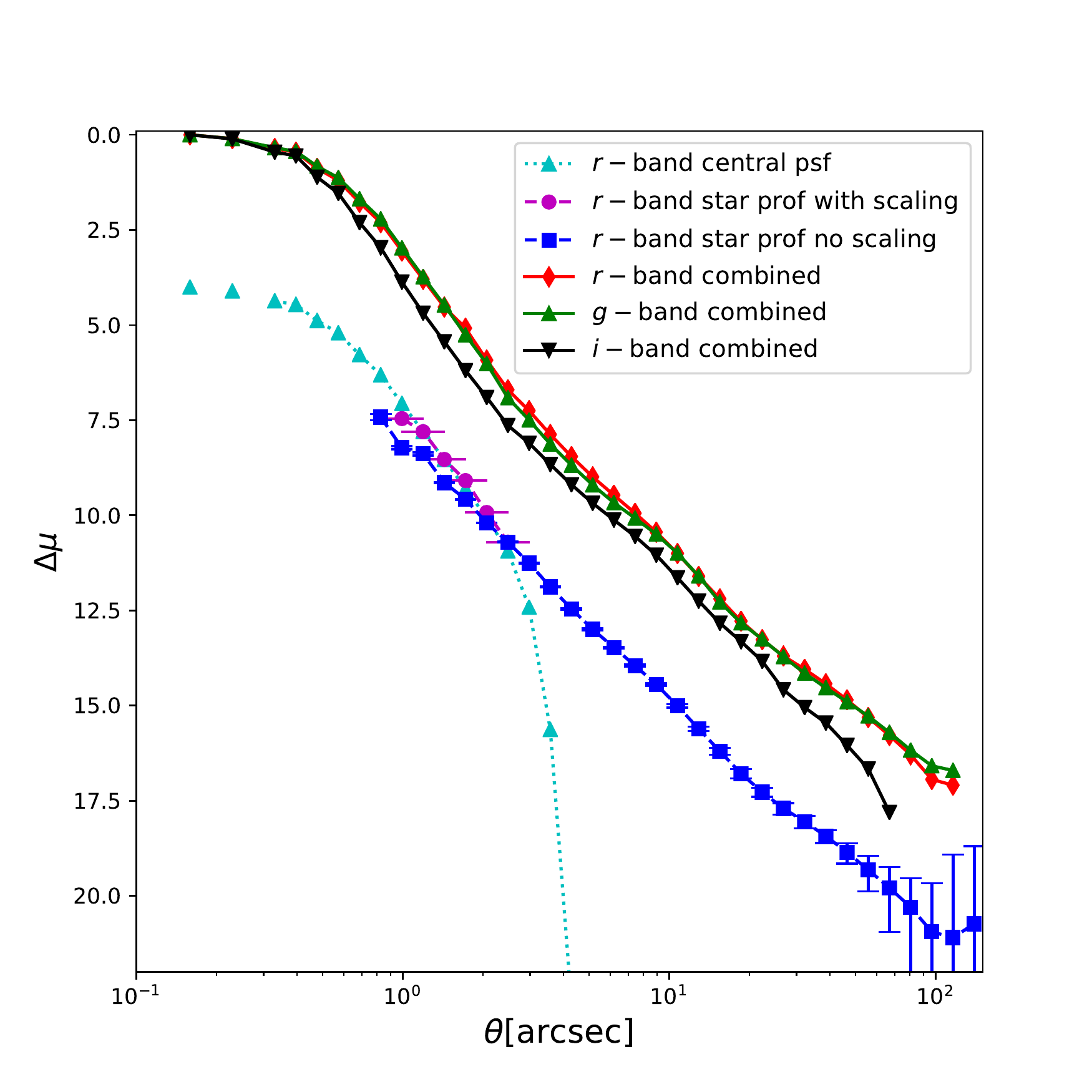}%
\caption{Cyan triangles connected by the dotted curve is the central PSF returned by the HSC pipeline in HSC $r$-band. 
Blue squares are stacked PSF profiles in HSC $r$-band centred on a sample of stars from {\it Gaia} DR2. Errorbars are 
the standard deviation based on 50 boot-strap realisations. Magenta dots with errors are stacked PSF using the 
same sample of stars, but each image has been scaled by the total flux of stars before stacking. Combinations of 
these data lead to the complete PSF profile out to $\sim 100 \arcsec$, which are shown as red diamonds 
for HSC $r$-band, green upper triangles for HSC $g$-band and black lower triangles for HSC $i$-band, with the 
inner most point scaled to zero.
}
\label{fig:psf}
\end{figure}

\begin{figure*} 
\includegraphics[width=0.9\textwidth]{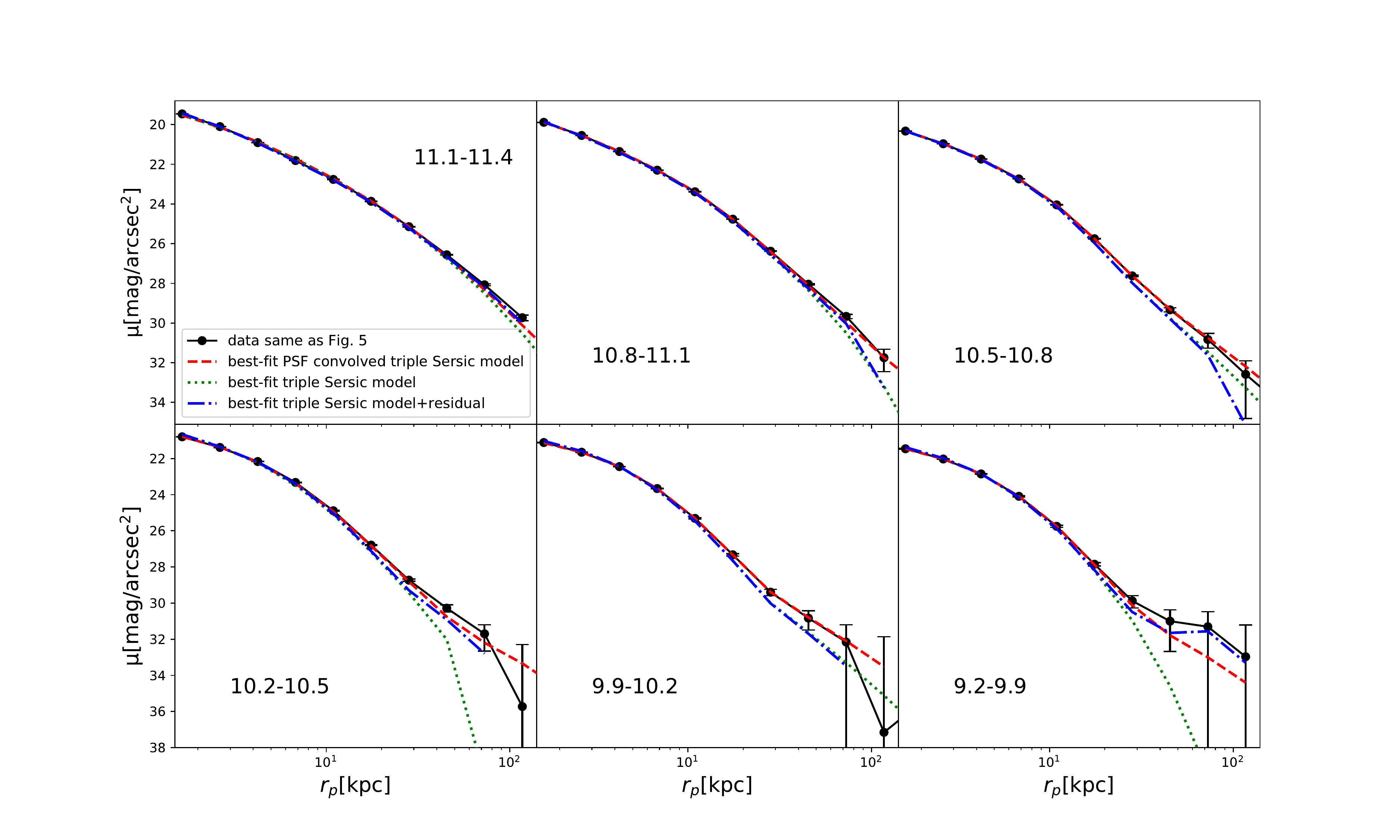}%
\caption{Black dots with errors in each panel are surface brightness profiles in HSC $r$-band centred 
on isolated central galaxies in different stellar mass bins (see the text in each panel), which are exactly the same 
as those in Fig.~5. Triple Sersic model profiles are convolved with the extended PSF profiles in Fig.~7, and fit to 
the measured surface brightness profiles. The best-fit model profiles convolved with the PSF are overplotted as dashed 
red curves. Green dotted curves are the PSF-free best-fit model profiles. The difference between the data and the 
best-fit PSF-convolved profiles are added to the green dotted lines to produce the blue dashed lines, to effectively 
correct for the imperfection of the proposed model after PSF deconvolution.
}
\label{fig:psffit}
\end{figure*}

\begin{figure} 
\includegraphics[width=0.49\textwidth]{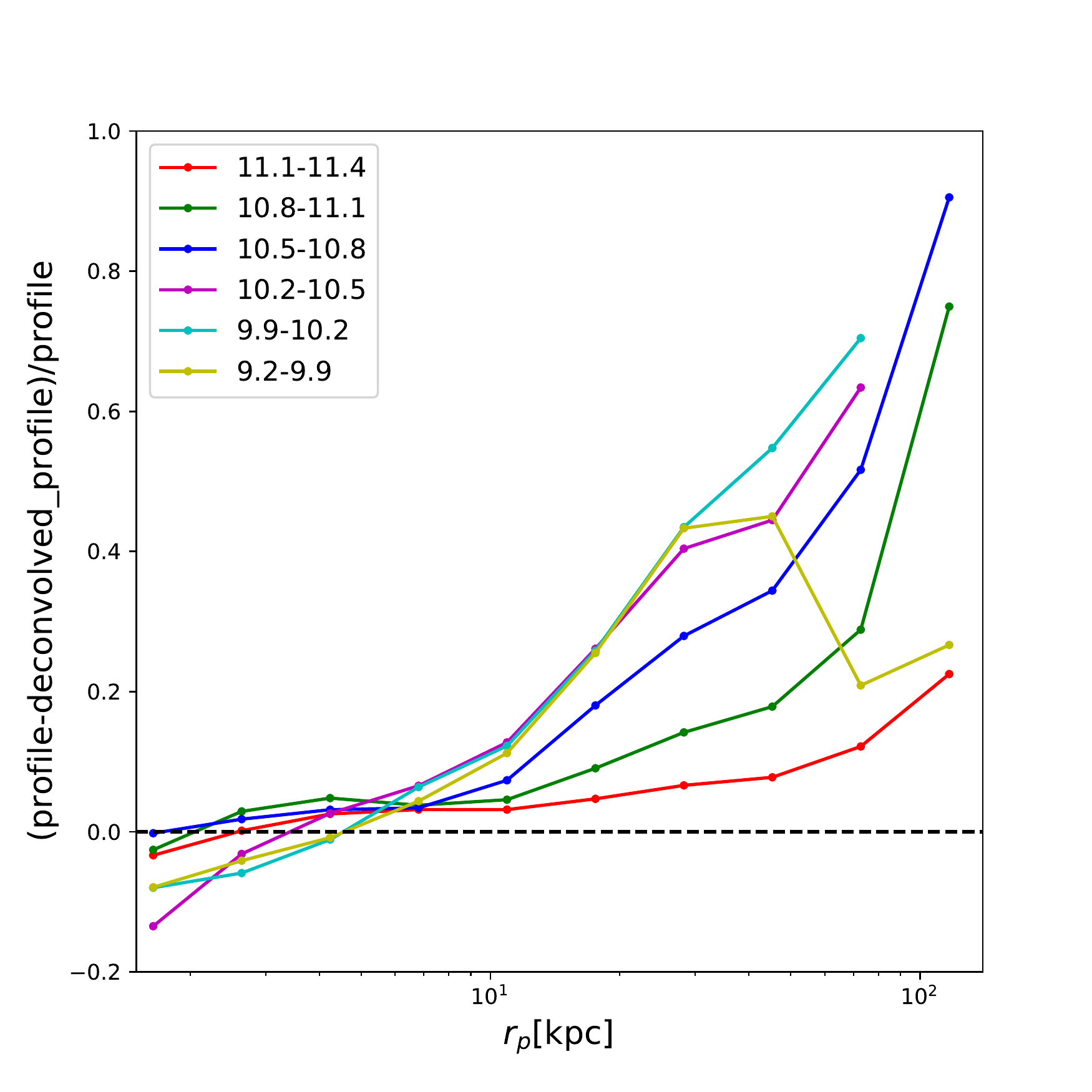}%
\caption{The fraction of surface brightness which can be contributed by the PSF, as a function of the 
projected distance to the central galaxy, for isolated central galaxies of different stellar masses (see the legend).
}
\label{fig:psffrac}
\end{figure}

The measured surface brightness profiles are true light distributions convolved with the PSF. HSC measured and 
interpolated the central PSF over the whole image, based on non-saturated stars. In principle, one can use 
the pipeline to return the central PSF for given coordinates within the HSC footprint. However, the 
central part of the PSF is dominated by atmosphere turbulence. As have been discussed in many previous studies 
\citep[e.g.][]{2002A&A...384..763M,2007ApJ...666..663B,2008MNRAS.388.1521D,2011ApJ...731...89T,2012MNRAS.426.2300L,
2014A&A...567A..97S,2015MNRAS.446..120D,2015ApJ...800..120Z,2018ApJ...859...85T}, the extended PSF wings, which 
are due to CCD, instrument, atmospheric aerosols and dust scattering, can potentially contaminate the light of 
the outer stellar halo. 

To obtain the extended PSF, we select a subsample of stars from the second data release (DR2) of {\it Gaia} 
\citep{2016A&A...595A...1G,2018A&A...616A...1G}, and stack HSC image cutouts centred on these stars in the same 
way as we stack galaxies. With {\it Gaia} we are able to target more faint stars, alleviating the problem of 
saturated pixels during stacking (see discussions below). The {\it Gaia} parallax and proper motion also enable 
more robust star-galaxy separations. We require stars to be in the magnitude and colour ranges of $16<G<13$ and 
$0.3<BP-RP<1.5$. In addition, we restrict ourselves to stars with measured parallax and proper motions larger 
than 5 times the corresponding errors, to exclude possible contamination of small compact galaxies. The magnitude 
range of $16<G<13$ is chosen so that we do not include very faint objects which are more likely contaminated by 
galaxies, while stars brighter than $G=13$ are also not used because of very large fractions of saturated pixels 
in the centre. 

Although the problem of saturation for stars with $16<G<13$ is less severe than that of brighter objects, 
saturated pixels still exist. If we directly stack these stars by masking saturated pixels, the central part of 
the stacked PSF is mainly contributed by non-saturated faint stars, which flattens the stacked profile. This is 
shown by the blue squares in Fig.~\ref{fig:psf}, which is obtained by directly stacking HSC $r$-band image 
cutouts centred on stars. Cyan triangles connected by the dotted curve is the averaged PSF returned by the 
pipeline, measured and stacked on the position of our sample of stars. It is clear that the stacked PSF on 
stars (blue squares) drop below the central PSF returned by the pipeline (cyan triangles) within 2$\arcsec$.

In principle, such flattening can be corrected, if we scale each image cutout by the total flux of the star before 
stacking, to remove the difference in relative flux. Unfortunately, we cannot correctly measure the total flux based 
on HSC images due to saturation, but we can convert {\it Gaia} $G$-band flux with the default {\it Gaia} DR2 zero 
point \citep{2018A&A...616A...4E} to HSC $r$-band flux, based on an empirical relation linking the two, which is a 
function of {\it Gaia} $BP-RP$ colour\footnote{We select stars with $0.3<BP-RP<1.5$, due to the valid colour 
range of this empirical relation.}.

Explicitly, the empirical relation is obtained by matching non-saturated HSC stars to {\it Gaia} DR2. We use 
the angular separation of 1~$\arcsec$ for the matching. Stars in HSC used for matching are required to have 
at least 2 photometric observations in each of the HSC $g$, $r$, $i$ and $z$-bands and must have no photometric flags 
(saturation, cosmic rays, interpolation etc.) in the Calexp processing. {\it Gaia} stars for the matching must be 
detected at least 5 times in each of $BP$ and $RP$. After the scaling, the stacked PSF profile between 1$\arcsec$ 
and 3$\arcsec$ is shown as magenta circles in Fig.~\ref{fig:psf}, which is steeper than blue squares and agree well 
with cyan triangles.

The shape of the outer PSF should not be affected by saturation. The entire PSF profile in HSC $r$-band out to 
$\sim$100~$\arcsec$ is obtained by combining the pipeline returned central PSF within $\sim$2.7~$\arcsec$, with 
the more extended PSF obtained by directly stacking stars. The same can be repeated for HSC $g$ and $i$-bands (green 
upper triangles and black lower triangles). The PSF profiles in Fig.~\ref{fig:psf} are very similar between HSC 
$g$ and $r$-bands, with a smaller and slightly less extended PSF in HSC $i$-band.

It is necessary to check the fraction of light scattered through the extended PSF. However, deconvolution of the 
observed light profile is difficult, because our measurements are noisy, which prevents the deconvolution from 
converging properly. We therefore follow a similar approach as \cite{2010ApJ...714L.244S} and \cite{2011ApJ...731...89T}. 
We fit triple Sersic model profiles convolved with the stacked PSF profile to the measured surface brightness of 
Fig.~\ref{fig:profmagr}. Explicitly, we Fourier transform the model and PSF map, the product between the two are 
transformed back to real space, which is more efficient than real space convolution. To convolve the model profiles 
as a function of physical separations with the PSF, we choose to use the median redshift of each stellar mass bin for 
conversions between angular scales and physical separations. We have also considered the full redshift distribution of 
galaxies in each bin, by calculating the PSF convolved profile at the redshift of each galaxy and averaging them, which 
gives very similar results.

The best-fit models are shown in Fig.~\ref{fig:psffit}. Measurements from Fig.~\ref{fig:profmagr} are overplotted as 
black dots. We calculate the residual profiles between the best-fit PSF convolved models (red dashed curves) and the 
measurements. We add the residuals to the best-fit PSF-free models. It has been shown by \cite{2010ApJ...714L.244S} that the 
residual added models (blue dashed curves in Fig.~\ref{fig:psffit}) are not sensitive to variations in the model parameters 
and is less model dependent. 

The residual-added and PSF-free best-fit model profiles can be used as estimates of the PSF-deconvolved profiles, 
and hereafter we call them the PSF-deconvolved profiles. Note the residuals have not been deconvolved with the PSF, but 
as pointed out by \cite{2010ApJ...714L.244S}, the residuals help to make a first-order correction for incorrect profile 
choice when the model does not properly describe the data. It is easy to show that the difference between the measured 
profile and the PSF-deconvolved profile defined this way is equivalent to the difference between the PSF-convolved and 
the PSF-free model profile.

To quantify the effect of the PSF, we plot the difference between the measured data points and the PSF deconvolved profiles, 
which are then divided by the measured data points and reported as a function of the projected distance to the central 
galaxy (Fig.~\ref{fig:psffrac}). On small scales, PSF flattens the measured profiles, but the flattening is not as obvious 
as those based on galaxies at $z\sim 0.25$ \citep[e.g.][]{2011ApJ...731...89T,2019ApJ...874..165Z}. This is due to the 
difference in redshift and PSF size. Our sample of isolated central galaxies are at redshift $z\sim0.1$. The average FWHMs 
of the central PSF are 0.756$\arcsec$, 0.759$\arcsec$, 0.752$\arcsec$, 0.755$\arcsec$, 0.7508$\arcsec$ and 0.7501$\arcsec$, 
for the most to least massive bins. The corresponding mean physical separations at the median redshift of different bins 
are 1.759~kpc, 1.693~kpc, 1.483~kpc, 1.278~kpc, 1.126~kpc and 0.984~kpc, respectively. Note the redshift distribution for 
less massive galaxies are on average lower, so a given angular scale corresponds to a smaller physical scale. The typical 
physical scale of the central PSF is smaller than the inner most point presented in Fig.~\ref{fig:profmagr}, and hence 
the radial range over which the PSF significantly flattens the inner profile is also smaller than the inner most data point. 

With the increase of the projected distance to the central galaxy, the contamination by scattered 
light from the central galaxy through extended PSF becomes more significant, and is more severe for stellar haloes of 
smaller galaxies. On scales of $\sim$100~kpc, PSF scattered light contributes less than 20\% for galaxies with 
$11.1<\log_{10} M_\ast/M_\odot<11.4$. It quickly increases to more than 50\% for less massive galaxies. This is mostly 
because the light profile of low mass galaxies drops faster with radius, and is thus more affected by PSF convolution. 
For galaxies smaller than $\log_{10} M_\ast/M_\odot=10.8$, PSF scattered light contributes more than 50\% at 70~kpc. 
Because the last three data points of the least massive stellar mass bin show flattening behaviour, the corresponding 
fraction in Fig.~\ref{fig:psffrac} drops on such scales. 

Before PSF corrections, we are able to detect the stellar halo down to $31.5\,\mathrm{mag}/\mathrm{arcsec}^2$ at 
a significance of $2\sigma$ for galaxies with $10.8<\log_{10} M_\ast/M_\odot<11.1$, as shown in Fig.~\ref{fig:profmagr}. 
PSF corrections remove $\sim$60\% to 70\% of this detected emission, leading to a net surface brightness of 
$\sim 31\,\mathrm{mag}/\mathrm{arcsec}^2$ at $\sim$85~kpc (corrected from $\sim 30\,\mathrm{mag}/\mathrm{arcsec}^2$), 
corresponding to a significance of $\sim 3\sigma$ with respect to the boot-strap error.

\subsection{Colour profiles split by stellar mass}

\begin{figure*} 
\includegraphics[width=0.49\textwidth]{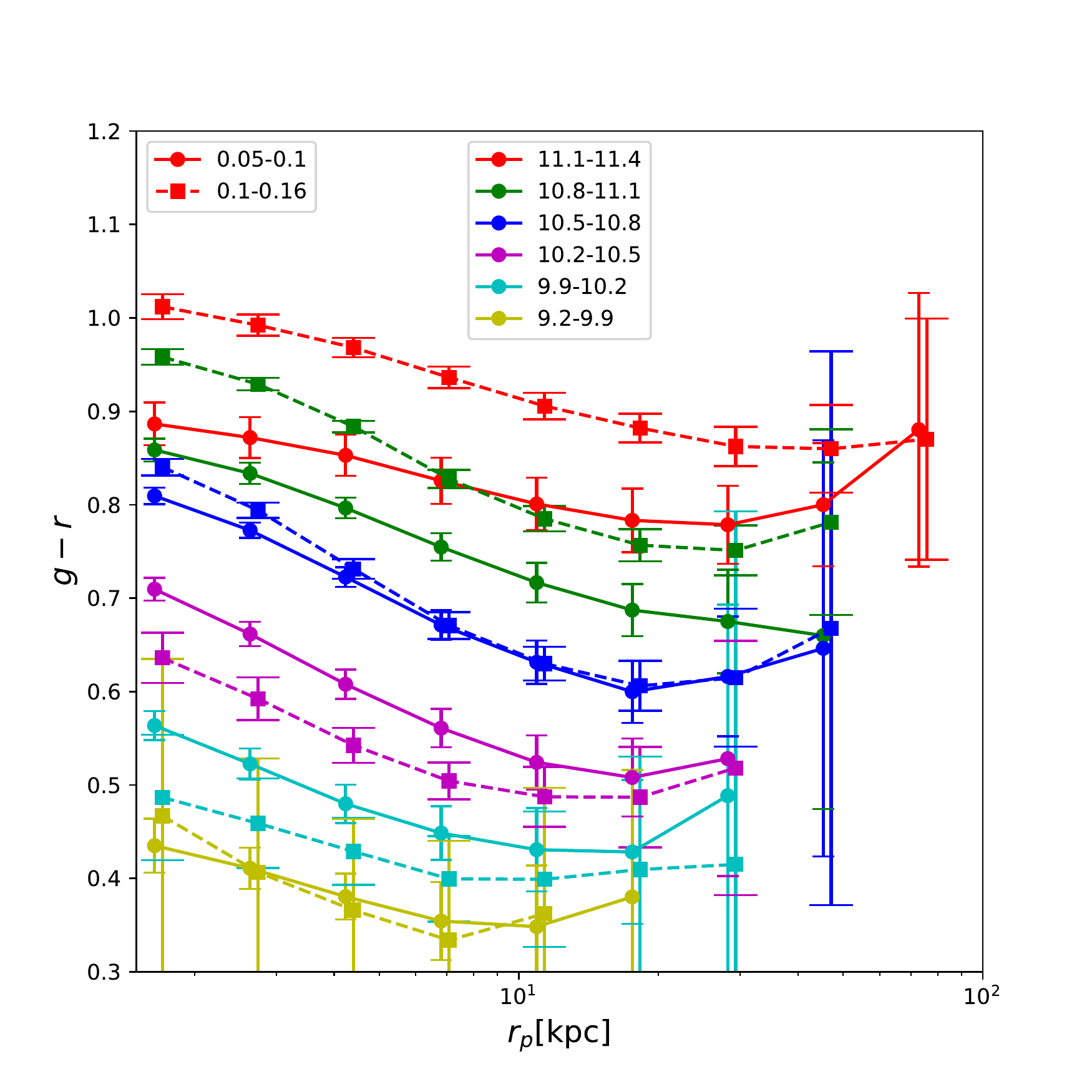}%
\includegraphics[width=0.49\textwidth]{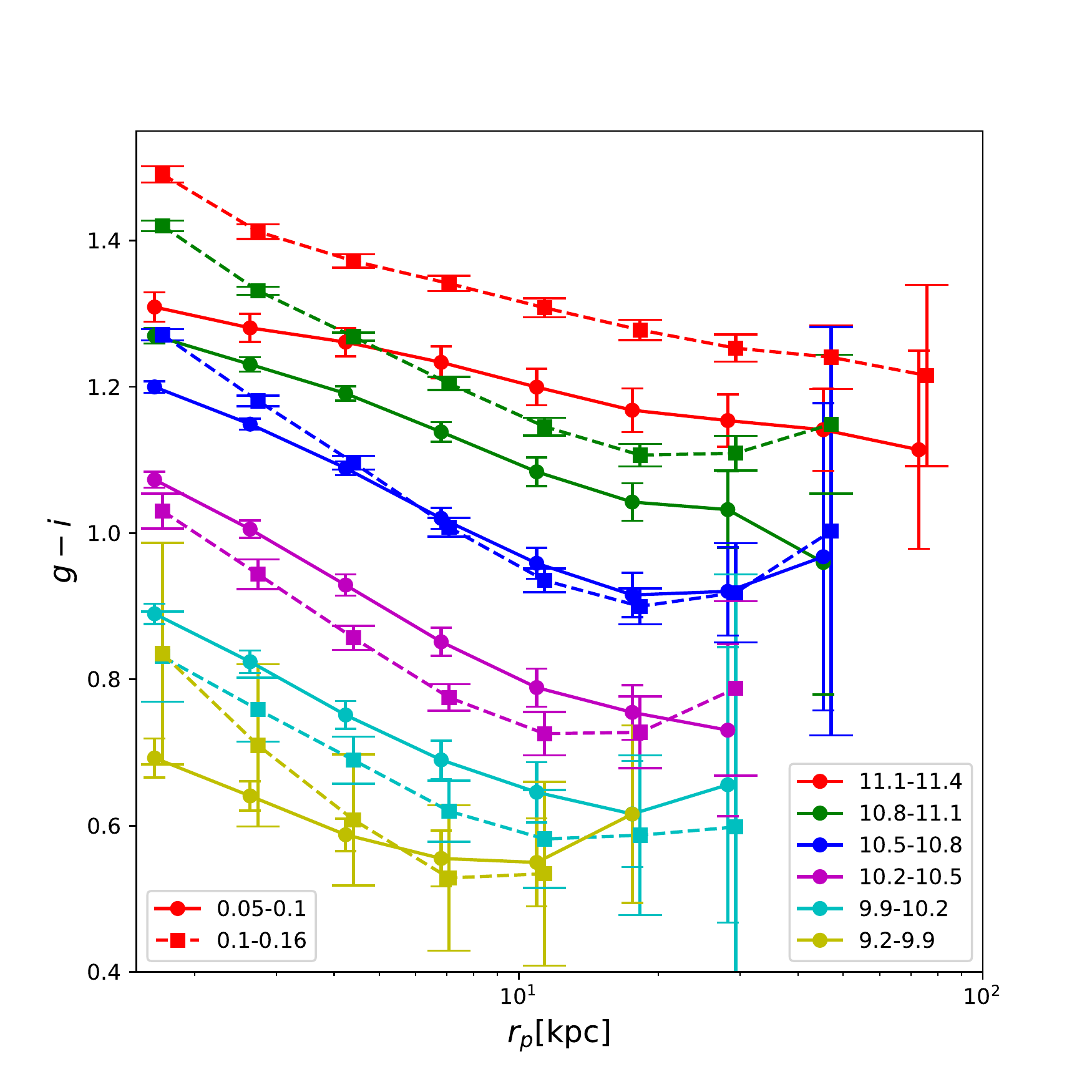}
\caption{$g-r$ (left) and $g-i$ (right) colour profiles of isolated central galaxies and their stellar haloes. The 
galaxy sample is split into two subsamples based on their redshifts, in order to have negligible amount of K-corrections. 
Solid squares connected by dashed lines with longer error-bar caps are based on galaxies in the redshift range of $0.1<z<0.16$, 
whereas round dots connected by solid lines with shorter error-bar caps are based on galaxies with $0.05<z<0.1$. Errorbars 
are 1-$\sigma$ errors obtained from 50 boot-strap resampled realisations. Small horizontal shifts have been added 
to  the high redshift subsample, to better display the errorbars. No PSF deconvolutions or corrections have been made.
}
\label{fig:profcolor}
\end{figure*}

\begin{figure*} 
\includegraphics[width=0.49\textwidth]{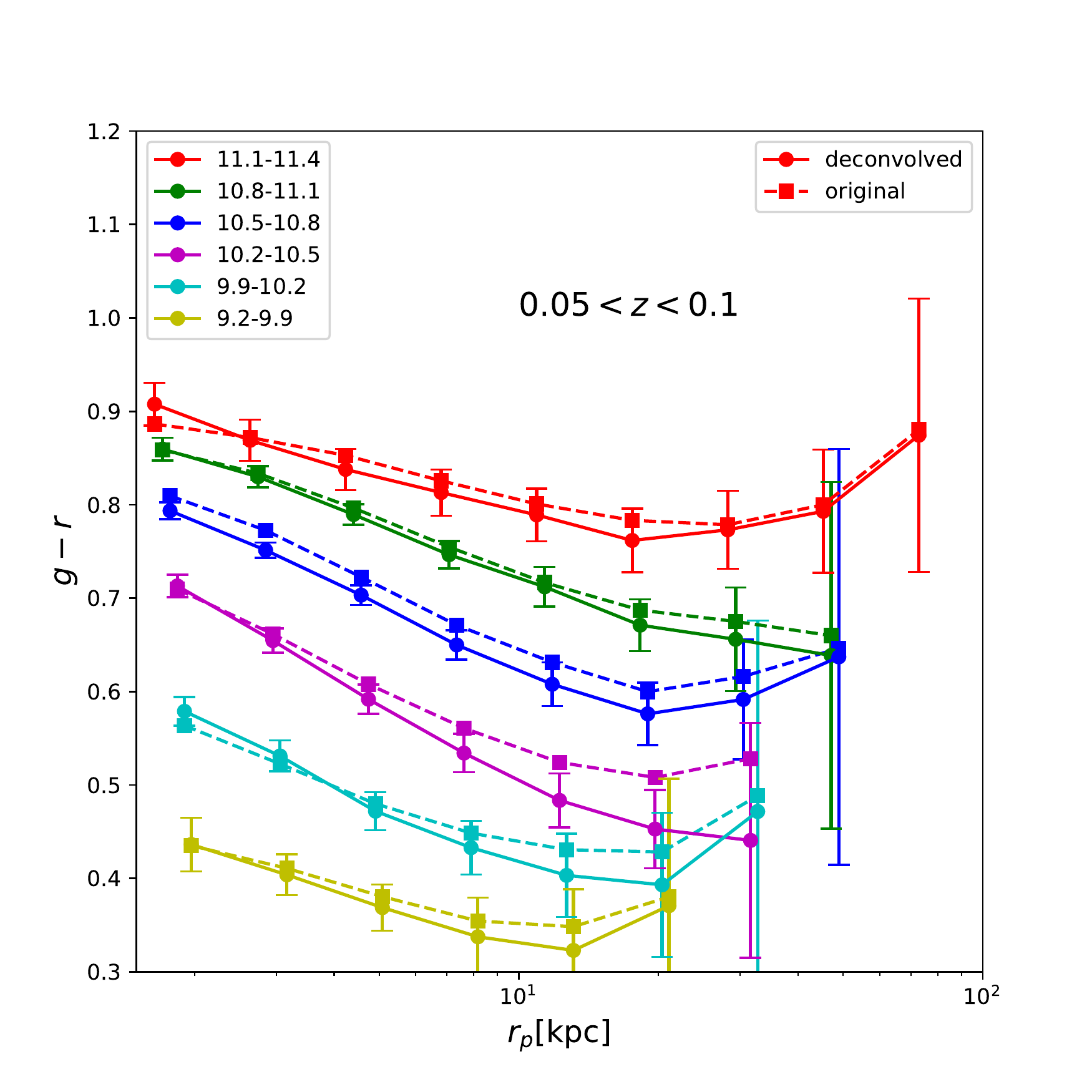}%
\includegraphics[width=0.49\textwidth]{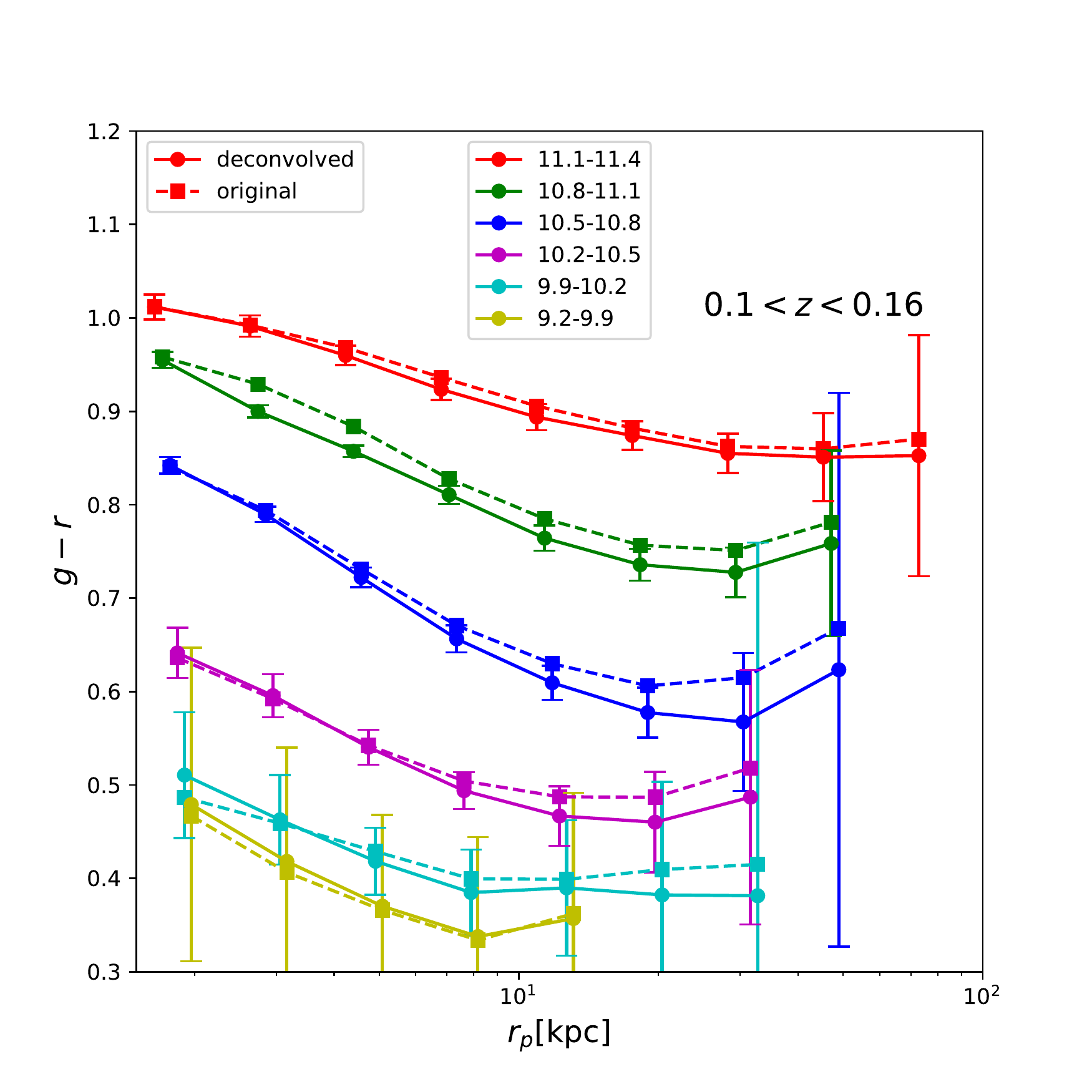}
\caption{$g-r$ colour profiles stacked on isolated central galaxies with redshifts of $0.05<z<0.1$ (left) and $0.1<z<0.16$ (right). 
Squares connected by dashed lines are exactly the same as those in Fig.~10. Dots connected by solid lines with errors are 
PSF-deconvolved colour profiles following the approach in Sec.~\ref{sec:psfdeconv}. Small shifts have been added to the 
$x$-coordinates of the second to least massive bins, to better display the errorbars.
}
\label{fig:profcolor_deconv}
\end{figure*}

The difference between surface brightness profiles measured in different filters bring the colour profiles. Note the 
PSF can vary among filters, but as we have shown in Fig.~\ref{fig:psf}, the PSF in HSC $g$ and $r$-bands are 
very similar to each other. Hence the $g-r$ colour profiles are unlikely to be significantly affected by the 
PSF difference crossing HSC $g$ and $r$-bands. 

Compared with surface brightness profiles, the colour profiles of galaxies and their stellar haloes span a much smaller 
range of magnitudes, which could be comparable to the amount of K-corrections \citep{2007AJ....133..734B} over the 
redshift range of $0.05<z<0.16$. Since we choose not to include K-corrections, we further split our sample of isolated 
central galaxies into two subsamples based on their redshifts ($0.05<z<0.1$ and $0.1<z<0.16$), to minimise the effect of 
ignoring K-corrections. 

The measured $g-r$ and $g-i$ colour profiles are shown in Fig.~\ref{fig:profcolor}. Symbols connected by solid and dashed 
lines correspond to the lower and higher redshift subsamples. It is clear that the colour profiles of galaxies at different 
redshifts are not the same, and the difference is stellar mass dependent. Massive galaxies at lower redshifts have bluer colours, 
whereas smaller galaxies tend to have redder colours at lower redshifts. $10.5<\log_{10}M_\ast/M_\odot<10.8$ is the transition 
stellar mass bin where the trend starts to reverse.

The trend is mainly due to the combined effect of K-corrections and the survey flux limit. Massive galaxies are mostly 
red, and their passive evolution with redshift is weak. The main effect is the ignorance of K-correction. The prominent 4000{\AA} 
break feature of massive galaxies due to absorption shifts to redder bands with the increase in redshift. Hence massive 
galaxies at $0.1<z<0.16$ are redder in observed frame than galaxies at $0.05<z<0.1$.

On the other hand, the spectra of small star-forming galaxies do not have prominent 4000{\AA} absorption features, the 
colour trend with redshift is a result of the selection bias given the survey flux limit. The stellar mass-to-light ratio is 
strongly correlated with the intrinsic colour of galaxies. For fixed stellar mass, blue star forming galaxies have smaller 
stellar mass-to-light ratios, and are brighter. Thus for a fixed flux limit at a given redshift, more small blue star forming 
galaxies can be observed, compared with red galaxies with the same stellar mass, which could have already become fainter 
than the flux limit. The effect is stronger at higher redshifts when the sample is more incomplete \footnote{We do not attempt 
to correct for the incompleteness in our sample due to the survey flux limit, but we have tested our results by weighting 
each image with the inverse of the maximum volume calculated based on the luminosity of the galaxy and the flux limit before 
stacking. The weighting barely changes our conclusions.}, which explains why less massive galaxies with $0.1<z<0.16$ in our 
flux-limited galaxy sample are bluer.

The colour difference between low and high redshift subsamples is less obvious in the least massive bin. This is 
probably due to the very small sample size of isolated central galaxies in that bin with $z>0.1$ 
(see Fig.~\ref{fig:redshifthist}).

The HSC colour profiles can be measured out to about 70, 50, 50, 30, 30 and 20~kpc for the six stellar mass bins. Two 
general trends are clearly revealed from Fig.~\ref{fig:profcolor}. Firstly, massive galaxies are redder, which 
is in good agreement with Fig.~\ref{fig:imagedeep}. Moreover, for galaxies with fixed stellar mass, the colour is 
redder in the inner region and bluer outside. However, the few outer points show signs of positive gradients in their 
$g-r$ colour, and the same is true for the $g-i$ colour. On such scales, the associated uncertainty levels as shown by 
the size of errorbars in Fig.~\ref{fig:profcolor} are also very large.

\cite{2014MNRAS.443.1433D} investigated the colour profiles around isolated central galaxies at $0.06<z<0.1$. Their results 
also show reddest colour in the very central regions of galaxies and bluer colours outside. There are indications of colour 
minimums in their measured profiles, beyond which there are positive colour gradients as well.

We will show in Appendix~\ref{app:maskrandom} that the indications of colour minimums and positive gradients are sensitive 
to the choice of source detection thresholds, which are used to create masks for companions sources. Lower detection thresholds 
help to mask more extended emissions of companion sources and weaken the indications of positive gradients. Besides, based on 
our discussions in Sec.~\ref{sec:psfdeconv}, the redder colour in the outer stellar halo is likely contaminated by scattered 
light from the central galaxy due to the extended PSF wings. In the next subsection, we move on to investigate the effect 
of extended PSF wings on colour profiles.

\subsection{PSF effect on colour profiles}
\label{sec:psfdeconvcolor}

Redder colour in the outer stellar halo can be related to instrumental effects. For example, SDSS used thinned 
CCDs, and the extended wings of the red-band PSF are wider than that of bluer bands for thinned CCDs, which is called 
the ``red halo'' effect \citep[e.g.][]{1998SPIE.3355..608S} that can affect the colour in outer parts of 
bright objects. \cite{2005ApJ...622..244W} pointed out that the SDSS $i$-band PSF is more extended than bluer bands, 
which scatters more light and brings redder colour on scales of $\sim$10$\arcsec$. 

Beyond the CCD scattering-specific red halo effect, the scattering by atmospheric aerosols, dust and the 
reflection within the instrument/telescope can cause the so-called ``aureole''\citep[e.g.][]{2014A&A...567A..97S} 
in the PSF on scales larger than the red halo. The aureole typically has size from a few tenth of arcsecond to 
about 1$^\circ$. At redshift $z=0.07$, the corresponding physical scales for 20$\arcsec$ are $\sim$30~kpc. The 
scales are roughly consistent with the radius where we start to see flattening and positive 
gradients in the colour profiles. 

HSC uses thicker CCDs. The red halo effect is expected to be less prominent, as is supported by 
Fig.~\ref{fig:psf} that the HSC $i$-band PSF is less extended. We also note that both \cite{2014MNRAS.443.1433D} 
and our study measure the $g-r$ colour, for which the red halo effect could be less prominent, given the fact 
that the PSF in $g$ and $r$-bands are more similar to each other for both HSC and SDSS. However, despite the 
similar PSF shapes in HSC $g$ and $r$-bands and the smaller PSF in HSC $i$-band, the scattered light from the 
central galaxy through the extended PSF might still be a source of contamination to the $g-r$ and $g-i$ colour 
of the outer stellar halo, because the central parts of galaxies have stronger emissions in redder bands.

We can easily check the effect of the PSF on the colour profiles following the approach in Sec.~\ref{sec:psfdeconv}. 
This is presented in Fig.~\ref{fig:profcolor_deconv}. Compared with the original measurements (dashed lines), the 
PSF-deconvolved colour profiles drop slightly faster and have steeper slopes, with bluer colours in the outskirts. 
It seems the original colour of the stellar halo is slightly contaminated by the PSF scattered light from the 
central part of the galaxy. As a result, the indications of positive colour gradients are only slightly weakened, 
but do not disappear. 

\subsection{surface brightness and colour profiles split by galaxy concentration and stellar mass}
\label{sec:splitC}

\begin{figure*} 
\includegraphics[width=0.98\textwidth]{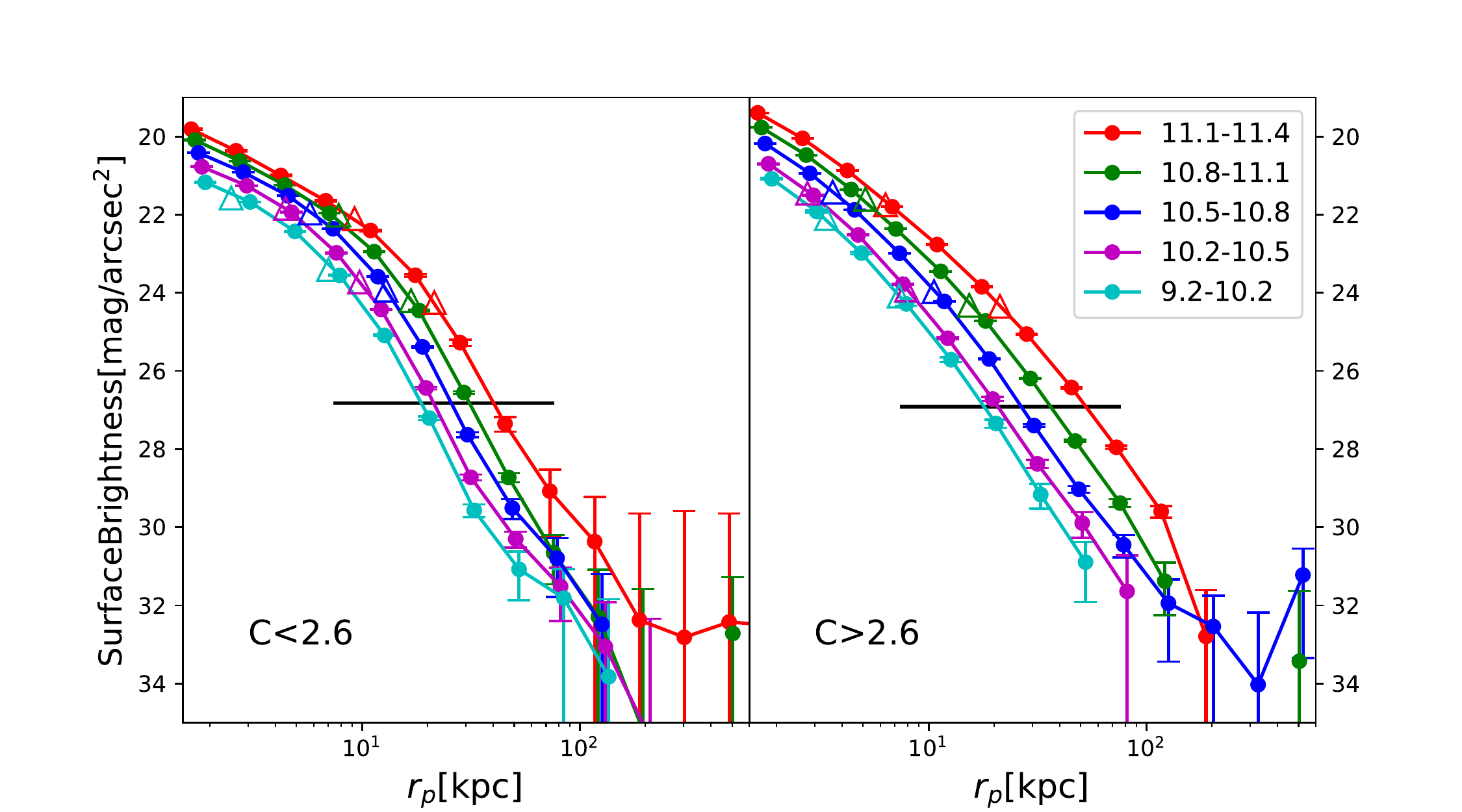}%
\caption{{\bf Left:} Surface brightness profiles in HSC $r$-band stacked on isolated central galaxies in five stellar 
mass bins, whose concentration, $C$, is required to be smaller than 2.6. {\bf Right:} Similar to the left panel, but 
shows surface brightness profiles stacked on isolated central galaxies with concentration larger than 2.6 . Errorbars 
in both panels are the 1-$\sigma$ scatter of 50 boot-strap realisations. The black horizontal lines mark the average 
rms of the background noise for single images before stacking. The two empty triangle associated with each curve 
with the corresponding colour shows the mean radii that contain 90\% and 50\% of the Petrosian flux for galaxies 
in each bin. Small horizontal shifts have been made to the second to least massive bins. No PSF corrections 
have been made.
}
\label{fig:profcon}
\end{figure*}

\begin{figure} 
\includegraphics[width=0.49\textwidth]{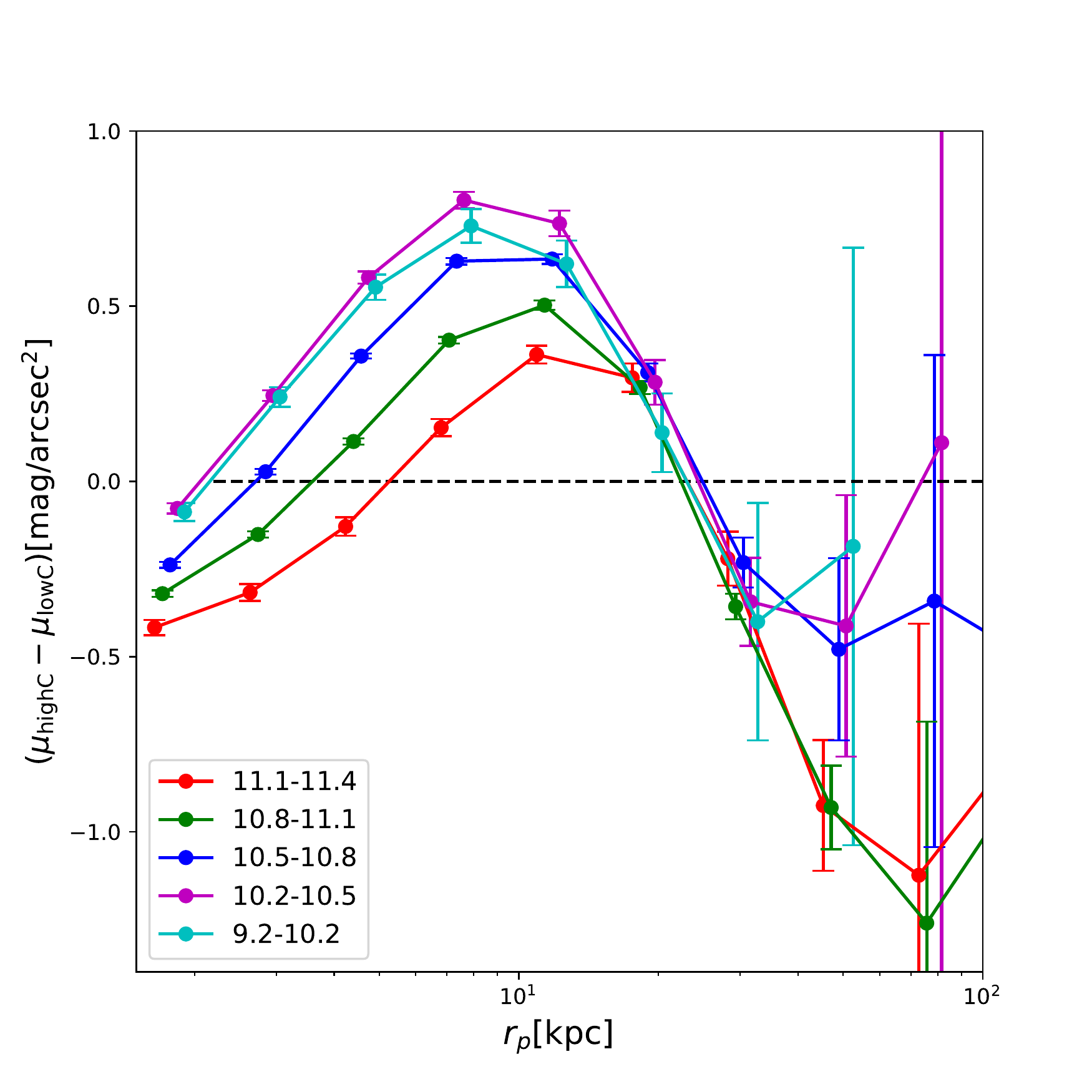}%
\caption{Differences between surface brightness profiles of isolated central galaxies with high and 
low concentrations in HSC $r$-band. Negative/positive values tell the profiles of high concentration 
galaxies are brighter/fainter. Small horizontal shifts have been added to the second to least massive 
bins. 
}
\label{fig:diffcon}
\end{figure}

\begin{figure*} 
\includegraphics[width=0.48\textwidth]{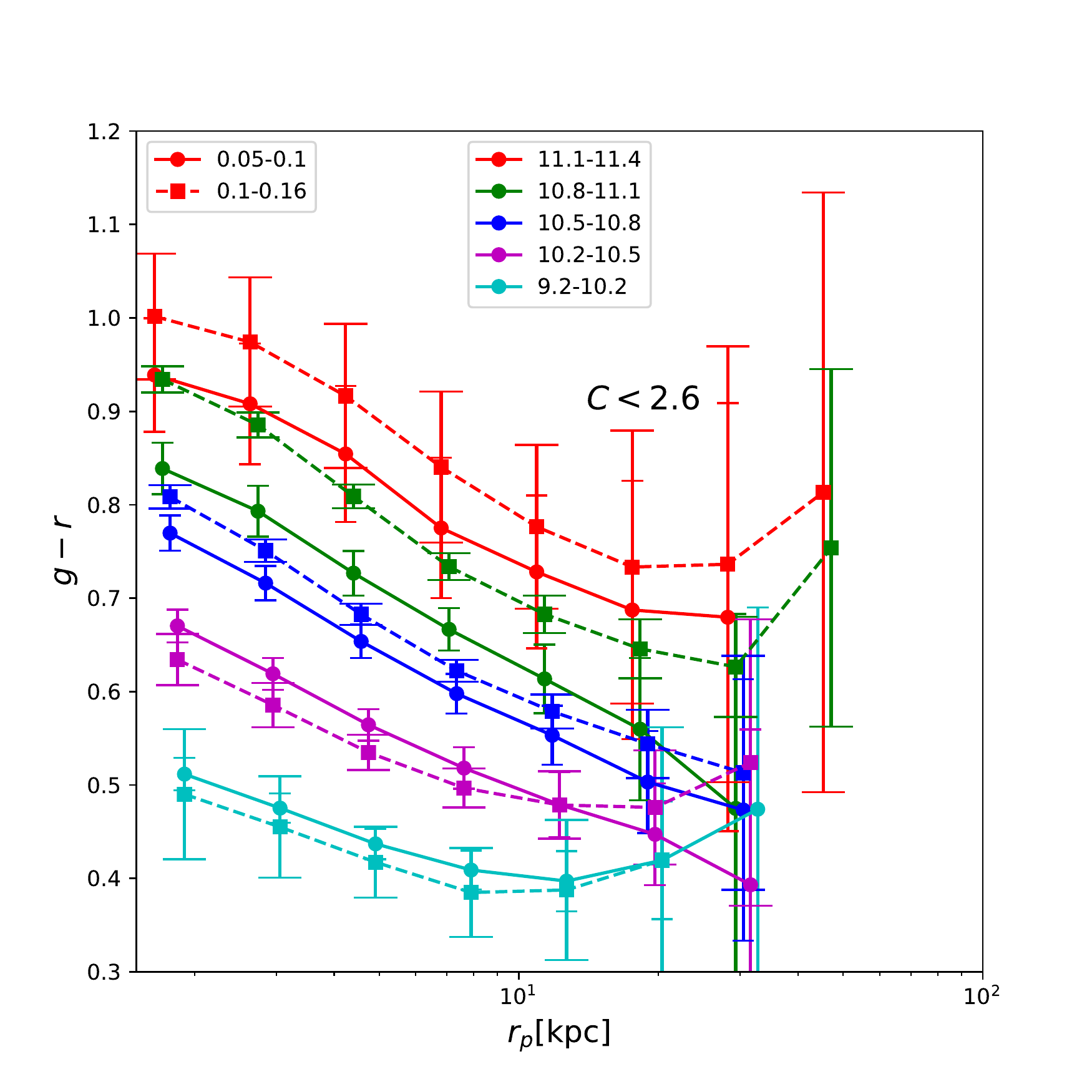}%
\includegraphics[width=0.48\textwidth]{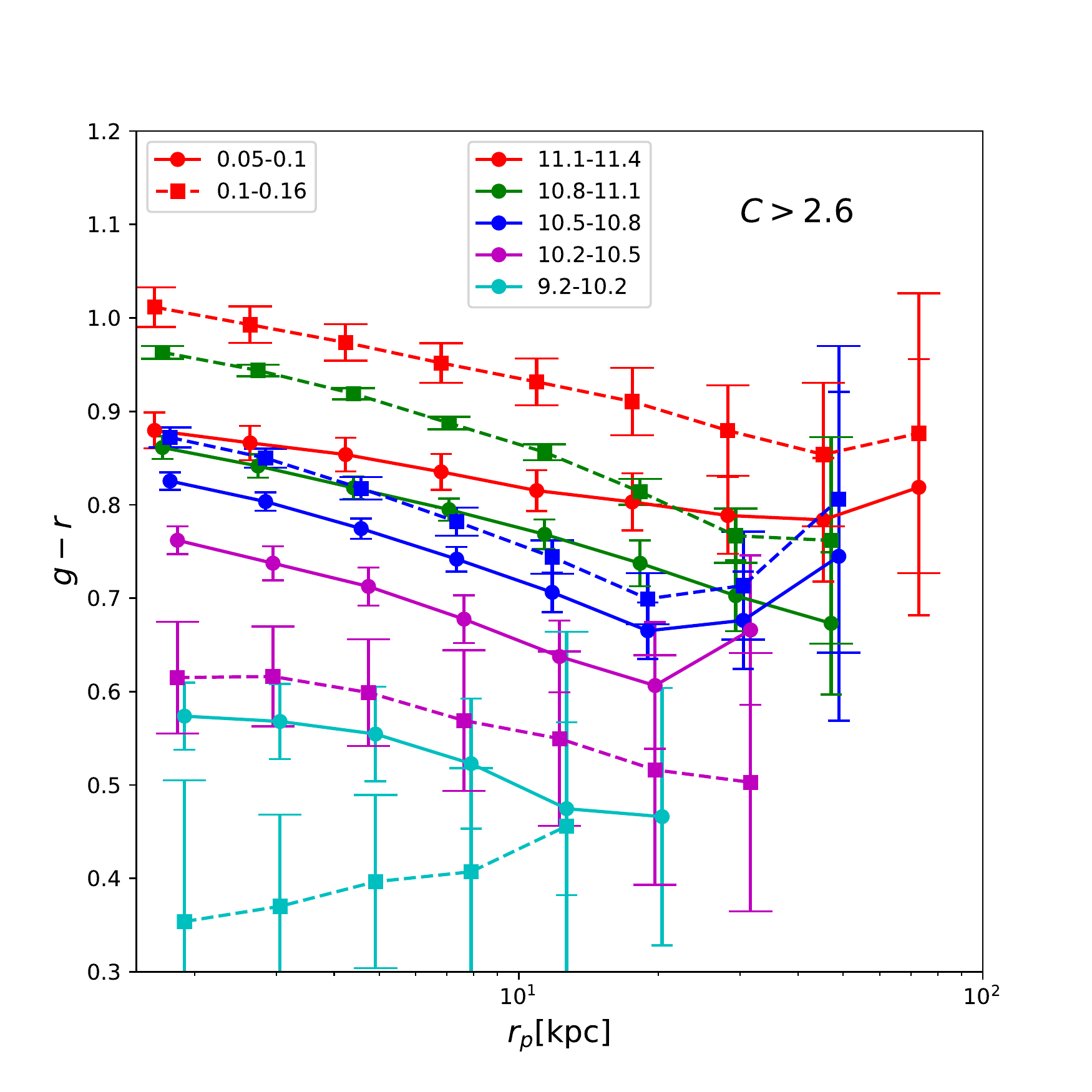}
\caption{{\bf Left:} $g-r$ colour profiles stacked on isolated central galaxies with concentration, $C$, smaller 
than 2.6, and in five stellar mass bins (see the legend). Galaxies in each bin are further divided into two subsamples 
based on their redshifts, i.e., $0.05<z<0.1$ (dots with solid lines) and $0.1<z<0.16$ (squares with dashed lines). No 
PSF corrections have been made. {\bf Right:} Similar to the left panel, but shows $g-r$ colour profiles stacked on 
isolated central galaxies with concentration larger than 2.6. Errorbars in both panels are the 1-$\sigma$ scatter 
of 50 boot-strap realisations. Small horizontal shifts have been made to the second to least massive bins. 
}
\label{fig:profcolorcon}
\end{figure*}

\begin{table}
\caption{Number of isolated central galaxies with low ($C<2.6$) and high ($C>2.6$) concentrations}
\begin{center}
\begin{tabular}{lrrrrrrr}\hline\hline
$\log M_*/M_\odot$ & \multicolumn{1}{c}{$C<2.6$} & \multicolumn{1}{c}{$C>2.6$} \\ \hline
11.1-11.4 & 169 & 1269 \\
10.8-11.1 & 1416 & 3652 \\ 
10.5-10.8 & 2638 & 2934 \\
10.2-10.5 & 2223 & 1108 \\
9.2-10.2 & 1994 & 343 \\
\hline
\label{tbl:lowChighC}
\end{tabular}
\end{center}
\end{table}

Following \cite{2014MNRAS.443.1433D}, we investigate the surface brightness and colour profiles for isolated central galaxies 
and their stellar haloes split into two subsamples with high concentrations ($C>2.6$) and low concentrations ($C<2.6$). Here the 
galaxy concentration is defined as the ratio of the radii that contain 90\% and 50\% of the Petrosian flux in $r$-band\footnote{The 
radii that contain 90\% and 50\% of the Petrosian flux are downloaded from the SDSS database. Deep HSC images can potentially 
improve the measured Petrosian radius and flux, but we just focus on measurements from SDSS. The SDSS spectroscopic sample is 
bright enough to ensure robust measurements, which already satisfies our science purposes.}, i.e., $C=R_{90}/R_{50}$. We choose 
the cut at $C=2.6$ to be consistent with \cite{2014MNRAS.443.1433D}. The number of low and high concentration isolated central 
galaxies in different stellar mass bins are provided in Table~\ref{tbl:lowChighC}. 

After dividing galaxies into subsamples of high and low concentrations, it is difficult to maintain enough signal-to-noise 
ratios for the two least massive bins ($9.9<\log_{10}M_\ast/M_\odot<10.2$ and $9.2<\log_{10}M_\ast/M_\odot<9.9$), especially 
for high concentration galaxies because low mass galaxies are less concentrated. Hence we choose to merge them into a single 
stellar mass bin (the bottom row of Table~\ref{tbl:isos}). The surface brightness and colour profiles are shown in 
Fig~\ref{fig:profcon} and Fig.~\ref{fig:profcolorcon}, respectively. We delay discussions about PSF-deconvolved surface 
brightness profiles of low and high concentration galaxies to Sec.~\ref{sec:universal}. For colour profiles, the effect 
of PSF is similar to our findings in Sec.~\ref{sec:psfdeconvcolor} that PSF tends to slightly flatten the profiles, and 
thus we avoid repeatedly showing the PSF-deconvolved results. 

Left and right panels of Fig~\ref{fig:profcon} are for low and high concentration galaxies, and for each, they are further 
separated into stellar mass bins. Low and high concentration galaxies show clear differences in their surface brightness 
profiles. Less concentrated galaxies in the left panel of Fig~\ref{fig:profcon} are more extended beyond 100~kpc. Despite 
the large errors, the surface brightness profiles extend up to about 500, 200, 120, 120 and 120~kpc for the five 
stellar mass bins. Moreover, the profiles of low concentration galaxies are more flattened in the very central region as 
a result of the definition of being less concentrated, but then drop faster beyond 20~kpc. This is more clearly revealed 
in Fig.~\ref{fig:diffcon} that within 20~kpc, the surface brightness differences between high and low concentration galaxies 
are mostly positive, which means low concentration galaxies are brighter on such scales. For scales between 20 and 
100~kpc, the differences are dominated by negative values, showing high concentration galaxies are brighter. This agrees 
with the conclusion of \cite{2014MNRAS.443.1433D} that the stellar halo of low concentration galaxies have steeper slopes 
beyond $r_p=25$~kpc. 

To guide the eye, we plot on top of each curve as empty triangles the mean Petrosian radii that contain 90\% and 
50\% of the Petrosian flux ($R_{90}$ and $R_{50}$). $R_{50}$ of high concentration galaxies are clearly smaller. 
The black horizontal lines in both panels mark the mean background noise of individual images before stacking.
Triangles are all above the noise level for individual images, which is reasonable because $R_{50}$ and 
$R_{90}$ are measured based on individual galaxy images.

Our colour profiles for low and high concentration galaxies, however, are not fully consistent with \cite{2014MNRAS.443.1433D}. 
\cite{2014MNRAS.443.1433D} found indications of flattened colour profiles for the outer stellar halo of high concentration 
galaxies, whereas for low concentration galaxies, their colour profiles are the reddest in the very central parts 
of galaxies, and become bluer up to some typical radius between 5 and 20~kpc. The typical radius corresponds to a colour 
minimum, beyond which their colour profile of low concentration galaxies tends to show indications of positive gradients 
or turn redder again. 

In Fig.~\ref{fig:profcolorcon}, we do see differences in the $g-r$ colour profiles for low and high concentration 
galaxies. High concentration galaxies have redder colours and shallower negative gradients. However, for low 
concentration galaxies there are no signs of colour minimums on 5 to 20~kpc scales in the four most massive 
bins of the low redshift subsample, which have comparable stellar mass and redshift range as \cite{2014MNRAS.443.1433D}. 

Cyan squares connected by the dashed line in the right panel of Fig.~\ref{fig:profcolorcon} show positive gradients 
within 10~kpc, which is very likely due to the small sample size. There are only $\sim$20 isolated central galaxies 
in that bin, because low mass galaxies with high concentrations at $z>0.1$ is rare. In fact, the colour profile for 
this bin is consistent with being flat given the large errors. Moreover, the indications of positive colour gradients 
could also be consistent with being flat given the large errors on corresponding scales. 

Despite the large errors of the colour profiles of the outer stellar halo, we provide discussions about 
possible systematics (the robustness of colour profiles to the chosen source detection thresholds in 
masking companions) and physical mechanisms that might cause positive colour gradients in Appendix~\ref{app:maskrandom} 
and Sec.~\ref{sec:positivegrad}. We also check in Appendix~\ref{app:isocmp} that the colour profiles are not sensitive 
to the change of isolation criteria adopted to select isolated central galaxies. In addition, since we adopted circular 
radial binning, while results in many previous studies are based on elliptical isophotal binning, we repeat our 
analysis with elliptical binning by rotating galaxy images along their major axes before stacking. Details are provided 
in Appendix~\ref{app:ell}. We show that circular radial binning is unlikely to significantly change 
the location of colour minimums.

\subsection{Universality of stellar haloes}
\label{sec:universal}

\begin{figure*} 
\includegraphics[width=0.98\textwidth]{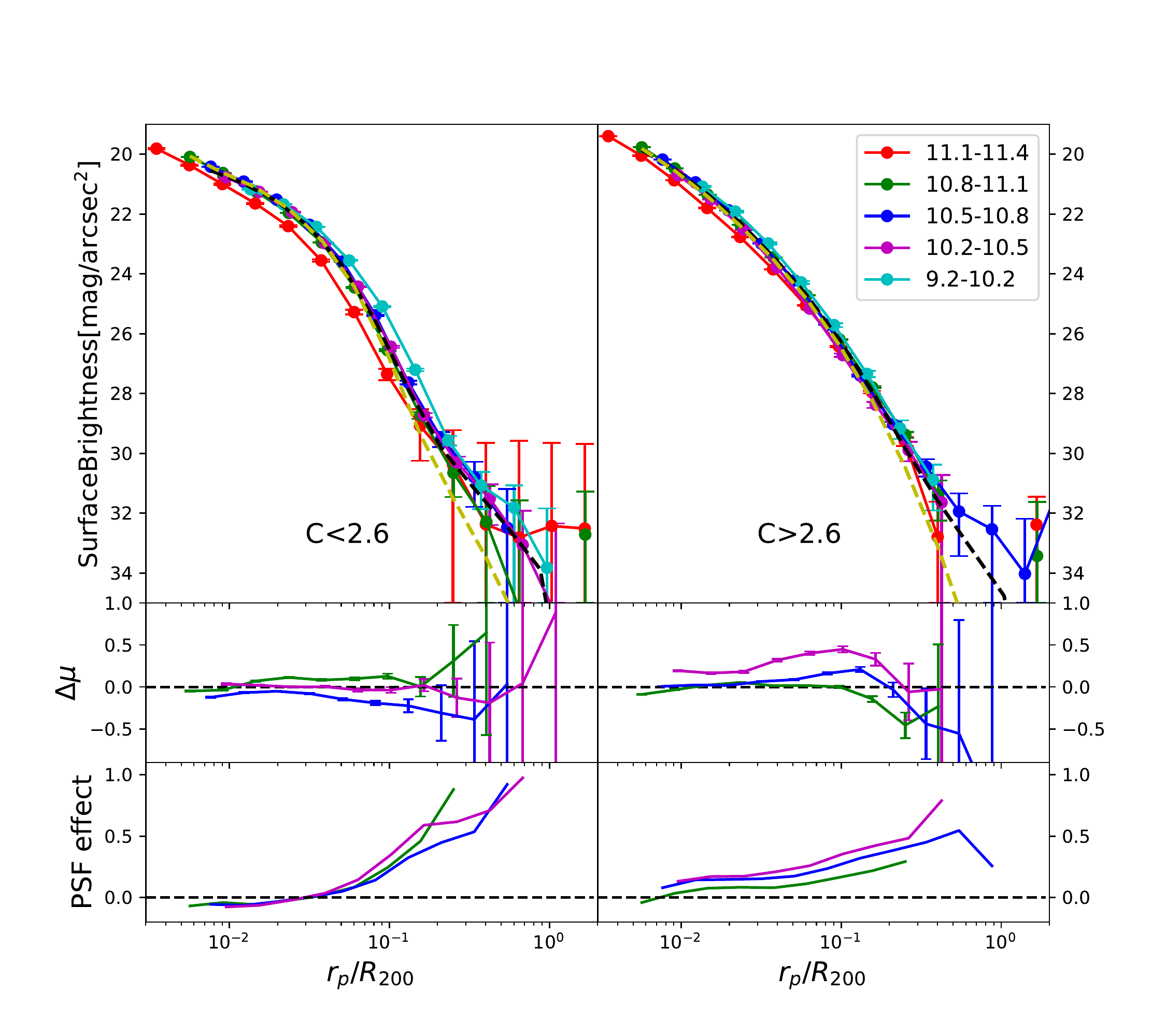}%
\caption{{\bf Top:} Surface brightness profiles for isolated central galaxies with low (left) and high (right) concentrations
and their stellar haloes in HSC $r$-band. The $x$-axis quantity is the projected radial distance scaled by the halo virial 
radius. Black dashed lines are triple (left) and double (right) Sersic models (PSF convolved), jointly fit to the three 
middle bins ($10.2<\log_{10}M_\ast/M_\odot<11.1$). Yellow dashed lines show the PSF-free best-fit Sersic models. Errorbars 
are boot-strap errors, which reflect the 1-$\sigma$ scatter of galaxies within each stellar mass bin. {\bf Middle:} Differences 
between best fits and true profiles of the three stellar mass bins used for fitting. {\bf Bottom:} Estimated fractions 
which are contaminated by scattered emissions through the PSF, for the three stellar mass bins used for fitting. 
}
\label{fig:profconscale}
\end{figure*}

\begin{figure*}
\includegraphics[width=0.98\textwidth]{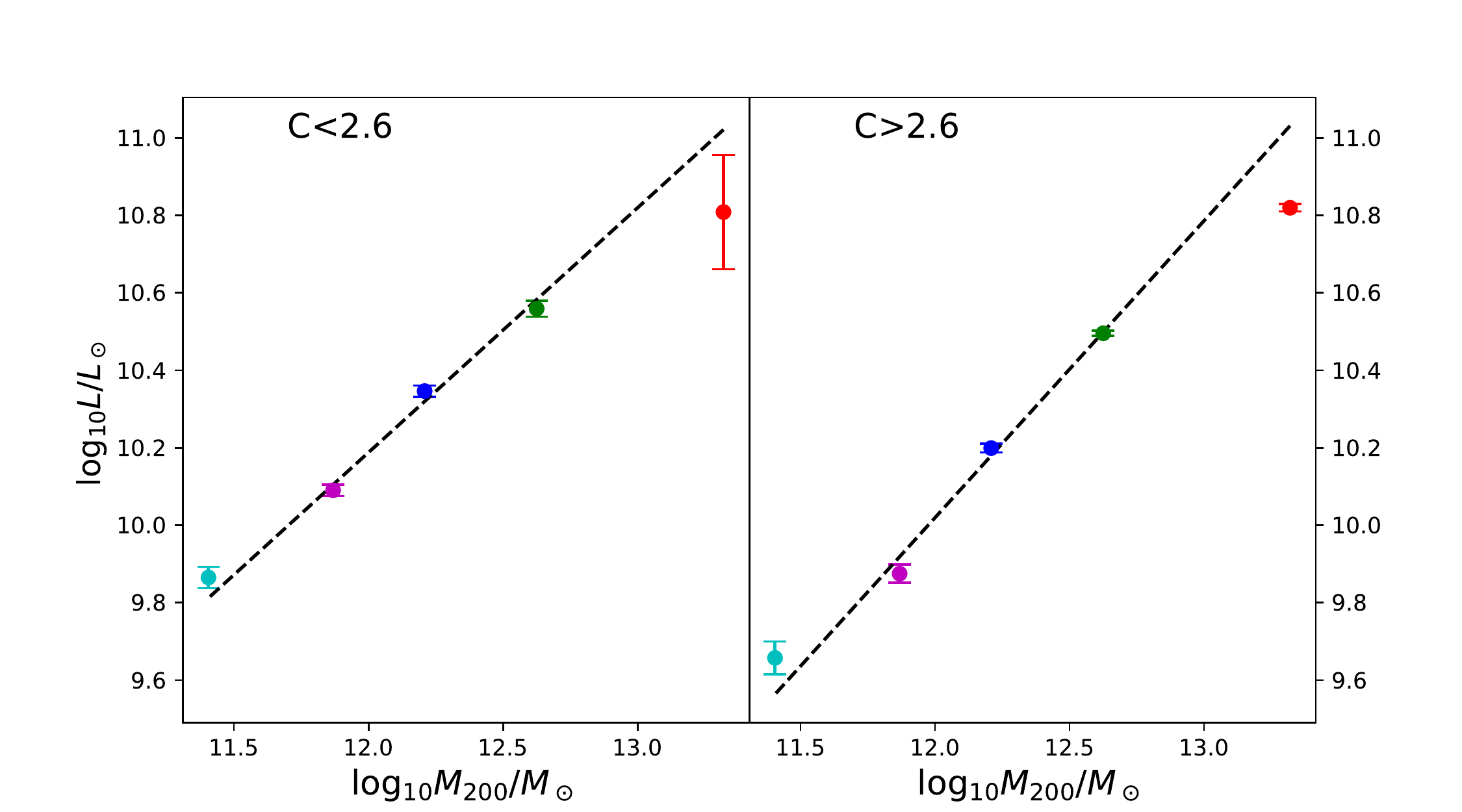}%
\caption{Integrated luminosity (out to the last positive data point of surface brightness profiles) versus the mean 
halo mass (based on a mock galaxy catalogue) for isolated central galaxies with low (left) and high (right) concentrations. 
Black dashed lines are best fits of $\log_{10}L/L_\odot\propto (0.6319\pm0.0505) \log_{10}M_{200}/M_\odot$ (left) and 
$\log_{10}L/L_\odot\propto (0.7369\pm0.0370) \log_{10}M_{200}/M_\odot$ (right). Only the three middle data points are 
used for fitting. 
}
\label{fig:LMh}
\end{figure*}

The density profiles of dark matter haloes, their mass accretion histories, the spatial, mass and phase-space distributions 
of their accreted subhaloes and satellite galaxies can all be described by unified models once they are scaled by some 
characteristic quantities \citep[e.g.][]{Navarro_1996, Navarro_1997, 2003ApJ...597L...9Z,2009ApJ...707..354Z,2016MNRAS.457.1208H,
2017ApJ...850..116L,2019MNRAS.484.5453C}. In this section, we investigate whether the surface brightness profiles of the stellar 
halo can be unified as well. We look into this in Fig.~\ref{fig:profconscale}, where the surface brightness profiles of low and 
high concentration galaxies in different stellar mass ranges are plotted as a function of the projected distance $r_p$ scaled 
by the halo virial radius $R_{200}$. 

Note, although we estimate $R_{200}$ based on abundance matching to select our sample of isolated central galaxies, the halo 
mass versus stellar mass relation of selected galaxies can be biased from that of all central galaxies, and thus we highlight 
again that $R_{200}$ values of isolated central galaxies in Table~\ref{tbl:isos}, which are used throughout the paper to 
determine the image size, and will be used to scale profiles in this section, are all based on isolated central 
galaxies selected in the mock galaxy catalogue of \cite{2011MNRAS.413..101G}. Hence $R_{200}$ in Table~\ref{tbl:isos} 
are slightly different form those estimated from abundance matching. However, we should bear in mind the uncertainty of 
$R_{200}$, which can in principle be quantified by comparing with true weak lensing measurements. We move on with this 
uncertainty, and postpone more accurate analysis of $R_{200}$ through weak lensing to future studies.

It is very encouraging that after plotting the profiles as a function of $r_p/R_{200}$ instead of $r_p$, the amplitudes 
and shapes of profiles for galaxies with different stellar masses tend to be very similar to each other. Profiles of the 
most and least massive bins show some discrepancies from the others after the scaling, but for the other stellar mass bins, 
the profiles are very similar to each other. 

We fit the following triple and double Sersic profiles to the three middle stellar mass bins of low and high concentration 
galaxies 

\begin{align}
 I/I_0=10^{I_{e,1}} \mathrm{exp}\{ -b_{n1}(x/x_{e,1})^{1/n_1}-1]\}\nonumber \\
 +10^{I_{e,2}} \mathrm{exp}\{ -b_{n2}[(x/(x_{e,1}+x_{e,2}))^{1/n_2}-1]\}\nonumber \\
 +10^{I_{e,3}} \mathrm{exp}\{ -b_{n3}[(x/(x_{e,1}+x_{e,2}+x_{e,3}))^{1/n_3}-1]\}, 
\end{align}

\begin{align}
 I/I_0=10^{I_{e,1}} \mathrm{exp}\{ -b_{n1}(x/x_{e,1})^{1/n_1}-1]\}\nonumber \\
 +10^{I_{e,2}} \mathrm{exp}\{ -b_{n2}[(x/(x_{e,1}+x_{e,2}))^{1/n_2}-1]\}, 
\end{align}
where we let $x=r_p/R_{200}$. $x_{e,1/2}$, $x_{e,1}+x_{e,2}$ and $x_{e,1}+x_{e,2}+x_{e,3}$ are the three effective radii, 
scaled by the halo virial radius, $R_{200}$. We formulate the effective radii in the incremental way to ensure a monotonic 
ordering among them during fitting. $b_{n,1/2/3}$ is defined through $x_{e,1/2/3}$.

Again the model profiles are convolved with the PSF of Fig.~\ref{fig:psf}. Measurements of the three middle stellar mass 
bins are jointly used for the fitting. The best fits are presented as black dashed lines, with the PSF-free Sersic model 
demonstrated as yellow dashed curves. The best-fit parameters and associated errors are provided in Table~\ref{tbl:bestfits}.

\begin{table}
\caption{Best-fit parameters of the double Sersic profile}
\begin{center}
\begin{tabular}{lrrr}\hline\hline
 & \multicolumn{1}{c}{$C<2.6$} & \multicolumn{1}{c}{$C>2.6$} \\ \hline
$I_{e,1}$ & -7.3500$\pm$0.0577 & -8.4710$\pm$0.0741 \\
$n_1$ & 0.0101$\pm$0.0707 &  1.5320$\pm$0.2115 \\
$x_{e,1}$ & 0.0015$\pm$0.0012 & 0.0056$\pm$0.0002 \\ 
$I_{e,2}$ & -8.7930$\pm$0.0153 & -8.9330$\pm$0.0423 \\
$n_2$ & 1.143$\pm$0.0372 & 2.6190$\pm$0.0804 \\
$x_{e,2}$ & 0.0231$\pm$0.0012 & 0.0165 $\pm$0.0008 \\
$I_{e,3}$ & -9.9220$\pm$0.1447 & - \\
$n_3$ & 3.691$\pm$0.1859 & - \\
$x_{e,3}$ & 0.0001$\pm$0.0019 & - \\
\hline
\label{tbl:bestfits}
\end{tabular}
\end{center}
\end{table}

Based on the best-fit unified model, we can estimate the effect of PSF in the same way as Sec.~\ref{sec:psfdeconv}. Bottom 
panels of Fig.~\ref{fig:profconscale} show the fractions in surface brightness profiles which are contributed by the scattered 
light through the extended PSF, for the three middle stellar mass bins used for fitting. The colour legend is the same as 
for top panels. Note the PSF-free best-fit model and the PSF are the same, but the PSF effect as a function of $r_p/R_{200}$
varies with the redshift distribution and the value of $R_{200}$ in each stellar mass bin\footnote{Best fits plotted in top 
panels are averaged over the three bins.}. The fraction of PSF scattered light is smaller than 50\% within 0.1$R_{200}$ for 
low concentration galaxies, and increases very rapidly on larger scales. Interestingly, the fraction is lower for high 
concentration galaxies beyond 0.1$R_{200}$. This indicates the outer stellar halo of less concentrated galaxies, which 
are dominated by star-forming late type galaxies, are contaminated more by the PSF scattered light.

The residuals compared with the best fits are demonstrated in the two middle panels of Fig.~\ref{fig:profconscale}, 
and for the three stellar mass bins used for fitting only.  On scales larger than $0.3R_{200}$, the residuals increase 
significantly, reflecting large scatters on such scales. On smaller scales, the residuals are about 0.1 to 0.2~dex. It 
is clear that the best fits tend to reconcile among measurements of different stellar mass bins, which are jointly used 
for the fitting. We conclude that, although the profiles become very similar to each other after scaled by $R_{200}$, 
discrepancies still remain even on scales not sensitive to the effect of PSF. The discrepancies might be partly related 
to the fact that we have ignored K-corrections. Moreover, such discrepancies might be caused by uncertainties of $R_{200}$. 
More accurate determination of $R_{200}$ has to depend on weak lensing measurements, and we will make further investigations 
in future studies. Lastly but importantly, we have ignored the scatter of the host halo mass distribution at fixed stellar 
mass, and the large scatter with respect to the mean halo mass versus stellar mass relation might be responsible for the 
discrepancies as well. 

The fact that the surface brightness profiles of the stellar halo can be approximately modelled by the same functional 
form once scaled by the halo virial radius is very interesting. In a recent study of \cite{2019ApJ...874..165Z}, 
which was submitted slightly later than the submission of this paper, the self-similarity is reported for massive 
galaxy clusters at $z\sim0.25$, after scaling clusters with different richness by the virial radius of the cluster. 

The self-similarity indicates that, the formation of the faint outer stellar halo is likely dominated by 
physical processes that can be modelled through halo-related properties. However, the seemingly unified stellar 
halo profile is only valid in terms of the ``averaged'' profiles for a large sample of galaxies. The shapes and 
amplitudes of stellar haloes for individual galaxies show large diversities \citep[e.g.][]{2017MNRAS.466.1491H,2018MNRAS.475.3348H}, 
reflecting the stochastic of merging histories. Using the high-resolution Illustris simulation, \cite{2014MNRAS.444..237P} 
investigated the logarithmic slopes of spherically averaged stellar density profiles for galaxies at $z=0$. The slopes 
are at first measured for individual galaxies in a radial range of $R_\mathrm{vir}/50$ to $R_\mathrm{vir}$. While 
individual slopes show large radial-dependence and large galaxy-to-galaxy scatters, the median slopes show strong 
trends with halo mass. At fixed halo mass, the slopes also depend on the colour, morphology, age and stellar mass 
of galaxies. Our stacked surface brightness profiles show a strong dependence on galaxy type (low and high 
concentration subsamples), but there is no clear indication for the slope of averaged profiles to depend on 
stellar mass or halo mass.

In fact, if the universality of the stellar halo profiles strictly holds, it means that the luminosity should be 
proportional to $M_{200}^{2/3}$. The total luminosity is obtained through $L=\int I(r,R_{200}) 2\pi r {\rm d}r$ or 
$L=\int I(x,R_{200}) 2\pi (x R_{200}) {\rm d}(x R_{200})$, where $x=r/R_{200}$. Now, $I(r,R_{200})$ can be modelled 
as $I(x)$, i.e., the surface brightness profile only depends on $x=r/R_{200}$. Thus the integral of $L=\int I(x) 2
\pi (x R_{200}) {\rm d}(x R_{200})$  naturally leads to the conclusion that the luminosity, $L$, is proportional to 
$R_{200}^2$ and hence $M_{200}^{2/3}$. The best-fit slopes of the integrated luminosity versus halo mass are 
$0.6319 \pm 0.0505$ and $0.7369 \pm 0.0370$ for low and high concentration galaxies with $10.2<\log_{10}M_\ast/M_\odot<11.1$ 
respectively (Fig.~\ref{fig:LMh}). The slopes are close to 2/3. The most massive data point shows more significant 
deviation from the scaling relation. This is in very good agreement with \cite{2008ApJ...676..248Y}. \cite{2008ApJ...676..248Y} 
measured the central galaxy luminosity versus halo mass relation through abundance matching using the SDSS galaxy 
group catalogue. They reported a best-fit relation of $L_C\propto M_h^{0.68}$ for $10^{11.6}h^{-1}M_\odot \leq M_h
\leq 10^{12.5}h^{-1}M_\odot$, whereas for $M_h> 10^{12.5}h^{-1}M_\odot$, the slope is significantly flat. Both the 
best-fit slope and the halo mass range where the slope is close to 0.68 or gets flattened are in very good agreement 
with our independent measurements here. 

The halo mass of \cite{2008ApJ...676..248Y} is obtained through abundance matching to the characteristic luminosity 
of galaxies in the SDSS group catalogue \citep{2007ApJ...671..153Y}, and hence maybe the relation between halo mass 
and luminosity is a result from abundance matching. In addition, although $R_{200}$ and $M_{200}$ for our isolated 
central galaxy sample is not exactly obtained from abundance matching, it only slightly deviates from the abundance 
matching relation. Since stellar mass is strongly correlated with luminosity, the scaling between luminosity and halo 
mass might be a reflection of how $M_{200}$ is determined. However, we emphasise that, we are able to self-consistently 
explain why the slope is close to the value of $2/3$, which cannot be obviously deduced through abundance matching. 
With $R_{200}$ determined through the stellar mass and the aid of a mock galaxy catalogue, we are at least able to 
bring the stellar halo profiles for galaxies spanning a wide range of stellar masses to be close to universal.

\section{Discussions}
\label{sec:discussion}

\subsection{Fraction of missing light}

Single component models are often used to fit the surface brightness distribution of galaxy images, such as the  
de Vaucouleurs profile for elliptical galaxies and the exponential profile for spiral galaxies. A pure de Vaucouleurs 
or a pure exponential model profile was used to fit galaxy images in SDSS, with the integrated flux called model 
magnitude. The fits are dominated by the central part of the galaxy, and hence for bright galaxies ($r<18$), model 
magnitudes underestimate the total flux \citep[e.g.][]{2001AJ....122.1861S}. As a comparison, the composite model 
(cModel) fits a combination of de Vaucouleurs and exponential profiles to the observed surface brightness 
of galaxies (Eqn.~\ref{eqn:cmodel}), which gives significant improvements in terms of modelling the total 
flux and also agrees well with the SDSS Petrosian magnitude

\begin{equation}
 I_\mathrm{composite}=\mathrm{frac_{dev}}I_0 e^{-7.67(r/R_e)^{1/4}}+(1-\mathrm{frac_{dev}})I_0 e^{-1.68(r/R_e)}.
 \label{eqn:cmodel}
\end{equation}

Even though cModel magnitudes perform better in modelling the total flux, there might be missing light in outskirts 
of the faint stellar halo, which cannot be detected given the background noise level of single images. Our stacked 
surface brightness profiles, however, are deeper than what can be obtained through individual images and can be used 
to test how cModel magnitude performs to recover the total flux, if the best fits are only achieved using measurements 
above the background noise level of single images. 

The average background noise level of individual images is represented by the black horizontal lines in Fig.~\ref{fig:profcon}. 
We fit PSF convolved Eqn.~\ref{eqn:cmodel} to data points above this level. Since only the central PSF has been used by the 
pipeline for cModel profile fitting, we use the central PSF for the convolution as well. 

\begin{figure*} 
\includegraphics[width=0.98\textwidth]{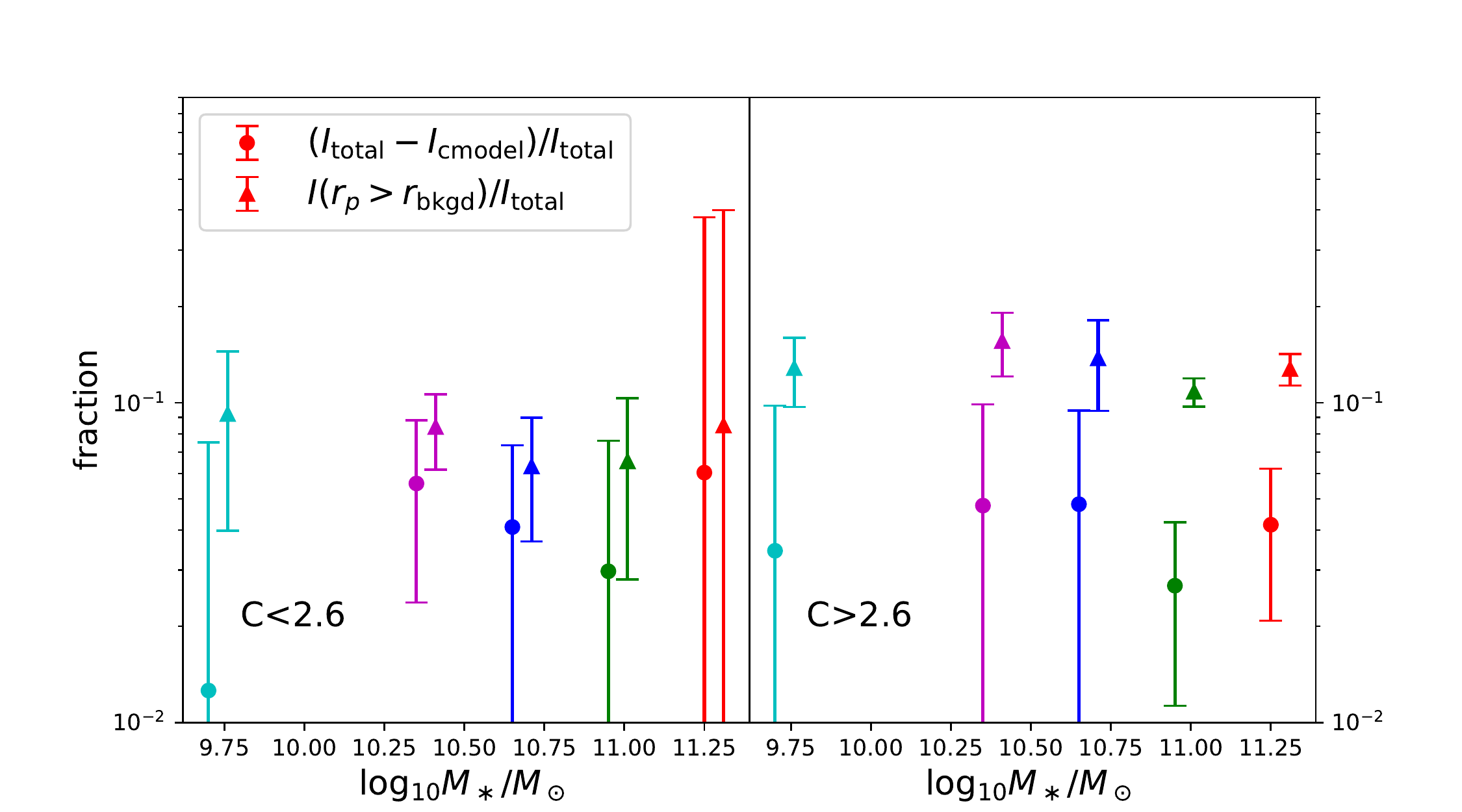}%
\caption{Round dots are the fractions of missing light based on cModel fits to HSC $r$-band surface brightness 
profiles above the average background noise level, and extrapolated to larger scales. Triangles are the fractions 
of light which are under the average background noise level of individual images. Left and right panels are for 
low and high concentration galaxies respectively.
}
\label{fig:fracmiss}
\end{figure*}

Dots in Fig.~\ref{fig:fracmiss} show the fraction of missing light if one relies on the integration over the best-fit 
cModel profiles. Both the best fits and the true profiles are integrated to the last data point of valid non-negative 
measurement. Fractions of emissions that are below the background noise level are overplotted as triangles. Despite 
the large volume in outskirts, the fraction of unresolved emissions falling below the background noise level of 
individual images is on average subdominant compared with the total integrated light (10\% to 20\%). This is because 
the surface brightness profiles drop very quickly as a function of radius. 

The extrapolation of cModel profiles helps to compensate part of the light under background noise level, so the 
fraction is lower than that of the triangles. Given the depth of HSC, it is encouraging that the missed fractions 
are below 10\%. High concentration galaxies show slightly higher fractions of unresolved light below the noise 
level of individual images, which is probably due to the higher fraction of accreted stars in outskirts of high 
concentration galaxies, with respect to the amount of stars formed in-situ through gas cooling 
\citep[e.g.][]{2014MNRAS.443.1433D}. 

\subsection{Possible explanations for the positive colour gradients}
\label{sec:positivegrad}

We have shown in Sec.~\ref{sec:overallprof} that the measured $g-r$ colour profiles have indications 
of colour minimums, beyond which there are signs of positive colour gradients in the outer stellar halo. 
Given the measured extended PSF profiles, we checked the effect of PSF. We found in Sec.~\ref{sec:psfdeconvcolor} 
that the PSF-deconvolved profiles tend to be slightly steeper and hence the indications of colour minimums 
and positive gradients are a bit less prominent, but still exist. 

Colour profiles beyond the minimums have very large errors. Within the errorbars, they are consistent 
with being flat. In the following, our discussions are simply based on the assumption that the colour 
minimums and positive gradients are real, and we discuss possible causes for them. 

Both \cite{2014MNRAS.443.1433D} and our current study found indications of positive colour gradients, though on 
different scales and stellar mass ranges. \cite{2014MNRAS.443.1433D} adopted a source detection threshold of 1.5 
times the background noise for detecting and masking companion sources. In our analysis, we successively applied 
1.5, 2 and 3 times the background noise level as our detection thresholds and masked companion sources for each 
threshold accordingly. The inclusion of higher thresholds helped to mask companions when source deblending fails 
with lower thresholds, but the lowest threshold adopted in our analysis (1.5 times the background noise) is the 
same as \cite{2014MNRAS.443.1433D}. 

We provide detailed investigations in Appendix~\ref{app:maskrandom} to test whether our results are robust given 
different choices of source detection and masking thresholds. Interestingly, we have found, with the decrease in 
thresholds, the positive colour gradients are weakened, while the inner colour profiles within the minimums are 
almost the same. Although such differences are still smaller than the large errors on corresponding scales, the nearly 
monotonic trend indicates the measured positive colour gradients are very likely due to incomplete masking of physically 
associated companions\footnote{For detection thresholds below unity, we only mask footprints associated with 
sources which are also detected by the threshold of 1.5 times the background noise, and thus faked detections of 
background fluctuations are unlikely to affect our conclusions.}. 

It has been shown in many previous studies that the colour of satellite galaxies are redder than their central 
galaxies of the host halo \citep[e.g.][]{2012MNRAS.424.2574W,2014MNRAS.442.1363W}. Incomplete masking of the 
extended emissions from satellite galaxies is hence likely to contaminate the colour of the outer stellar halo. 

Fig.~\ref{fig:profcolorcmpiso} shows that even with the source detection threshold of 0.5 times the background 
noise, the indications of positive colour gradient does not entirely disappear for galaxies with $9.9<\log_{10}M_\ast/M_\odot<10.2$, 
though corresponding errors on such scales are very large. If the positive gradient of that bin could be real, it might 
be indicative of the existence of older stellar populations in the stellar halo. In fact, \cite{2008ApJ...683L.103B} 
have robustly detected colour minimums and upturns in nearby late-type spiral galaxies, which have breaks in their 
surface brightness profiles (truncated or anti-truncated).

It is natural to expect that the colour of galaxies is the reddest in the very central region, where the density 
is extremely high, the possible existence of black holes heats the surrounding gas and the disc instability 
triggers the formation of bulges. These processes can all potentially act to prohibit star formations. 
The feedback effect drops in the outskirts of galaxies and the amount of cold gas increases, which allows for 
more star formations and hence bluer colours.

On larger scales, the extended stellar halo is mainly built up through accretion of merged satellites in the 
current standard structure formation theory of our universe \citep[e.g.][]{2010ApJ...725.2312O,2013MNRAS.434.3348C}. 
After falling into the host halo, satellite galaxies undergo environmental effects such as tidal and ram-pressure 
stripping, which gradually remove their gas reservoir, i.e., the fuel for their star-formation activities. So 
there is less star formation in satellite galaxies than galaxies with comparable stellar masses in the field. 
The stripped population of stars from satellites is thus likely redder than the central galaxy-especially redder 
than the disc of star-forming galaxies. The colour minimum can thus be caused by star formation happening in the 
disc of the central galaxy.

\section{Conclusions}
\label{sec:conclusion}
Stacking images of the Hyper Suprime-Cam Subaru Strategic Program Survey (HSC-SSP), we investigate the properties 
of the faint stellar halo of isolated central galaxies, which are the brightest within the projected halo virial radius 
and three times the virial velocity along the line of sight. Tested against a mock galaxy catalogue of the Munich 
semi-analytical model, we find above 85\% of the isolated central galaxies are true central galaxies of dark matter 
haloes. We estimate and further subtract the residual sky background using similar stacks centred on random points. 
Encouragingly, the random stacks look flat and uniform. 

The deep HSC images and the isolated central galaxy sample allow us to measure the average surface brightness profiles 
of galaxies spanning a wide range in stellar mass ($9.2<\log_{10}M_\ast/M_\odot<11.4$), and out to a projected radius 
of $\sim$120~kpc, with indications of signals extending further behind. The surface brightness in HSC $r$-band 
(without PSF corrections) can be marginally measured down to $\sim$31.5~$\mathrm{mag/arcsec^2}$ (2-$\sigma$ above 
the boot-strap error) for galaxies with $10.8<\log_{10}M_\ast/M_\odot<11.1$. 

The extended PSF wings play an important role, which contaminate the faint stellar halo of galaxies smaller 
than $10^{11.4}M_\odot$. The fraction of contamination ranges from less than 20\% for galaxies with $11.1<\log_{10}M_\ast/M_\odot<11.4$ 
to nearly 90\% for galaxies with $10.5<\log_{10}M_\ast/M_\odot<10.8$ at 120~kpc. For galaxies smaller than $10^{10.5}M_\odot$, 
the measurement at 120~kpc is too noisy, while the extended PSF wings contribute $\sim$60\% at 70~kpc. After 
PSF corrections, we can measure a net surface brightness down to 31~$\mathrm{mag/arcsec^2}$ with 3-$\sigma$ 
significance (corrected from 30~$\mathrm{mag/arcsec^2}$).

More massive galaxies have brighter and more extended surface brightness, and redder colour profiles. The colour 
profiles have negative gradients to $\sim$30~kpc for galaxies more massive than $10^{10.8}M_\odot$, to 
$\sim$20~kpc for galaxies with $9.9<\log_{10}M_\ast/M_\odot<10.8$, and out to $\sim$10~kpc for galaxies smaller 
than $10^{9.9}M_\odot$. There are indications of colour minimums, beyond which the colour profiles tend to have 
positive gradients, though within the errors the profiles on such scales could be consistent with being flat. The 
positive colour gradients, however, are slightly weakened in PSF-deconvolved colour profiles, and are sensitive 
to the chosen detection and masking thresholds for companion sources. Low thresholds help to mask more extended 
emissions from satellite galaxies, and weaken the positive gradients.

After further dividing galaxies into low and high concentration subsamples, we detect distinct features in their 
surface brightness and colour profiles. Low concentration galaxies are more flattened within 20~kpc, but their outer 
stellar halo profiles drop more quickly between 20 and 100~kpc. High concentration galaxies have redder and shallower 
colour gradients. These detections are in good agreement with \cite{2014MNRAS.443.1433D}. However, we fail to see 
indications of colour minimums on similar scales and stellar mass ranges as \cite{2014MNRAS.443.1433D}.

Despite the fact that the surface brightness profiles of individual galaxies show large scatters, we find the average 
surface brightness profiles for galaxies with $10.2<\log_{10}<11.1$ can be modelled by a unified functional form, 
after scaling the projected radius, $r_{p}$, by the halo virial radius, $R_{200}$. The discovery naturally leads to 
the conclusion that the total luminosity, $L$, of galaxies and the stellar halo scales with halo mass, $M_{200}$, in 
the manner of $L\propto M_{200}^{2/3}$, in very good agreement with \cite{2008ApJ...676..248Y}. The result also suggests 
that on top of the large scatter, the formation of galaxy stellar haloes can be on average modelled through physical 
processes that are close to universal once scaled by the host halo scale.

Stacking a large number of images enables us to push below the background noise level for individual images. We quantify 
the fraction of missing light based on the stacked surface brightness profiles which drop below the average background 
noise of single images in HSC. The fractions as a function of stellar mass are between 6\% and 15\% and are slightly 
higher for high concentration galaxies. We fit a composite model of exponential and de Vaucouleurs profiles (cModel) to 
measurements above the background noise of single images, and extrapolate to large scales to obtain the total integrated 
flux. Compared with the true measurements, the integrated cModel fluxes fall below the true measurements by fractions of 
less than 10\%.

Our results are robust to variations in the isolation criteria for sample selection, and we have made detailed investigations 
of the robustness of our results to source detection thresholds used for creating masks. We can validate the good performance 
of the pipeline in terms of removing the sky background and instrumental features in the internal S18a data release.

\section*{Acknowledgements}
Kavli IPMU was established by World Premier International Research 
centre Initiative (WPI), MEXT, Japan. This work was supported by 
JSPS KAKENHI Grant Number JP17K14271. WW is extremely grateful for 
extensive helps and discussions with Zhenya Zheng, Peter Draper, 
Shaun Cole, Nigel Metcalfe and John Lucy, which opened the door to 
image processing. WW is grateful for helps provided by John Good for 
details about image reprojection and the usage of Montage during the 
early stage of this paper, though in the end results are all decided 
to be based on the image resampling module of HSC pipeline instead of 
Montage. WW is also extremely grateful for useful email discussions 
with Richard D'Souza, and discussions with Chunyan Jiang, 
Jun Zhang and Chengze Liu.   
This work has made use of data from the European Space Agency (ESA) mission
{\it Gaia} (\url{https://www.cosmos.esa.int/gaia}), processed by the {\it Gaia}
Data Processing and Analysis Consortium (DPAC,
\url{https://www.cosmos.esa.int/web/gaia/dpac/consortium}). Funding for the DPAC
has been provided by national institutions, in particular the institutions
participating in the {\it Gaia} Multilateral Agreement.
We thank the anonymous referee for his/her careful reading of this paper, 
and rigorous and thoughtful comments on the science, which have led to 
significant improvements compared with the first version.

\bibliography{master}

\appendix

\section{Robustness of our results to variations in source detection thresholds and residual sky background subtraction}
\label{app:maskrandom}

\begin{figure*} 
\includegraphics[width=0.9\textwidth]{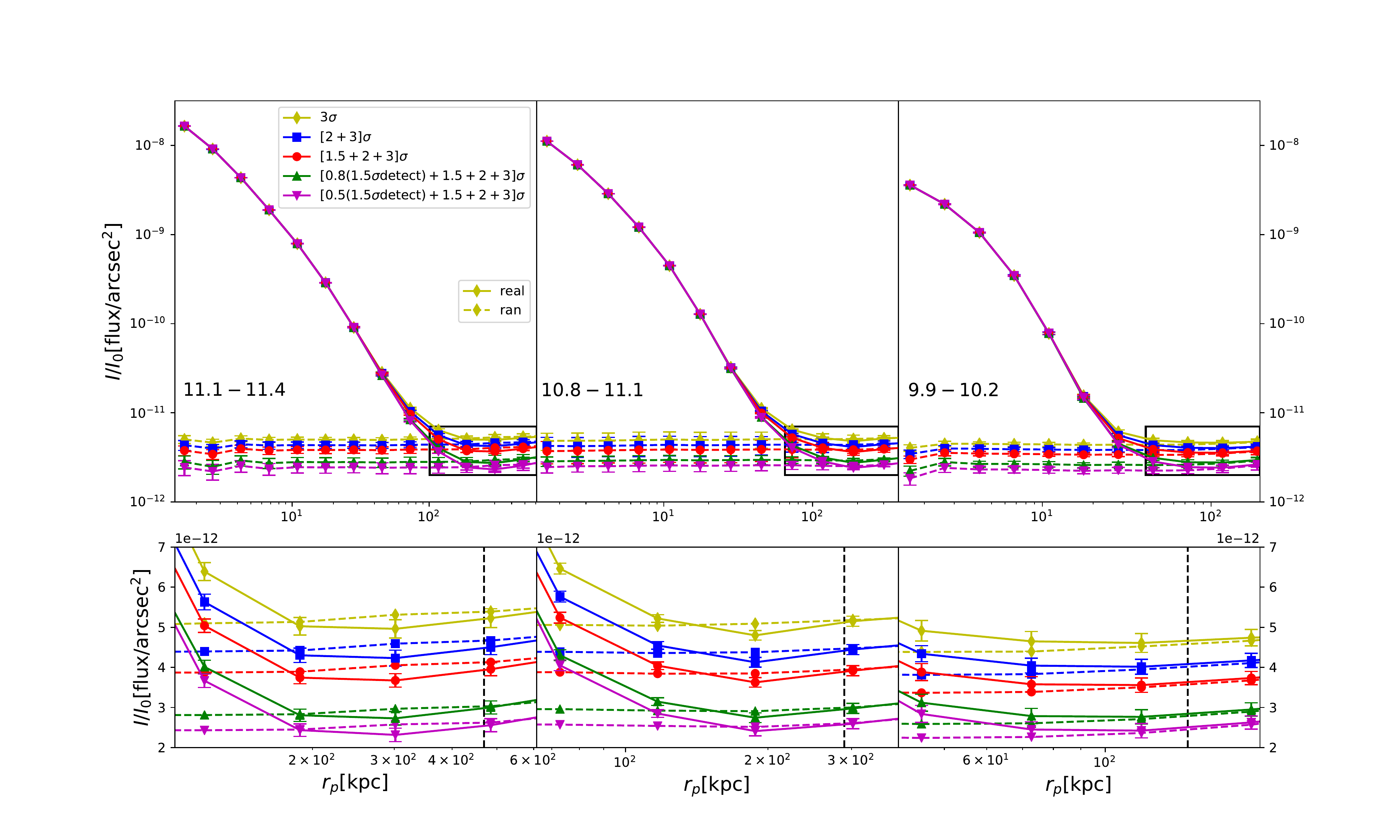}%
\caption{{\bf Top panels:} Symbols connected by solid lines are the stacked surface brightness profiles on isolated central 
galaxies in three different stellar mass bins, and without residual sky background subtractions. Symbols connected by dashed lines 
are the stacked surface brightness profiles on random points.  Errorbars are 1-$\sigma$ scatters among 50 bootstrap samples. 
For each bin, a series of different source detection thresholds have been adopted to mask pixels associated with companion sources, 
as indicated by the legend. In the legend, a number plus $\sigma$ means the adopted source detection threshold with respect to the 
image background noise. $[1.5+2+3]\sigma$ means we successively masked detected footprints of companion sources above 1.5, 2 and 3 
times the background noise. For thresholds below unity ($0.8\sigma$ and $0.5\sigma$), we only use footprints which are also 
associated with detected sources above the $1.5\sigma$ threshold, to avoid masking faked detections falling below the background 
noise. Errorbars are 1-$\sigma$ scatters among 50 bootstrap samples. {\bf Bottom panels:} Regions within the black boxes in top 
panels are zoomed in to highlight the difference between the stacked profiles on real galaxies and random points. We only show 
errorbars for real galaxies (solid curves). The vertical dashed lines are locations of the virial radius, $R_{200}$. 
}
\label{fig:thresran}
\end{figure*}

\begin{figure*} 
\includegraphics[width=0.9\textwidth]{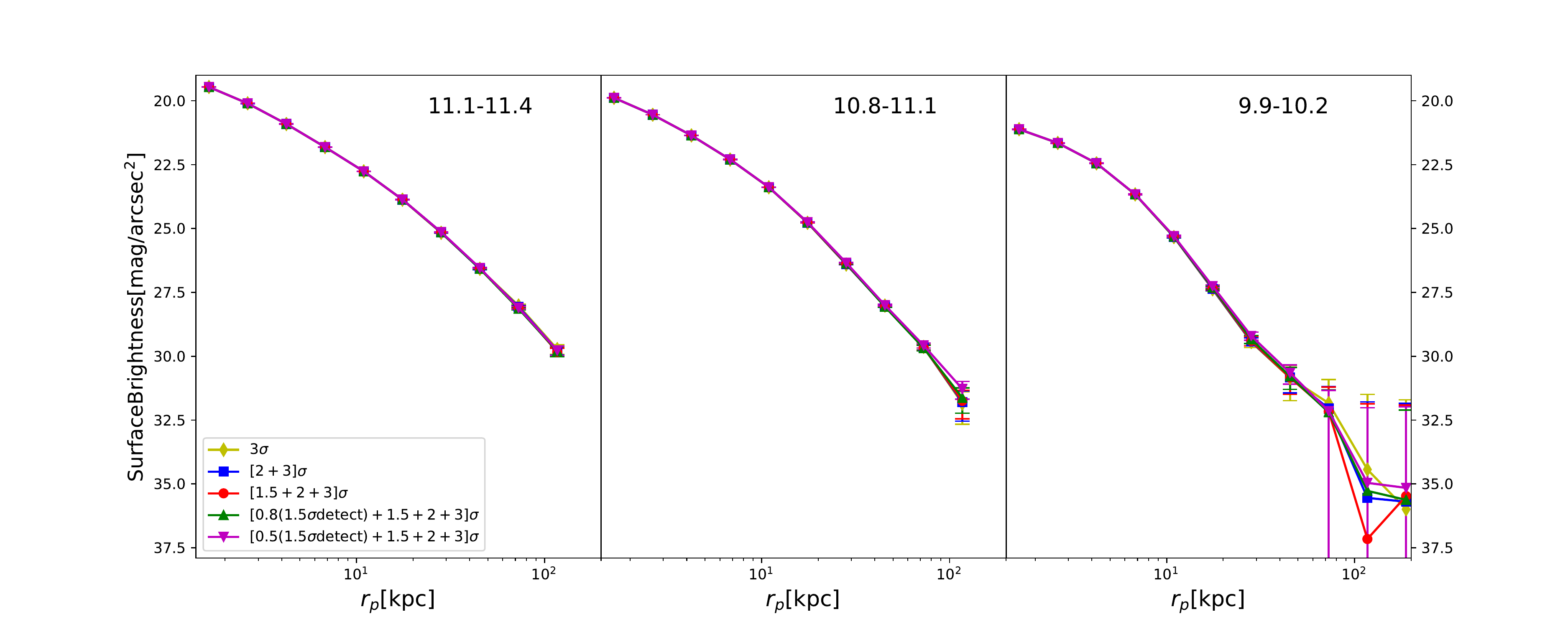}%
\caption{Surface brightness profiles in three stellar mass bins (see the text in each panel). The residual sky background 
estimated from stacked images on random points have been subtracted. Symbols and lines with different colours are based on 
different combinations of source detection thresholds used to create masks for companions (see the legend). The 
differences between measurements in the same stellar mass bin are mostly smaller or comparable to the symbol and error size. 
}
\label{fig:thres}
\end{figure*}

\begin{figure*} 
\includegraphics[width=0.9\textwidth]{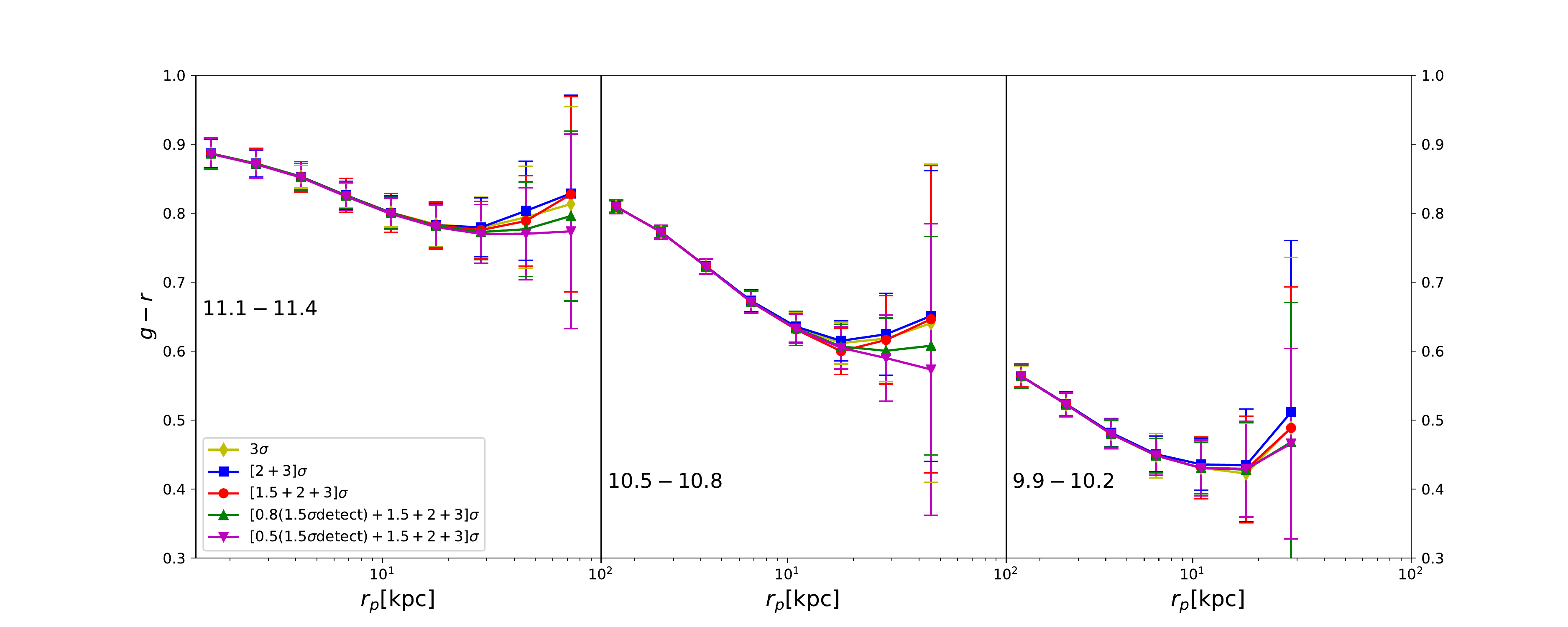}%
\caption{$g-r$ colour profiles in three stellar mass bins (see the text in each panel). Symbols and lines with different colours 
are based on different combinations of source detection thresholds used to create masks for companion sources (see the legend).
}
\label{fig:thres2}
\end{figure*}

The sky background and instrumental features of HSC coadd images have been modelled and subtracted by the pipeline, while the 
residual sky background is not precisely zero. Proper subtractions of the residual sky background are very important for scientific 
studies of diffuse objects. Instead of correcting the residual sky background for each individual image before stacking, we choose 
to do the correction statistically, by applying exactly the same procedures of processing real galaxy images to a sample of random 
points which share exactly the same redshift and image size distributions as real galaxies. The stacked images and surface brightness 
profiles centred on random points can be used as estimates of the residual sky background levels. 

The estimated residual sky background in this way, however, is very sensitive to how sources are detected and masked. Incomplete 
masking of the extended emissions of bright sources can contaminate the true residual sky background. Complete masking of very 
bright stars and their extended emissions is very difficult. We start by trying a few different source detection thresholds to 
test the robustness of our results to the estimated residual sky background. 

Fig.~\ref{fig:thresran} shows the stacked surface brightness profiles before any residual sky background is subtracted, 
for real galaxies (solid curves) and random points (dashed lines). A few different source detection thresholds are adopted 
as indicated by different symbols and colours. Out of the many stellar mass bins analysed in the main part of the paper, we 
choose to show the results for only the two most massive bins and the second to least massive bin. The two most massive bins 
are chosen because we want to investigate the robustness of the faintest magnitudes that we can reach (see Fig.~\ref{fig:profmagr} 
in the main text), and we will also show that there are some slight over-subtractions of the extended emissions in outskirts 
of massive/bright objects. The second to least massive bin is chosen to demonstrate one more case for smaller galaxies. The 
other stellar mass bins do not involve much additional information, so we avoid showing all of them. 

Five different source detection thresholds are adopted. In principle, a lower detection threshold yields more detected 
sources and larger footprints associated with these sources. It seems the lower the detection threshold, the better one can 
estimate the true residual sky background with less contamination from incomplete masking of extended and unresolved sources. 
However, in reality the situation is more complicated because we need to properly mask all companion sources but keep the 
correct footprint for the central galaxy. When the source detection threshold gets lower, the footprint of companions  
can merge with the central galaxy and have to be properly deblended. Unfortunately there is no perfect deblending. If the 
companion fails to be deblended from the central galaxy, emissions from the companion can contaminate the diffuse stellar 
halo. Very strict deblendings can help to reduce such cases, but may also introduce additional problems. For example, some 
substructures (e.g. star-forming regions and spiral arms), which are part of the central galaxy, can be misidentified as 
companions following a more aggressive deblending procedure. 

To avoid the possibilities that companions fail to be deblended and contaminate the stellar halo, we successively run 
\sextractor with a series of different detection thresholds, and mask the corresponding footprints of all detected companions. 
When a lower detection threshold leads to confusion between footprints of centrals and companions, a higher detection threshold 
can help to mask this companion and maintain the correct footprint of central galaxies. Note the masked footprint of companions
with high thresholds can be smaller than its true extended size, which we will discuss later in this section.

In addition to the issue of deblending, when a detection threshold close to or lower than the background noise level 
is adopted, it may introduce many fake detections that are background fluctuations. To avoid such problems, when thresholds 
of 0.8 and 0.5 times the background noise are adopted, we only mask detected footprints which also associate with detections 
made by the threshold of 1.5 times the background noise. 

Compared with real galaxies, it is very encouraging to see that the profiles centred on random points in Fig.~\ref{fig:thresran} 
are close to being flat, indicating that our steps of processing images are successful. However, the random profiles are not exactly 
zero. Because lower detection thresholds lead to more aggressive masking of sources, the profiles centred on random points, which 
are based on unmasked pixels, are as a result lowered. The amplitude keeps dropping with the decrease in source detection thresholds, 
indicating even with 0.8 times the background noise as the threshold, sources are still not fully masked and can contaminate the 
true residual sky background . Encouragingly, though contamination by extended sources to the estimated residual sky background is 
hard to avoid, it is clear from the zoomed-in regions in the bottom panels of Fig.~\ref{fig:thresran} that the random profiles 
trace well the large scale profiles centred on real galaxies. 

However, the profiles centred on real galaxies drop slightly below the random profiles at 300~kpc and at 200~kpc 
in the two left panels of Fig.~\ref{fig:thresran}, which reflects the slight over-subtractions of the extended emissions. 
As have been introduced in Sec.~\ref{sec:S18baksub}, the pipeline adopts a box size of $\sim$1000 pixels to model the sky 
background. For bright objects larger than 1000 pixels, there could still be some over-subtractions. Smaller galaxies, on 
the other hand, do not show evidences of over-subtractions, which is supported by the right panel of Fig.~\ref{fig:thresran}. 
In fact, we will show in Appendix~\ref{app:comprelease} that the amount of over-subtraction has been significantly improved 
compared with older data releases. 

Despite the slight over-subtractions, we can still use the random profiles to correct for the residual sky background. 
We subtract those random stacks from the stacks centred on real galaxies to obtain residual-subtracted images. After 
correcting for the residual sky background in this way, we further rescale the image by adding/subtracting a constant 
value to make the mean value of unmasked pixels at the halo boundary ($R_{200}<r<1.3R_{200}$) equal to zero.

Although the estimated absolute residual sky background level is sensitive to how sources are masked, it is fortunate that 
the differences between stacks centred on random points and those centred on real galaxies on large scales are not sensitive 
to the choice of source detection thresholds or to how sources are masked, as long as we adopt exactly the same source detection 
thresholds for real galaxies and random points. This method is analogous to the correction of systematic shear using random 
stacks in galaxy-galaxy lensing analysis~\citep[e.g.][]{Hirata04,Han15}, and to the correction of selection and boundary effects 
using random catalogues in correlation function analysis of large scale structure. Throughout the paper, we have chosen to adopt 
the $[1.5+2+3]\sigma$ combination of detection thresholds for source masking. Note, the estimated residual sky background levels 
vary between $I/I_0\sim2.5\times 10^{12}$ to $I/I_0\sim5\times 10^{12}$ for different source detection thresholds, which corresponds 
to $\sim$29~$\mathrm{mag}/\mathrm{arcsec}^2$ and $\sim$28~$\mathrm{mag}/\mathrm{arcsec}^2$. If not correcting for the residual 
sky background, 29~$\mathrm{mag}/\mathrm{arcsec}^2$ is about the faintest surface brightness that we can reach. 

While we can use random stacks to account for the residual sky background and incomplete masking of foreground/background 
sources including stars, our measured surface brightness profiles might still be sensitive to how physically associated 
satellites are masked, as the emissions from outer parts of satellites can contaminate the stellar halo. This cannot be 
corrected by using random stacks, because random points do not have physical companions. 

In a recent study of \cite{2019ApJ...874..165Z}, efforts have been spent to estimate the fraction of satellite 
contaminations, by calculating the satellite profiles extending to faint magnitudes, assuming a functional form 
of satellite luminosity function and adopting the contamination fraction due to incomplete masking of Sersic 
profiles \citep{2005PASA...22..118G}. In this paper, we do not attempt to make similar corrections or provide any 
optimal methodology of masking physically associated satellites, but we quantify the underlying uncertainties 
by comparing the surface brightness and colour profiles based on different source detection and masking thresholds. 
The comparisons are shown in Fig.~\ref{fig:thres}, that the surface brightness profiles measured with different 
source detection thresholds almost entirely overlap with each other, and the discrepancies are visible only for 
the last few data points, which are at most about 0.1 to 0.2~dex. 

The colour profiles are more sensitive to the chosen source detection threshold. As shown in Fig.~\ref{fig:thres2}, 
the positive colour gradients become weaker with lower source detection thresholds. Note for the middle panel, we 
choose to show the result for galaxies with $10.5<\log_{10}M_\ast/M_\odot<10.8$ instead, because these galaxies have 
more prominent positive colour gradients in Fig.~\ref{fig:profcolor}. Lower detection thresholds help to mask more 
extended footprints of satellite galaxies. If the colour of satellites are redder than the central galaxy and the 
underlying diffuse stellar halo, more complete masking of satellites can help to bring bluer colour profiles and 
weaken the positive colour gradients. 


\section{A comparison between S15b and S18a}
\label{app:comprelease}

\begin{figure} 
\includegraphics[width=0.49\textwidth]{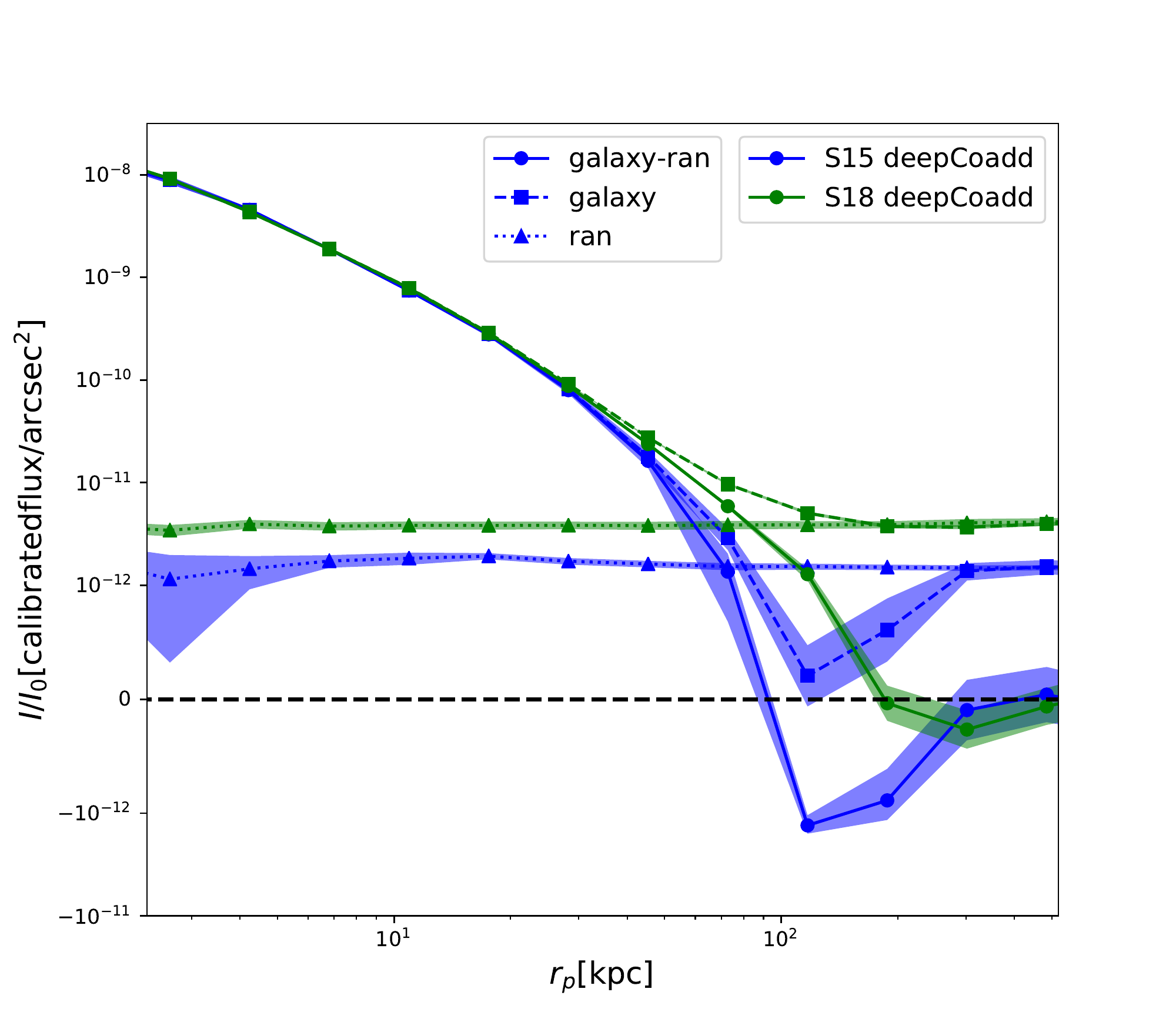}%
\caption{Stacked surface brightness profiles on isolated central galaxies with stellar mass in the range of $11.1<\log_{10}M_\ast/M_
\odot<11.4$. Companion sources detected above 1.5, 2 and 3 times the background noise level have been masked. Blue and green 
lines and symbols correspond to stacks based on S15b and S18a coadd images, respectively. For a given colour, solid line with 
dots, dashed line with squares and dotted line with triangles refer to the final random corrected surface brightness profile, 
the profile centred on real galaxies and on random points, respectively. Errorbars are Poisson errors of photon counting. 
Shaded regions show the 1-$\sigma$ scatter based on 50 boot-strap resampled realisations. The quantity of $y$-axis is 
intensity, $I$, divided by the corresponding zero point intensity, $I_0$. To demonstrate negative values, $y$-axis regions
with $I/I_0>10^{-12}$ and $I/I_0<-10^{-12}$ are displayed in log scales for the absolute values, while linear scales are 
adopted for $-10^{-12}<I/I_0<10^{-12}$. The black dashed horizontal line at $I/I_0=0$ is to guide the eye. 
}
\label{fig:profcmp}
\end{figure}

A comparison of the stacked surface brightness profiles using coadd images of the S15b and S18a internal releases is 
provided in Fig.~\ref{fig:profcmp}, following exactly the same image processing steps in Sec.~\ref{sec:method}. Dashed 
lines with squares are surface brightness profiles centred on real galaxies without correcting for residual sky backgrounds. 
The horizontal dotted lines with triangle symbols are surface brightness profiles centred on random points, which have 
different amplitudes for S18a and S15b. We have already shown in Appendix~\ref{app:maskrandom} that the estimated 
residual sky background of S18a is sensitive to how sources are detected and masked. Here we have chosen exactly the 
same source detection and masking thresholds for S15b and S18a, and the estimated residual sky background in S15b tends 
to be lower than S18a\footnote{In fact, if using detection thresholds smaller than unity to create masks for sources, 
we find the estimated residual sky background of S15b goes negative, whereas it is still positive for S18a. This is due 
to the significant over-subtraction of the extended emissions for bright sources in S15b (see Fig.~\ref{fig:profcmp}). 
When a detection threshold of 0.8$\sigma$ the background noise leaves some unmasked emissions of extended sources 
that go into the estimated residual sky background, such extended emissions have been significantly over-subtracted 
in S15b. The over-subtraction also explains why the estimated residual sky background is lower in S15b.}.

The stacked profile on real galaxies in S15b has a prominent dip at 100~kpc (blue squares connected by the dashed line), 
which is even more prominent after random correction (blue dots connected by the solid line). This is due to the over-subtraction 
of extended emissions for bright objects, and it is very promising to see that the dip is absent in S18a (green symbols and 
lines). For S15b, we interprete the feature as possible over-fitting in the background, which is due to the high order 
polynomial model used for background modelling (or equivalently the fitting  scale). In comparison, the result based 
on S18a is significantly better (see Sec.~\ref{sec:S18baksub}), though as we have mentioned in Appendix~\ref{app:maskrandom}, 
a small level of over-subtraction ($I/I_0\sim10^{-13}$) is still visible for the two most massive bins of isolated 
central galaxies in our analysis.

\section{Robustness of our results to the removal of instrumental patterns}
\label{app:alternative}

\begin{figure} 
\includegraphics[width=0.49\textwidth]{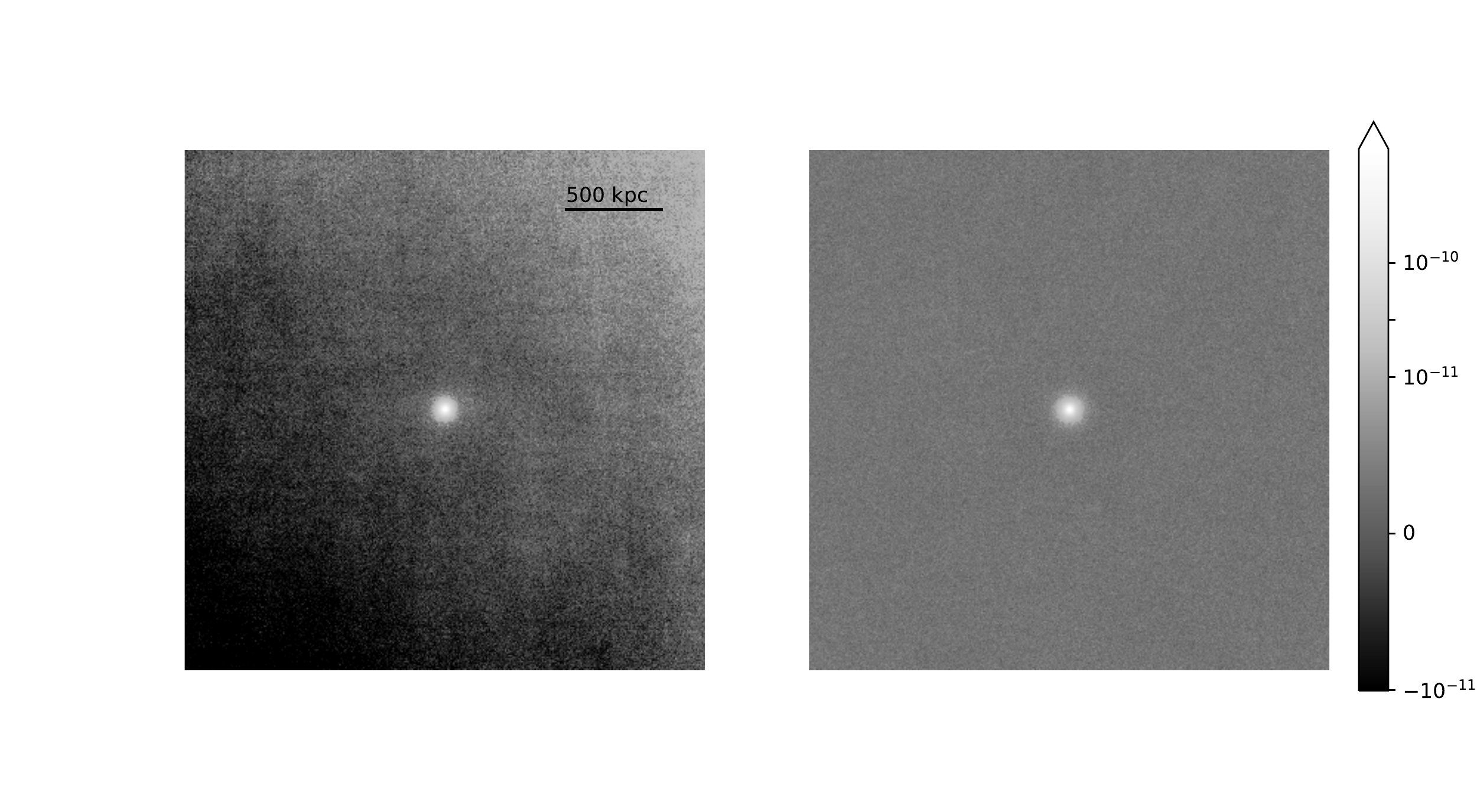}%
\caption{Stacked images (HSC $r$-band) centred on isolated central galaxies in the mass and redshift range of 
$11.1<\log_{10}M_\ast/M_\odot<11.4$ and $0.05<z<0.1$. The left plot is based on Calexp images of the S15b data 
release, before subtracting the sky background and instrumental features. For each Calexp image, we estimate and 
subtract a mean background before stacking. The right plot is based on the sky background and instrumental feature 
subtracted coadds in S18a, and to make fair comparisons, we only use s18a coadd images which are within the 
S15b footprint. Pixel values are intensity divided by zero point intensity ($I/I_0$), i.e., $-2.5\log_{10}I/I_0$ 
gives the surface brightness in unit of magnitude. The colour-map is exactly the same for the two panels, and is 
in log scale for $I/I_0>10^{-11}$ and linear for $I/I_0<10^{-11}$, in order to show negative pixel values. Image 
side length (diameter) is $\sim$2812~kpc, i.e., about six times of the virial radius. 
}
\label{fig:2Dcmp}
\end{figure}

Throughout the paper, we use coadd images of the internal HSC S18a data release for our analysis. The sky 
background and instrumental features have been removed by the pipeline (Sec.~\ref{sec:S18baksub}), which is 
proven to be successful in Fig.~\ref{fig:plate} of the main text. In this section, we provide one more test 
on the robustness of the removal of instrumental patterns. 

We turn back to stack single exposure Calexp images, before any sky background and instrumental features are 
removed. Image cutouts of these single exposure images are at first created centred on isolated central galaxies 
with $11.1<\log_{10}M_\ast/M_\odot<11.4$, and with the image side length of $6R_{200}$.

The general steps of processing, source masking and stacking of coadd and single exposure images are the 
same. The main differences come from the following aspects: 
\begin{enumerate}
\item  A given sky 
region in HSC can be visited for multiple times. The pipeline has coadded all single exposure images together. 
For each galaxy in our analysis, it only corresponds to one coadd image, so the number of galaxies and the 
number of coadd image cutouts are the same. For single exposure Calexp images, one galaxy may correspond to 
multiple images, and we have to stack all these single exposure images for a given sample of galaxies together. 
The number of Calexp images is hence larger than the number of galaxies. 
\item Using Calexp images, we have ignored the second joint calibration step of photometry and astrometry (see 
Sec.~\ref{sec:step} for details). The source position accuracy after joint calibration is improved by about 
25~mas, which is much smaller than the HSC pixel size (0.168$\arcsec$), so ignoring the joint calibration step 
will not significantly affect the science of this study. 
\item The step of mosaic correction (Sec.~\ref{sec:step}) for the nonuniform flat field and for pixel area 
variations is not included in Calexp images, so we need to run the mosaic correction step at first. 
\end{enumerate}

No background models have been removed from these single exposure images, but we estimate and subtract a mean 
background value using unmasked pixels within an annulus of $R_{200}$ and $3R_{200}$ from the galaxy or image 
centre. The true sky background and instrumental features are, however, not a constant, so the mean-subtracted 
images have both positive and negative pixel values for regions with higher and lower background levels. After 
mean background subtraction, we stack all single-exposure image cutouts in exactly the same way as for coadd 
images. Note the images are not rotated before stacking. The result is shown in the left panel of Fig.~\ref{fig:2Dcmp}. 
The final stacked image without removal of sky background and instrumental features tend to have a large scale 
gradient, that the image is fainter in the bottom left corner and brighter in the top right. 

The gradient is mainly a result of the instrumental features in Fig.~\ref{fig:plate}. To prove it, we spread a 
large realisation of random points over the plate of the left panel of Fig.~\ref{fig:plate}, and centred on each point, 
we extract a cutout, whose edge length is randomly drawn from the edge length distribution of our galaxies. We stack 
all the cutouts together, and the final stacked image shows a very similar gradient. Rotating images before stacking 
will help to ``hide'' such a gradient, but does not help to remove the residuals.

As a comparison, we show the stacked image centred on exactly the same sample of galaxies, but based on coadd 
images of S18a in the right panel of Fig.~\ref{fig:2Dcmp}. It is very encouraging to see that such a gradient 
is absent. Note we have used exactly the same colour map for the two panels. The test proves the robustness of 
the removal of instrumental features by the pipeline. We also note that though the comparison is made between 
Calexp images in S15b and a subset of coadd images in S18a which overlap with the S15b footprint, the same 
conclusion is valid if we compare S18a with the internal releases of S16 and S17, because S16 and S17 adopt 
the same approach of removing sky background and instrumental features as S15b.

\section{Robustness of our results to variations in isolation criteria}
\label{app:isocmp}

\begin{figure} 
\includegraphics[width=0.49\textwidth]{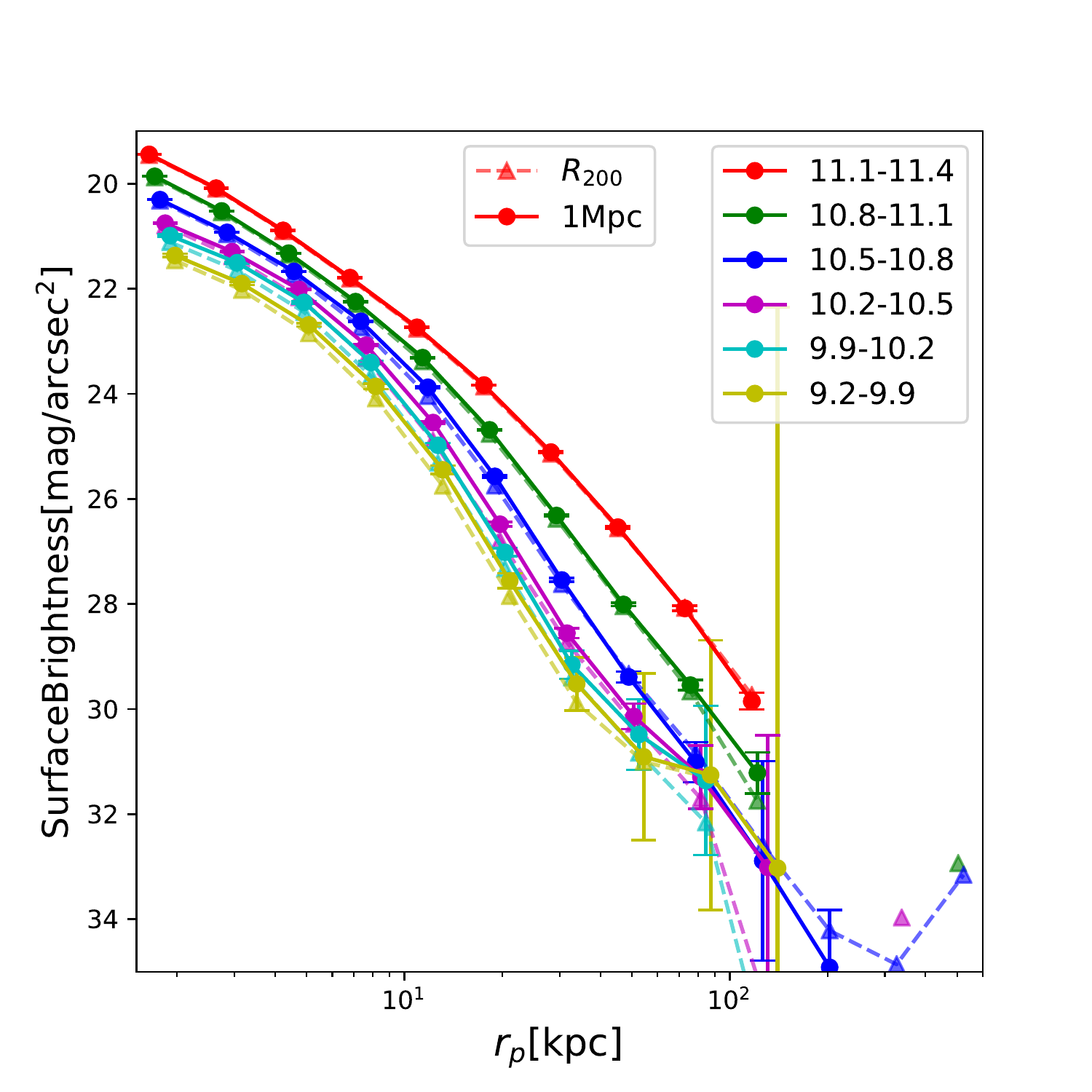}%
\caption{Solid lines and round dots are stacked surface brightness profiles in HSC $r$-band on a sample of isolated central 
galaxies that are brightest within 1~Mpc and 1000~km/s along the line of sight. Errorbars are boot-strap errors based on 50 
realisations. Triangles connected by dashed lines without errors are exactly the same as the solid lines in Fig.~\ref{fig:profmagr}, 
which are based on the sample of isolated central galaxies that are brightest within $R_{200}$ and three times virial velocity 
along the line of sight, and are overplotted for direct comparisons. Small horizontal shifts have been added to the second to 
least massive bins, to better display the errorbars.
}
\label{fig:profcmpiso}
\end{figure}

\begin{figure} 
\includegraphics[width=0.49\textwidth]{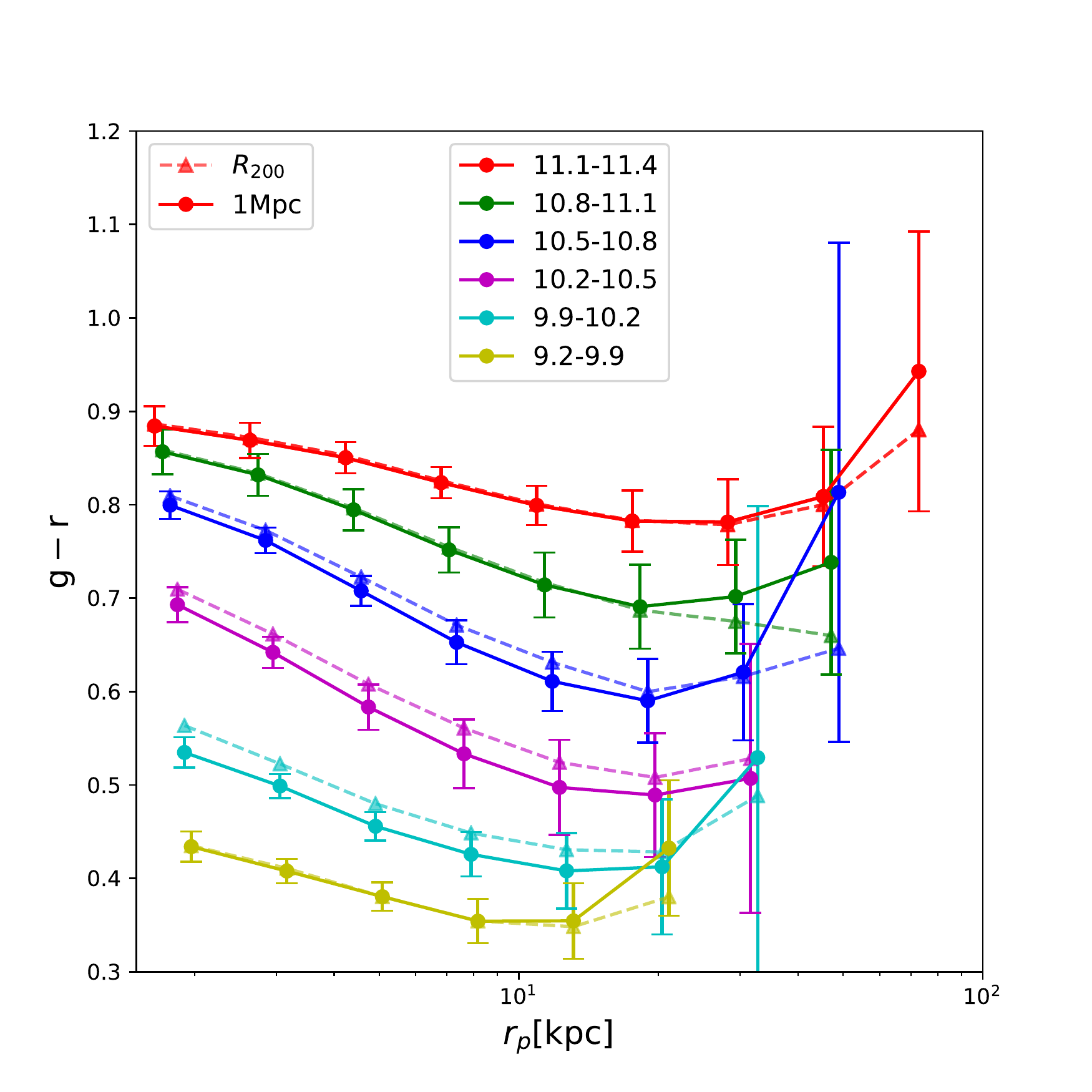}%
\caption{Solid lines and round dots are $g-r$ colour profiles stacked on isolated central galaxies that are brightest 
within 1~Mpc and 1000~km/s along the line of sight and are with $0.05<z<0.1$. Errorbars are boot-strap 
errors based on 50 realisations. Triangles connected by dashed lines without errors are exactly the same as the measurements 
in the left plot of Fig.~\ref{fig:profcolor}, which are based on the sample of isolated central galaxies that are brightest 
within $R_{200}$ and three times virial velocity along the line of sight. Small horizontal shifts have been added to the 
second to the least massive bins, to better display the errorbars.
}
\label{fig:profcolorcmpiso}
\end{figure}

\begin{table}
\caption{Total number of isolated central galaxies that are brightest within 1~Mpc in projected distance and 1000~km/s along 
the line of sight, and within the footprint of S18a internal data release. }
\begin{center}
\begin{tabular}{lrrrrrrr}\hline\hline
$\log M_*/M_\odot$ & \multicolumn{1}{c}{$N_\mathrm{galaxy}$} \\ \hline
11.1-11.4 & 1318  \\
10.8-11.1 & 3846  \\ 
10.5-10.8 & 3232  \\
10.2-10.5 & 1456  \\
9.9-10.2 & 518 \\
9.2-9.9 & 228\\

\hline
\label{tbl:iso1Mpc}
\end{tabular}
\end{center}
\end{table}

Throughout the main text of this paper, the sample of galaxies are selected to be the brightest 
within the projected virial radius, $R_{200}$, and three times the virial velocity along the line of sight. This 
gives us a large sample, but the purity of halo central galaxies is about 85\% at $9.2<\log_{10}M_\ast/M_\odot<11$. 
In this section we test our results using a sample of isolated central galaxies selected with more rigorous criteria 
at the small mass end and hence have significantly higher purity of halo central galaxies at $9.2<\log_{10}M_\ast/M_\odot<11$. 
This sample is selected to be the brightest within a projected distance of 1~Mpc and 1000~km/s along the line of sight. 

The reader can find details about the purity and completeness for this sample in \cite{2016MNRAS.457.3200M}.
Basically, the 1~Mpc selection gives a sample purity that is above 90\% for galaxies with $\log_{10}M_\ast/M_\odot<10.8$. 
At $9.2<\log_{10}M_\ast/M_\odot<10.$, the purity is nearly 100\%, whereas the completeness is only $\sim$40\%. This is 
because the scale of 1~Mpc is larger than the mean halo virial radius at $\log_{10}M_\ast/M_\odot<11.5$, and 1000~km/s 
is comparable to three times the mean virial velocity for galaxies with $\log_{10}M_\ast/M_\odot\sim11.1$. 
Selecting galaxies with this more stringent criteria significantly helps to increase the sample purity at the 
small mass end.

Fig.~\ref{fig:profcmpiso} shows the stacked surface brightness profiles in HSC $r$-band for this sample, and results 
in the main text are overplotted for direct comparisons. The numbers of isolated central galaxies selected with 
the more strict criteria and are within the S18a footprint are provided in Table~\ref{tbl:iso1Mpc}. Compared with 
Table~\ref{tbl:isos}, the sample size is smaller by factors of 1.1, 1.3, 1.7, 2.3, 3.0 and 3.5 from the most to 
least massive stellar mass bins. The two samples are similar at the massive end. Solid and dashed lines for the 
two most massive bins are almost identical. The discrepancy becomes slightly larger for smaller mass bins. The profile 
of the more strictly selected sample is a bit less extended in the third bin, and the amplitudes of the profiles are 
slightly higher for galaxies smaller than $10^{10.5}M_\odot$. We have checked the mock galaxy catalogue introduced 
in Sec.~\ref{sec:isogal}, and found more strictly selected isolated central galaxies tend to have slightly larger 
stellar masses and hence a bit higher amplitude in their surface brightness profiles. Despite these small differences, 
the two sets of results are very similar to each other. 

The colour profiles are provided in Fig.~\ref{fig:profcolorcmpiso} for this new sample of galaxies, 
to be compared with the results in the main text (overplotted as triangles with dashed lines). The agreement 
is good, indicating our measurements are robust to how the sample of isolated central galaxies are selected.

\section{Robustness of our results to circular and elliptical binning in radius}
\label{app:ell}

\begin{figure} 
\includegraphics[width=0.49\textwidth]{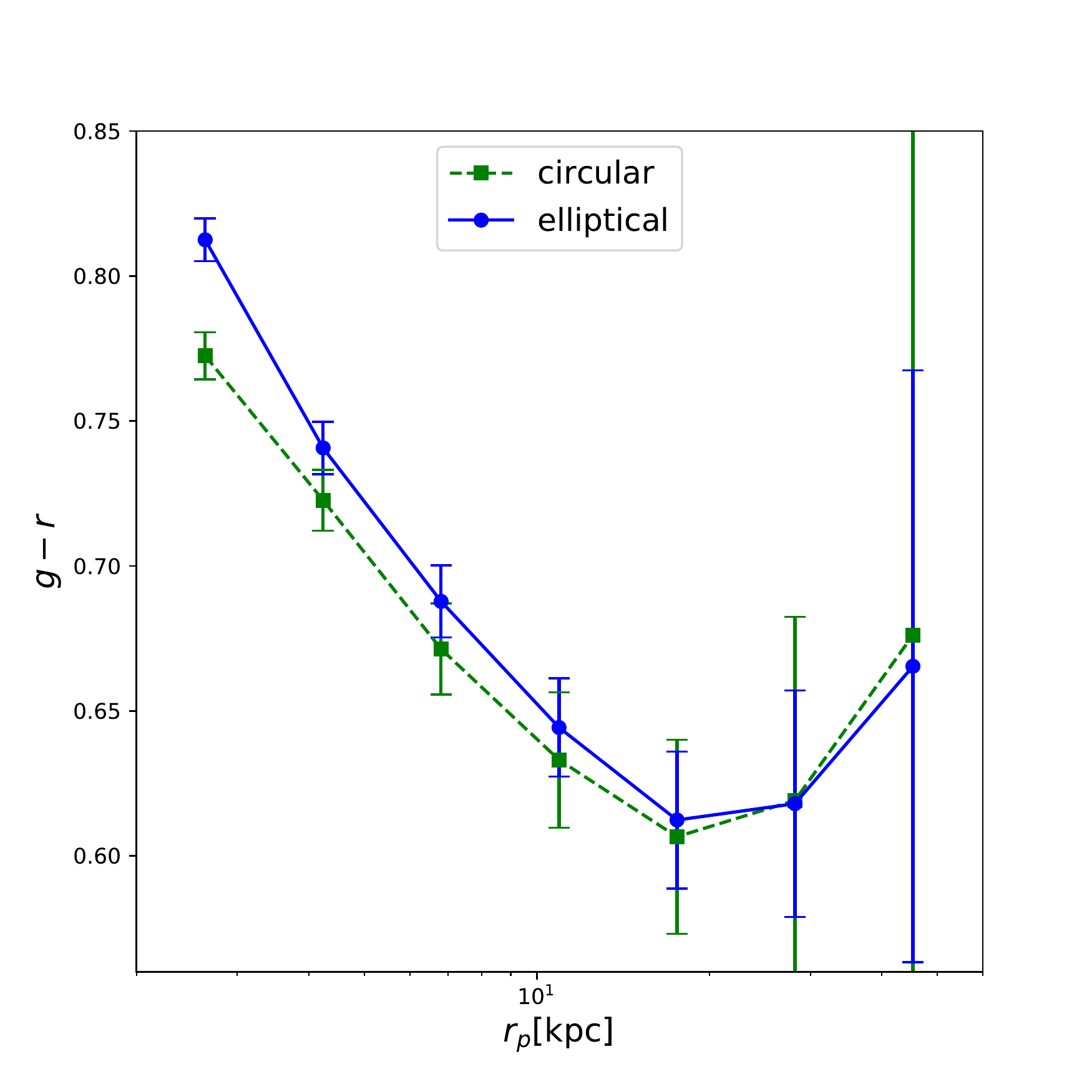}%
\caption{$g-r$ colour profiles stacked on isolated central galaxies with $10.5<\log_{10} M_\ast/M_\odot<10.8$ and in 
the redshift range of  $0.05<z<0.1$. Blue dots connected by the solid line are obtained by rotating galaxy images to 
make them aligned with their semi-major axis before stacking, and the measured profiles are binned using elliptical 
intensity contours centred on galaxies, reported as a function of the semi-major axis of these isophotal ellipses. 
Green squares connected by the dashed line are based on circular binning adopted throughout the paper. 
}
\label{fig:elltest}
\end{figure}

Throughout the paper, we adopted circular radial binning to calculate the surface brightness and colour profiles 
as a function of the radius. There is no preferable angular direction, and the obtained profiles are isotropic. Many 
previous studies, however, rotate galaxy images to make them aligned with the major axis before stacking, and the profiles 
are calculated based on intensity contours of elliptical shapes, reported as a function of the length of the major axis of 
these ellipses. It is thus necessary to check whether the slopes and minimums in $g-r$ colour profiles are sensitive to 
image alignment and the way of binning.

Fig.~\ref{fig:elltest} shows the $g-r$ colour profile for galaxies with $10.5<\log_{10} M_\ast/M_\odot<10.8$ and 
within the redshift range of $0.05<z<0.1$. Green squares connected by the dashed lines are the same results as in the 
main text, based on circular radial binning. Blue dots connected by the solid line is based on elliptical intensity 
contours, with galaxy images rotated along their major axis before stacking, based on the orientation angle returned 
by \sextractor. To draw the elliptical isophote, we use the python \textsc{photutils} package. The elliptical contours 
are at first made to HSC $r$-band, and applied to HSC $g$-band to ensure exactly the same set of elliptical isophote 
being applied to different bands. Note we do not smooth the image as \cite{2014MNRAS.443.1433D}. 

Because $r_p$ stands for the semi-major axis of ellipses, the effective radius averaged along the ellipse should 
be less than the value of $r_p$ when circular radial binning is adopted. Thus with elliptical binning, it is natural 
to see redder colour profiles when the profile gradient is negative, and bluer colours when the gradient is positive. 
The gradient within 10~kpc seems to be slightly steeper than that from circular radial binning. Note the innermost 
data point of Fig.~\ref{fig:profcolor} in the main text is not shown here, because we cannot have converged estimate 
of an elliptical contour at that radius, given a very small number of pixels. In addition, the elliptical binning seems 
to slightly weaken the positive colour gradient and the colour minimum, though the difference between blue dots and 
green squares is much smaller than the large errors on such scales.

We conclude that circular radial binning can bring slightly bluer colour and a bit flattened slopes compared 
with elliptical intensity contours, if the colour gradient is negative. In addition, since elliptical binning 
requires galaxies to be aligned along their major axis before stacking, this will reduce the amount of 
scatter among different galaxy images due to the different orientation of semi-major axes. As a result, the 
errorbars of elliptical binning are smaller beyond 10~kpc in Fig.~\ref{fig:elltest} than circular binning.

\end{document}